\documentclass[preprint,12pt, 3p,times]{elsarticle}

\usepackage{hyperref}

\usepackage{amsmath}
\usepackage{makeidx}
\usepackage{graphics}
\usepackage{color}
\usepackage{listings}
\usepackage{bm}
\usepackage{url}
\usepackage{booktabs}
\usepackage{paralist}
\usepackage{amssymb}
\usepackage{amsthm}
\usepackage{subcaption}
\usepackage{algorithm}
\usepackage{algorithmic}

\usepackage[percent]{overpic}

\usepackage{longtable}
\usepackage{dsfont}

\biboptions{sort&compress}
\graphicspath{ {fig/} }

\newcommand*\diff{\mathop{}\!\mathrm{d}}

\begin{document}

\begin{frontmatter}

\title{Relativistic Lattice Boltzmann Methods: 
       Theory and Applications.
       }

\author{A. Gabbana \fnref{corr-author} }
\address{Universit\`a di Ferrara and INFN-Ferrara, I-44122 Ferrara,~Italy}
\address{Bergische Universit\"at Wuppertal, D-42119 Wuppertal,~Germany}
\author{D. Simeoni}
\address{Universit\`a di Ferrara and INFN-Ferrara, I-44122 Ferrara,~Italy}
\address{Bergische Universit\"at Wuppertal, D-42119 Wuppertal,~Germany}
\address{University of Cyprus, CY-1678 Nicosia,~Cyprus}
\author{S. Succi}
\address{Center for Life Nano Science @ La Sapienza, Italian Institute of Technology, Viale Regina Elena 295, I-00161 Roma,~Italy}
\address{Istituto Applicazioni del Calcolo, National Research Council of Italy, Via dei Taurini 19, I-00185 Roma,~Italy}
\author{R. Tripiccione}
\address{Universit\`a di Ferrara and INFN-Ferrara, I-44122 Ferrara,~Italy}
\fntext[corr-author]{Corresponding author. E-mail address: \href{mailto:alessandro.gabbana@unife.it}{alessandro.gabbana@unife.it}}

\begin{abstract}
  We present a systematic account of recent developments of the relativistic
  Lattice Boltzmann method (RLBM) for dissipative hydrodynamics.
  We describe in full detail a unified, compact and dimension-independent
  procedure  to design relativistic  LB schemes capable of bridging the gap
  between the ultra-relativistic regime, $k_{\rm B} T \gg mc^2$, and the
  non-relativistic one, $k_{\rm B} T \ll mc^2$.
  We further develop a systematic derivation of the transport coefficients as a
  function of the kinetic relaxation time in $d=1,2,3$ spatial dimensions. The
  latter step  allows to establish a quantitative bridge between the parameters
  of the kinetic model and the macroscopic transport coefficients. 
  This leads to accurate calibrations of simulation parameters and is also relevant
  at the theoretical level, as it provides neat numerical evidence of the correctness 
  of the Chapman-Enskog procedure.
  We present an extended set of validation tests, in which simulation results
  based on the RLBMs are compared with existing analytic or semi-analytic
  results in the  mildly-relativistic ($k_{\rm B} T \sim mc^2$) regime for the case of
  shock propagations in quark-gluon plasmas and laminar electronic flows in
  ultra-clean graphene samples.    
  It is hoped and expected that the material collected in this paper may allow
  the interested readers to reproduce the present results and generate new
  applications of the RLBM scheme.
\end{abstract}

\end{frontmatter}

\newpage

\tableofcontents


\newpage

\section{Introduction}\label{sec:intro}

Relativistic hydrodynamics and kinetic theory are comparatively mature
disciplines, whose foundations have been laid down more than half a century
ago, with the seminal works of Cattaneo, Lichnerowitz, Landau-Lifshitz, Eckart, Mueller,
Israel and Stewart, to name but a few major pioneers 
\cite{cattaneo-book-2011,lichnerowicz-book-1967,eckart-prl-1940,muller-zphys-1967,landau-book-1987,israel-anp-1976,israel-prsl-1979}.  

Relativistic hydrodynamics deals with the collective motion of material bodies which move close
to the speed of light, hence they have been traditionally applied mostly to large-scale
problems in space-physics, astrophysics and cosmology \cite{degroot-book-1980,rezzolla-book-2013}.

In the last ten-fifteen years, however, relativistic hydrodynamics and kinetic
theory have witnessed a tremendous outburst of activity outside their traditional
context of astrophysics and cosmology, particularly at the fascinating
interface between high-energy physics, gravitation and condensed matter,
(for a very enlightening review, see the recent book by Romatschke\&Romatschke \cite{romatschke-book-2019}).

In particular, experimental data from the Relativistic Heavy-Ion Collider (RHIC) and the 
Large Hadron Collider (LHC), have significantly boosted the interest for the study of viscous relativistic 
fluid dynamics, both at the level of theoretical formulations and also for the development of efficient 
numerical simulation methods, capable of capturing the collective behaviour 
observed in Quark Gluon Plasma (QGP) experiments (see \cite{florkowski-rpp-2018} for a recent review), 
down to the "smallest droplet ever made in the lab", namely a fireball of QGP just 
three to five protons in size \cite{phenix-natphys-2019}.
Relativistic hydrodynamics has also found numerous applications in
condensed matter physics, particularly for the study of strongly correlated
electronic fluids in exotic (mostly 2-d) materials, such as graphene sheets and Weyl
semi-metals (see \cite{lucas-jopcm-2018} for a recent review).
Last but not least, gravitational wave observations from LIGO/VIRGO have added
to the picture by providing measurements of black-hole non-hydrodynamic modes, as
well as neutron star mergers, will likely play a key role in calibrating
future relativistic viscous fluid dynamics simulations of compact stars.
Hence, we appear to experience a truly golden-age of relativistic hydrodynamics!

From the theoretical side, a major boost has been provided
by the famous AdS/CFT duality, which formulates a constructive equivalence between gravitational
phenomena in $d+1$ dimensions and an associated field theory, living on the corresponding
$d$ dimensional boundary \cite{maldacena-ijtf-1999}.

More specifically, gravitational analogues of fluids have been
discovered, which permit to formulate strongly interacting d-dimensional fluid problems 
as weakly interacting (d+1)-dimensional gravitational ones, and viceversa, whence the name of holographic fluid dynamics.
Besides experimental excitement for genuinely new, extreme and exotic states of matter, this
state of affairs has also raised very fundamental theoretical challenges.

On the one side, holographic fluids stand as the graveyard of kinetic theory, since they interact
strongly to the point of invalidating the very cornerstone of kinetic theory, namely
the notion of quasiparticles as weakly interacting collective degrees of freedom.
Particles interact so strongly that they are instantaneously frozen to local equilibrium, 
thereby bypassing any kinetic stage.

On the other side, different studies, especially quark-gluon plasmas experiments with small
systems, have highlighted the unanticipated ability of hydrodynamics to describe extreme
nuclear matter under strong gradients, hence far from local equilibrium, which we
refer to as to the beyond-hydro (BH)regimes \cite{nagle-arofnps-2018}.

Both findings call for a conceptual extension of the notion
of hydrodynamics, to be placed in the more general framework of an 
effective field theory of slow degrees of freedom.
This implies a corresponding paradigm shift towards an effective kinetic theory,  
to be --designed-- top-down from the effective fields equations, rather than
being --derived-- bottom-up from an underlying microscopic theory of quantum relativistic fields.

By fine-graining the macro-equations instead of coarse-graining the microscopic ones, one can
indeed enrich the hydrodynamic formulation with the desired BH features “on demand”,
i.e tailored to the specific problem at hand, without being trapped by obscure and often
intractable microscopic details.

Among others, a major bonus of the kinetic approach is that dissipation comes with built-in causality, since, by
construction, kinetic theory treats space and time on the same footing, i.e. both first order.
This stands in stark contrast with a straightforward extrapolation
from non-relativistic fluid dynamics, in which, to leading order,
dissipation is represented by second order derivatives, which imply
infinite speed of propagation, hence breaking causality.
The problem was of course spotted since the early days of relativistic kinetic theory, and
mended by replacing static constitutive relations with first-order (hyperbolic) dynamic
relaxation equations for the momentum-flux and heat tensors, the Maxwell-Cattaneo formulation \cite{cattaneo-book-2011},
possibly the most popular version being the one due to Israel and Stewart \cite{israel-anp-1976,israel-prsl-1979}.

Nevertheless, such formulations remain empirical in nature, hence they leave some ambiguity
as to the actual relation between the relaxation time scales and
the actual value of the transport coefficients.

On the other hand, as we shall detail in this Review, kinetic theory is amenable to highly efficient
lattice formulations which allow to simulate complex relativistic flows, thereby
providing a numerical touchstone for the various analytical/asymptotic theories to compare with.

At this point, it is worth noting that the lattice kinetic program has been in action for nearly
three decades in the context of classical (non-relativistic) fluids, and with a rather
spectacular success across many scales of motion, from macroscopic turbulence, all the way
down to micro and nanofluids of biological interest \cite{succi-epl-2015,succi-book-2018},  
the name of the game being Lattice Boltzmann (LB).

LB provides a computationally efficient instance of effective field theory,
whereby the effective degrees of freedom are selected by taking full
advantage of the smoothness and symmetries of momentum space.

Lattice kinetic theory has indeed been extended to the relativistic case
through a series of papers, starting with Mendoza et al. \cite{mendoza-prl-2010}
and subsequently refined and extended in the last decade.

Yet, this is an unfinished program, and in this review, after discussing the
historical developments of relativistic lattice Boltzmann (RLB), we shall outline
future directions to go in order to accomplish the task of turning RLB into
an operational tool to advance knowledge in relativistic hydrodynamics, the
way that LB has been doing in the non-relativistic framework. 

This work is structured as follows: after this Introduction, 
Sec. \ref{sec:background} introduces at a more technical level the state-of-the-art 
and outlines the major results obtained in the last two decades.
Sections \ref{sec:ideal-hydro} and \ref{sec:dissipative-hydro} briefly review non-dissipative 
and dissipative relativistic hydrodynamics, with a close look at the link between meso-scale 
parameters and transport coefficients; Sec. \ref{sec:rlbm} describes in details the 
construction of our relativistic RLBMs, at the theoretical and numerical level. 
This is followed by Sec. \ref{sec:numericalRec}, presenting more practical details 
on the development of RLBM codes. We then proceed with two sections discussing numerical 
results: Sec. \ref{sec:calibration} presents numerical evidence in favour of the 
Chapman-Enskog approach, while \ref{sec:bench} discusses several validation exercises 
in several application areas and kinematic regimes. This is followed by concluding 
remarks and by several Appendices, with detailed mathematical derivations and results. 
The bulkiest mathematical results are made available as supplemental material.

\section{Background} \label{sec:background}
Following a standard approach, relativistic hydrodynamics can be formulated as a gradient expansion of 
relativistic kinetic theory, whose zeroth-order corresponds to ideal hydrodynamics. 
The formulation involving relativistic inviscid fluids is well established and widely used in astrophysics,
but is not adequate to explain the quantitative behaviour of e.g. experimental observables in the QGP evolution 
for which, due to the ultra-high density conditions, dissipative effects need to be taken into account.
First-order dissipative theory, known as relativistic Navier-Stokes (RNS) equations, are inconsistent with 
relativistic invariance, because second order derivatives in space and first order in time imply superluminal 
propagation, hence non-causal and unstable behaviour.

These problems have long been recognised, and several attempts to cure the
issues of the RNS formulation have been proposed to this day. 
Historically, the hyperbolic formulation proposed by Israel and Stewart (IS) \cite{israel-anp-1976,israel-prsl-1979} 
has been the first and most widely used formalism able to restore causal dissipation, and 
has served as the reference frame for several decades.
However, recent work has highlighted both theoretical shortcomings \cite{denicol-prd-2012} of 
the IS formulation, as well as poor agreement with numerical solutions of the Boltzmann equation 
\cite{huovinen-prc-2009, bouras-prc-2010,florkowski-prc-2013}.
As a result, in the recent years intense work has been directed to the definition of complete and self-consistent relativistic fluid-dynamic equations
\cite{muronga-prc-2007,muronga-prc-2007b,betz-epj-2009,el-prc-2010, denicol-prl-2010,betz-epj-2011,denicol-prd-2012,
jaiswal-prc-2013,jaiswal-prc-2013b,jaiswal-prc-2013c,bhalerao-prc-2013,bhalerao-prc-2014,chattopadhyay-prc-2015};
the debate is still very much open, no unique model having emerged to date.

In this context, the kinetic approach offers several advantages for the study of
dissipative hydrodynamics in relativistic regimes. One of its key assets is
that the emergence of viscous effects does not  break relativistic invariance
and causality, because space and time are treated on the same footing, i.e.
both  via first order derivatives (hyperbolic formulation). This overcomes many
conceptual issues associated with the consistent formulation of relativistic
transport phenomena.
The second key aspect is that non-linear advection in configuration space is replaced by linear streaming in
phase space, an operation that can be implemented error-free in the lattice, with major benefits for the
numerical treatment of low-viscous regime, such as the ones characterizing strongly-interacting fluids.
%
\begin{figure}[tbh]
  \centering
  \includegraphics[width=0.85\textwidth]{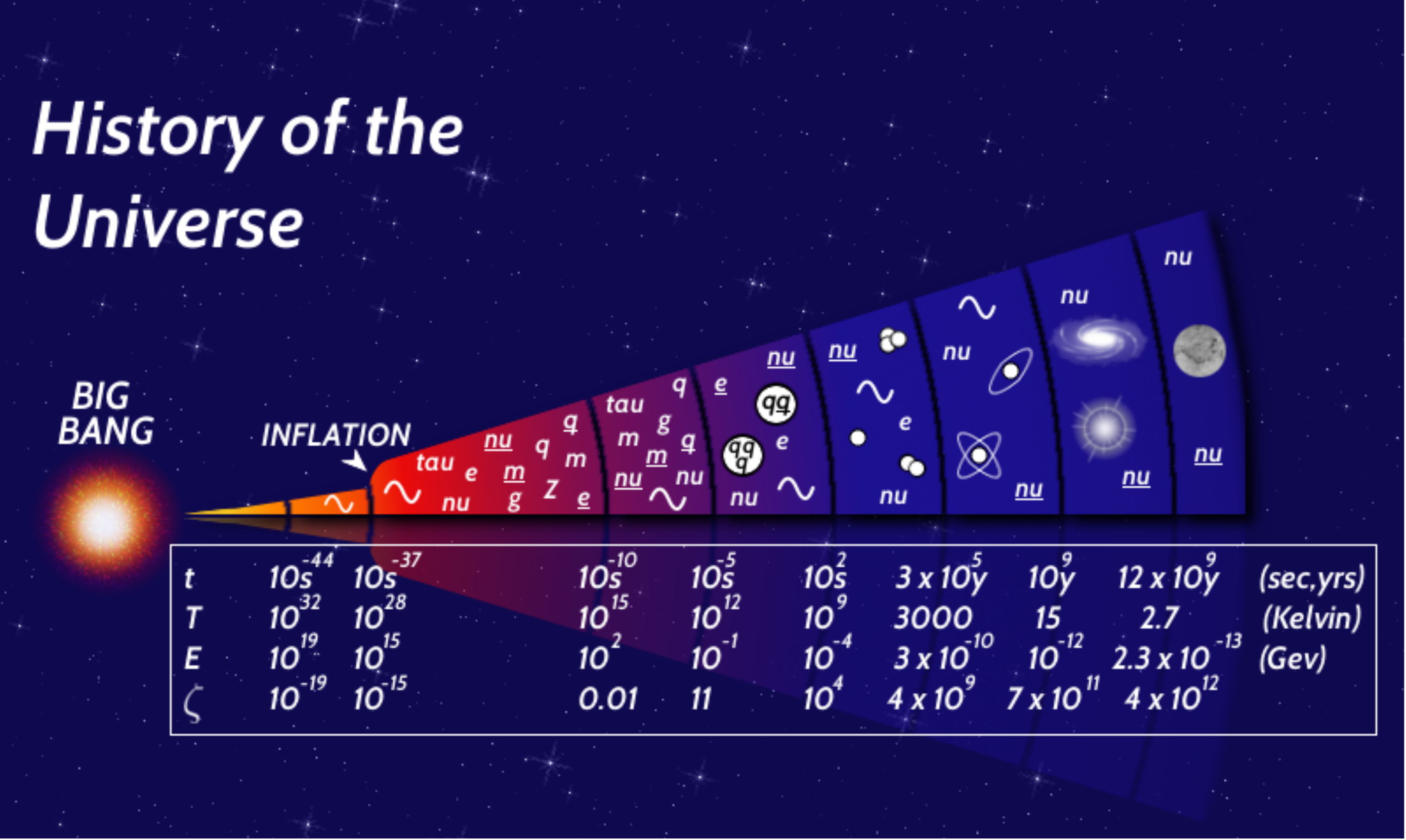}
  \caption{ 
            Sketch of the history of the universe as a function of time and temperature.
            As the universe evolves in time away from the big-bang, typical values of $\zeta$ for 
            the proton change from about $10^{-19}$ at the Planck time to $10^{12}$ at the present day.
            Besides, a sequence of thermodynamic transitions takes place, starting from 
            the electroweak transition ($\sim~10^{-9}~\rm s$), followed by the QCD transition 
            ($\sim~10^{-6}~\rm s$) and by the $e^{+} e^{-}$ annihilation ($\sim~10^{2}~\rm s$) \cite{schwarz-adp-2003}.
          }\label{fig:universe}
\end{figure}
%

Relativistic Lattice Boltzmann schemes (usually referred
to as Relativistic Lattice Boltzmann Methods (RLBMs) ) enter the game as a conceptual path to the 
construction of specific instances of kinetic models  and as a computer-efficient numerical approach 
to the problem, meeting the obvious need of developing efficient and accurate
numerical relativistic hydrodynamic solvers. 
These numerical tools are a clear must, as analytical methods suffer major limitations in describing
complex phenomena which arise from strong nonlinearities and/or non-ideal
geometries of direct relevance to experiments.

RLBMs have been developed in many variants starting from the beginning of the
present decade
and have emerged as a promising tool for the study of dissipative hydrodynamics
in relativistic regimes. In this approach, the time evolution of the system is
described by the one-particle distribution function, and macroscopic quantities
are obtained as moments of this function. 

The first model was developed by Mendoza et al. \cite{mendoza-prl-2010, mendoza-prd-2010}
as an extension of the standard LB equation.
The model is derived basing on Grad's moment-matching technique and 
uses two different distribution functions, one for the particle 
number and one for energy and momentum.

Romatschke et al. \cite{romatschke-prc-2011} developed a scheme for 
an ultra-relativistic gas based on the expansion on orthogonal 
polynomials of the Maxwell-J\"uttner distribution, 
following a procedure similar to the one used for non-relativistic LBM.
This model is not compatible with a Cartesian lattice, 
thus requiring interpolation to implement the streaming phase, 
but has the advantage of supporting the description of
systems in general space-time coordinates. 
Romatschke \cite{romatschke-prd-2012} has also shown that it is
possible to extend the method to support non-ideal equations of state.

Li et al. \cite{li-prd-2012} have extended the work of Mendoza et al. 
by using a multi relaxation-time collision operator.
The model uses standard Cartesian lattices, and it is found that by
independently tuning shear and bulk viscosity it is possible 
to cure numerical errors and discontinuities present in the original model.
However, the model recovers only the first two moments of the distribution
function, and does not allow accurate simulations of flows at large values of $\beta=u/c$, 
where $u$ is the fluid speed and $c$ the speed of light.

Mohseni et al. \cite{mohseni-prd-2013} have shown that it is possible to 
avoid multi-time relaxation schemes, still using a D3Q19 lattice 
and properly tuning the bulk viscosity for ultra-relativistic flows, 
so as to recover only the conservation of the momentum-energy tensor.
This is a reasonable approximation in the ultra-relativistic regime, 
where the first order moment plays a minor role, but leaves open the 
problem of recovering higher order moments.
A further step was taken in \cite{mendoza-prd-2013}, 
with a relativistic lattice Boltzmann method (RLBM) able to recover higher 
order moments on a Cartesian lattice. This model provides 
an efficient tool for simulations in the ultra-relativistic regime
in the Minkowski space-time.

All these developments use pseudo-particles of zero proper mass $m$.
The extension of the model to account for massive particles was 
presented in \cite{gabbana-pre-2017}, with the derivation of 
a {\em unified} scheme, allowing to conceptually bridge 
the gap between the ultra-relativistic regime 
($\beta \simeq 1$), all the way down to the non-relativistic one ($\beta \rightarrow 0$).

Another significant algorithmic development was presented in \cite{ambrus-prc-2018},
where the authors describe a systematic procedure to define quadrature rules
at high orders, giving the possibility to go well beyond hydrodynamics and 
to handle flows from a strongly interacting near-inviscid regime, all the way to the ballistic regime.
The model is restricted to flows of massless particles and has been so far 
applied only to one-dimensional flows. From a conceptual point of view, further
extensions of this model could be of great interest to analyse the transition
between hydrodynamic and non-hydrodynamic regimes in the framework of QGP.

Indeed QGP in possibly the most natural field of application for RLBM methods. 
In this context it has been used to investigate several problems of shock waves propagation
\cite{mendoza-prl-2010, mendoza-prd-2010,li-prd-2012,hupp-prd-2011,mohseni-prd-2013,mendoza-prd-2013,gabbana-pre-2017,gabbana-pre-2017b,ambrus-prc-2018} 
and other standard benchmarks, such as the 1-d Bjorken flow \cite{romatschke-prc-2011, ambrus-prc-2018, ambrus-aip-2019}. 

However, to the best of our knowledge, a fully-fledged implementation for simulating 
nuclear collisions has not been reported, as yet.

Another related application for RLBM is the theoretical study of relativistic transport coefficients.
Based on RLBM simulations, recent works \cite{gabbana-pre-2017b, coelho-cf-2018,ambrus-prc-2018b, gabbana-pre-2019} 
have reported an accurate analysis of the relativistic transport 
coefficients in the single-relaxation time approximation, 
presenting numerical evidence that the Chapman Enskog expansion 
accurately relates kinetic transport coefficients and macroscopic hydrodynamic parameters 
in dissipative relativistic fluid dynamics, confirming recent theoretical results. 

Finally, several authors have attempted to adapt RLBM schemes to 
the study of $(2+1)$-dimensional relativistic hydrodynamics, motivated
by the interest for the study of pseudo-relativistic systems such as the electrons flow in graphene.
A series of theoretical works have taken into consideration the possibility of observing 
Rayleigh-B\'enard instability \cite{oettinger-pre-2013, furtmaier-prb-2015}, 
Kelvin-Helmholtz instability \cite{coelho-prb-2017}, current whirlpools \cite{gabbana-cf-2018},
as well as preturbulent regimes \cite{mendoza-prl-2011, gabbana-prl-2018} in a electronic fluid.

Most of these works and the results here summarized, have been based, with few exceptions,
on formulations in three spatial dimensions with focus on the
ultra-relativistic regime, in that they consider massless one-particle
distribution functions; this is reflected at the macroscopic level  by the
emergence of an ultra-relativistic equation-of-state ($\epsilon = 3 P$, with
$\epsilon$ the energy density and P the pressure).  On the other hand, one would
like to explore all kinematic regimes, as characterized by the dimensionless
parameter 
\begin{equation}
  \zeta = \frac{m c^2}{k_{\rm B} T} \quad ,
\end{equation} 
($m$ is a typical particle mass and $T$ a typical temperature). We shall make extensive reference to this parameter throughout the paper; while in many high-energy astrophysics 
contexts $\zeta \approx 0$, mildly relativistic regimes ($\zeta \approx 1 \cdots 5$), 
which are typical of the QGP physics \cite{pasechnik-uni-2017,scardina-prc-2017,florkowski-rpp-2018}, 
indicate that ultra-relativistic treatments are not appropriate in this case. 
An interesting remark is that, as one follows the history of the Universe 
$\zeta$ gradually increases from $\zeta \approx 0$ towards $\zeta \rightarrow \infty$ (see Fig.~\ref{fig:universe}).

Another important area is the study of low-dimensional systems, as it has been recently realised that ``relativistic'' 
fluid dynamics in $2d$ is relevant to the dynamics of electrons in graphene sheets  
or wider classes of $2d$ ``exotic'' materials governed, to a good approximation, by the dispersion relation 
$ \epsilon_{\rm F} = v_{\rm F} | \bm{p} |$ (formally equivalent to that of a ultra-relativistic particle 
with $c$ replaced by the Fermi velocity $v_{\rm F}$). 

A further open problem has been the lack of (conceptually and numerically)-accurate calibration procedures, 
relating the mesoscopic parameters (relaxation time) to the macroscopic transport coefficients 
e.g., shear and bulk viscosity and thermal conductivity. 

The latter is not only a computational problem but a conceptual one as well, 
since the time-honored approaches to derive transport coefficients from the Boltzmann equation 
(such as Grad's method and the Chapman-Enskog expansion, 
see later for details) yield different results in the relativistic regime.  

These problems have been recently addressed in a series of papers 
\cite{gabbana-pre-2017,gabbana-pre-2017b,gabbana-cf-2018,gabbana-pre-2019}
that have: i) extended the kinematic regime from 
ultra-relativistic, all the way to near non-relativistic, using finite-mass pseudo-particles, 
ii) included the two-dimensional case as well and, iii) developed an accurate calibration procedure of the 
mesoscopic vs. macroscopic transport coefficients.

The present paper builds on  these results and considerably extends them as follows: i) collects and summarizes in a 
structured way all the formal developments of early works; ii) extends algorithmic developments to in 
principle any number of spatial dimensions (in practice, $1 \le d \le 3$) including external forcing as well, 
and using Gauss-type  quadratures on space-filling Cartesian lattices, preserving the computational advantages of the classic LBM; 
iii) recasts early results in a more compact mathematical format; iv) extensively and accurately compares the relationship between mesoscopic 
and macroscopic transport coefficients in $1-2-3$ spatial dimensions, across all kinematic regimes, and finally, v) presents a wider set of 
validation benchmarks.   

For validation purposes, we consider several flows in which approximate
analytical solutions can be worked out and compared with numerical simulations
based on the RLBMs described in this paper. In detail, we present results of
simulations solving the Riemann problem for a quark-gluon plasma, showing good
agreement with previous results obtained using other solvers present in the
literature. We also present simulation results of laminar flows in ultra-clean
graphene samples; we consider geometrical setups  actually used in
experiments, and provide numerical evidence of the formation of electron
back-flows (``whirlpools'', in the jargon of graphene practitioners) in the proximity of current injectors.
%

\section{Ideal Relativistic Hydrodynamics}\label{sec:ideal-hydro}

In this section we introduce the hydrodynamic equations of an ideal relativistic fluid
starting from the basic principles of relativistic kinetic theory, which will serve as the stepping stone for the derivation of the RLBM. A few fundamental references on the formulation of relativistic kinetic theory are the books by
De Groot \cite{degroot-book-1980}, Cercignani and Kremer \cite{cercignani-book-2002}, along with the recent monograph of Rezzolla and Zanotti \cite{rezzolla-book-2013} and the review by Paul and Ulrike Romatschke \cite{romatschke-book-2019}. 

We consider an ideal non-degenerate relativistic fluid, consisting at the
kinetic level of a system of interacting particles of mass $m$. 
The particle distribution function $f( (x^{\alpha}), (p^{\alpha}) )$, depending on space-time 
coordinates $\left( x^{\alpha} \right) = \left( ct, \bm{x} \right)$ and momenta 
$\left( p^{\alpha} \right) = \left( p^0, \bm{p} \right) = \left( E/c, \bm{p} \right)$ 
($c$ is the speed of light, $E$ the particle energy, with $E = c p^0 = c \sqrt{ |\bm{p}|^2 + m^2 c^2}$, 
and $\bm{x}$, $\bm{p} \in \mathbb{R}^d$), describes the probability of finding a particle 
with momentum $\bm{p}$ at a given time $t$ and position $\bm{x}$. 
We adopt Einstein's summation convention over repeated indexes, and use Greek indexes 
to denote $(d+1)$ space-time coordinates and Latin indexes for $d$ dimensional spatial coordinates.

The particle distribution function obeys the relativistic Boltzmann equation, here taken in 
the Anderson-Witting \cite{anderson-witting-ph-1974a, anderson-witting-ph-1974b} relaxation-time approximation: 

\begin{equation}\label{eq:rbe-rta}
  p^\alpha \frac{\partial f }{\partial x^{\alpha}} + m K^{\alpha}\frac{\partial f }{\partial p^{\alpha}}
  =  
  \frac{U^{\alpha} p_{\alpha} }{\tau c^2} (f^{\rm eq} - f)  \quad,
\end{equation}
with $\tau$ the relaxation (proper-)time, $U^{\alpha}$ the macroscopic relativistic $(d+1)$-velocity
(defined such that $U^{\alpha} U_{\alpha} = c^2$), 
and $K^{\alpha}$ the external forces acting on the system, assumed for simplicity 
not to depend on the momentum $(d+1)$-vector.
The local equilibrium $f^{\rm eq}$ is given by the Maxwell-J\"uttner distribution:
\begin{equation}\label{eq:maxwell-juttner}
  f^{\rm eq} = B(n, T) \exp{ \left( - \frac{U^{\alpha} p_{\alpha}}{k_{\rm B} T} \right) } \quad ,
\end{equation}
with $k_{\rm B}$ the Boltzmann constant and $B$ a $d$-dependent normalization factor to be defined later in the text.

The Anderson-Witting model ensures the local conservation of particle number, energy and momentum, 
meaning that the particle four flow $N^{\alpha}$ and the energy momentum tensor $T^{\alpha \beta}$,
defined respectively as the first and second moment of $f$
\begin{align}  
  N^{\alpha}       &= c \int f p^{\alpha}           \frac{\diff^d p}{p_0} \label{eq:pdf-moments-N} \quad , \\
  T^{\alpha \beta} &= c \int f p^{\alpha} p^{\beta} \frac{\diff^d p}{p_0} \label{eq:pdf-moments-T} \quad , 
\end{align}
are conserved:
\begin{align} 
  \partial_{\alpha} N^{\alpha}       &= 0 \label{eq:cons-equation-N} \quad , \\
  \partial_{\beta } T^{\alpha \beta} &= 0 \label{eq:cons-equation-T} \quad .
\end{align}
The conservation equations do not provide any dynamical property of the fluid until a
specific decomposition of $N^{\alpha}$ and $T^{\alpha \beta}$ is specified. 
For an ideal fluid at the equilibrium it can be shown that 
\begin{align}
  N^{\alpha}_E       = & ~ n U^{\alpha}                                                                         \label{eq:first-moment-eq}  \quad , \\
  T^{\alpha \beta}_E = & ~ \left( \epsilon + P \right) \frac{U^{\alpha} U^{\beta}}{c^2} - P \eta^{\alpha \beta} \label{eq:second-moment-eq} \quad , 
\end{align}
where $\epsilon$ ($n$) is the energy (particle) density, $P$ the hydrostatic pressure
and $\eta^{\alpha \beta}$ the Minkowski metric tensor.
In the following we will use $\eta^{\alpha \beta} = diag(1, -\mathds{1})$,
with $\mathds{1} = \left( 1, \dots, 1 \right) \in \mathbb{N}^d $.


The closure for the conservation equations is given by an appropriate Equation of State (EOS).
In order to derive the EOS for a perfect gas in $(d+1)$ space-time coordinates 
in a relativistic regime, we first define the normalization factor
$B(n,T)$ in Eq.~\ref{eq:maxwell-juttner} in order to satisfy the constraint given by Eq.~\ref{eq:first-moment-eq}.
Therefore we write:
\begin{equation}
  c \int f^{\rm eq} p^{\alpha} \frac{\diff^d p}{p_0} 
  =
  c B \int e^{-\frac{p_\mu U^\mu}{k_{\rm B} T}} p^{\alpha} \frac{\diff^d p}{p_0} 
  = 
  c~B~Z^{\alpha}
  = 
  n U^{\alpha} \quad ,
\end{equation}
and together with the analytical expression for the integral $Z^{\alpha}$
(see Appendix~\ref{sec:appendixB} for details), we can determine
the correct normalization factor for the equilibrium distribution function
\begin{equation}\label{eq:phys-normalization}
  B(n,T) = \left(\frac{c}{k_{\rm B} T}\right)^d \frac{n}{2^{\frac{d+1}{2}} \pi ^{\frac{d-1}{2}} \zeta ^{\frac{d+1}{2}} K_{\frac{d+1}{2}}(\zeta )} \quad ;
\end{equation}
the relativistic parameter $\zeta = \frac{m c^2}{k_{\rm B} T}$ has been already defined in the previous section, 
and $K_i(\zeta)$ is the modified Bessel function of the second kind of index $i$.
Next, we take the definition of the momentum-energy tensor (Eq.~\ref{eq:pdf-moments-T}), and use the normalization 
factor $B(n,T)$ together with the analytical expression for $Z^{\alpha \beta}$ (see again Appendix~\ref{sec:appendixB}), giving 
\begin{align}\label{eq:momentum-energy-tensor-integral}
  c \int f^{\rm eq} p^{\alpha} p^{\beta} \frac{\diff^d p}{p_0} 
  = 
  c~B~Z^{\alpha \beta}
  = 
  P G_{d} \frac{U^{\alpha} U^{\beta}}{c^2} - n k_{\rm B} T \eta^{\alpha \beta}  \quad ,
\end{align}
where we have introduced the dimensionless parameter
\begin{equation}\label{eq:defG}
  G_{d} = \frac{\epsilon + P}{P} = \zeta \frac{ K_{\frac{d+3}{2}}(\zeta) }{ K_{\frac{d+1}{2}}(\zeta) } \quad .
\end{equation}

In order to identify the equation of state it is sufficient to match the terms with the same tensor structure 
in Eq.~\ref{eq:momentum-energy-tensor-integral} and Eq.~\ref{eq:second-moment-eq}; one finally obtains:
\begin{equation}\label{eq:eos-ddim}
\begin{array}{rll}
  \epsilon &=& P \left(G_{d} - 1 \right) \quad , \\
        P  &=& n k_{\rm B} T             \quad .
\end{array}
\end{equation}
%
%
\begin{figure}[tbh]
  \centering
  \includegraphics[width=0.8\textwidth]{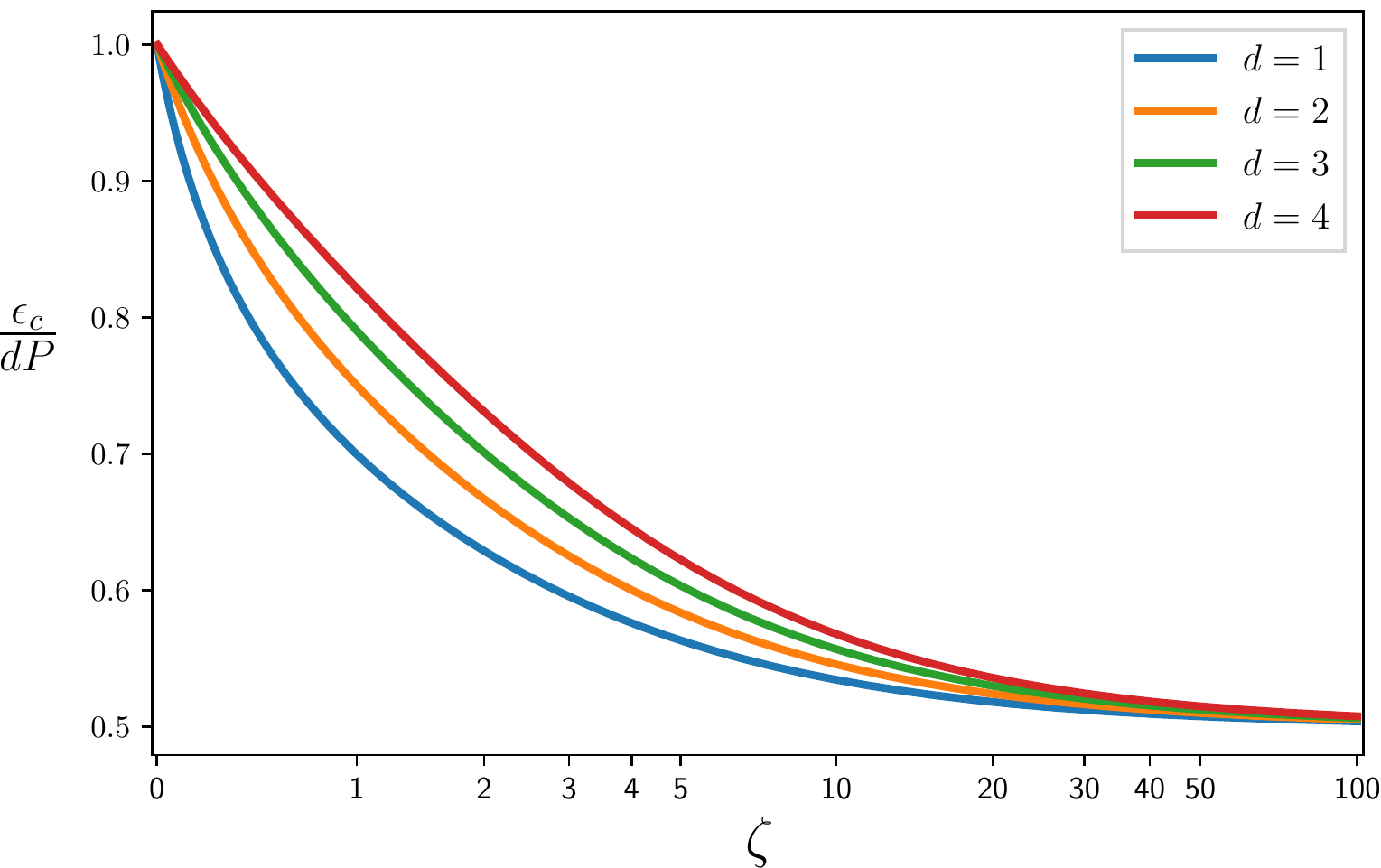}
  \caption{Ratio of kinetic energy density (normalized on the number of spatial dimensions $d$) 
           and pressure for an ideal gas. For better readability the coordinate on the horizontal axis 
           is rescaled as $\zeta \rightarrow \log(\zeta+\sqrt{1+\zeta^2})$. 
           The correct limiting value for kinetic energy density is recovered both in the classical 
           and in the ultra-relativistic regime. 
}\label{fig:edensity-ration}
\end{figure}

For example, in $(3+1)$ dimensions we have:
\begin{equation}\label{eq:relativistic-eos-3D}
\begin{array}{rll}
  \epsilon &=& P \left( \zeta \frac{K_{3}(\zeta)}{K_{2}(\zeta)} -1 \right) = P \left( 3 + \zeta \frac{K_{1}(\zeta)}{K_{2}(\zeta)} \right) \quad .\\
         P &=& n k_{\rm B} T \quad .
\end{array}
\end{equation}
a result already derived many years ago \cite{karsch-pra-1980}.

It is interesting to look at the asymptotic behaviour of Eq.~\ref{eq:eos-ddim}:
it is simple to show that taking the limit $\zeta \rightarrow 0$, for which $G_{d} \rightarrow d+1$,
we obtain the well known ultra-relativistic EOS:
\begin{align}\label{eq:ultra-relativistic-eos}
  \epsilon_{\rm ur} &= d P     \quad .
\end{align}
For the non-relativistic limit we define the kinetic energy density $\epsilon_c = \epsilon - n~m c^2$
and take the limit for $\zeta \rightarrow \infty$ .
Using the fact that $(x~K_{\alpha}(x)/K_{\alpha-1}(x) -x) \rightarrow \alpha -1/2$ as $x \rightarrow \infty$ one recovers
the well known non-relativistic expression for the EOS of an ideal gas:
\begin{align}
  \epsilon_c &= \frac{d}{2} P     \quad .
\end{align}
Finally, Fig.~\ref{fig:edensity-ration} plots the ratio of kinetic energy divided by pressure 
(and rescaled by the number of spatial dimensions) for several values of $d$ and in a wide kinematic range, 
showing a continuous crossover from the ultra-relativistic to the classical regimes.

From the EOS it is straightforward to derive a few thermodynamic quantities (see e.g. \cite{cercignani-book-2002} for
their formal definition) which will be useful in the coming sections, such as the heat capacity at constant volume $c_v$:

\begin{equation}\label{eq:heat-capacity-const-volume}
  c_v 
  = 
  \frac{\partial (\epsilon / n)}{\partial T} 
  = 
  k_{\rm B} \left[ (2 + d)G_{d} + \zeta^2 -G_{d}^2 -1 \right] \quad ,
\end{equation}
the heat capacity at constant pressure $c_P$ ( $h_e = (\epsilon + P) / n$ is the relativistic enthalpy per particle) :
\begin{equation}\label{eq:heat-capacity-const-pressure}
  c_P 
  = 
  \frac{\partial h_e}{\partial T} 
  = 
  k_{\rm B} \left[ (2 + d)G_{d} + \zeta^2 -G_{d}^2 \right] \quad ,
\end{equation}

and the adiabatic sound speed $c_s$:
\begin{equation}
  c_s 
  = 
  c \sqrt{\frac{P}{\epsilon + P} \frac{c_p}{c_v} } 
  = 
  c \sqrt{ \frac{(2 + d)G_{d} + \zeta^2 -G_{d}^2 -1}{G_{d} \left( (2 + d)G_{d} + \zeta^2 -G_{d}^2 \right) } } \quad .
\end{equation}

\section{Dissipative Effects and Transport Coefficients}\label{sec:dissipative-hydro}

When dissipative effects are taken into account, the definition of the non-equilibrium
component of $N^{\alpha}$ and $T^{\alpha \beta}$ is ambiguous as it depends on the choice of
the local rest frame, with the two most common choices being the one suggested by 
Eckart \cite{eckart-prl-1940} and by Landau and Lifshitz \cite{landau-book-1987}.
The Anderson-Witting model is based on the Landau-Lifshitz decomposition, where
the fluid velocity $U^{\alpha}$ is defined such as to satisfy 
\begin{equation}
  T^{\alpha \beta} U_{\beta} = \epsilon ~U^{\alpha} \quad ,
\end{equation}
and for which, assuming a linear combination of the contribution due to the equilibrium 
and the non-equilibrium part, it follows that
\begin{align}
  N^{\alpha}       &= N^{\alpha}_E  - \frac{1}{h_e} q^{\alpha}                                   \quad , \label{eq:first-moment-neq}  \\
  T^{\alpha \beta} &= T^{\alpha \beta}_E  - \varpi \Delta^{\alpha \beta} + \pi^{<\alpha \beta >} \quad , \label{eq:second-moment-neq}
\end{align}
where $q^{\alpha}$ is the heat flux,
$\pi^{<\alpha \beta>}$ the pressure deviator, $\varpi$ the dynamic pressure, and 
\begin{equation}
  \Delta^{\alpha\beta} = \eta^{\alpha\beta} - \frac{1}{c^2} U^\alpha U^\beta \quad , 
\end{equation}
is the (Minkowski-)orthogonal projector  to the fluid velocity $U^{\alpha}$ (see Appendix~\ref{sec:appendixA} for complete definition
of all tensorial objects that we use and \cite{cercignani-book-2002} for a full treatment of the problem).

The non-equilibrium contribution to $N^{\alpha}$ and $T^{\alpha \beta}$ can be used to define 
the transport coefficients which enter the linear relations between thermodynamic forces and fluxes:
\begin{align}
  q^{\alpha}             & =   \lambda  \left( \nabla^{\alpha} T - \frac{T }{n h_e} \nabla^{\alpha} P \right) \quad , \label{eq:heat-flux} \\
  \pi^{< \alpha \beta >} & =   \eta     \left( \Delta^{\alpha}_{\gamma} \Delta^{\beta}_{\delta} + \Delta^{\alpha}_{\delta} \Delta^{\beta}_{\gamma}  
                - \frac{1}{d}  \Delta^{\alpha \beta} \Delta_{\gamma \delta} \right) \nabla^{\gamma} U^{\delta} \quad , \label{eq:stress-tensor} \\
  \varpi                 & = - \mu      \nabla_{\alpha} U^{\alpha} \quad ; \label{eq:dynamic-pressure} 
\end{align}
$\lambda$ is the thermal conductivity, $\eta$ and $\mu$ the shear and bulk viscosities, 
and we have used the shorthand notation
\begin{equation}
\begin{array}{lcl} 
  \nabla^{\alpha}         & = & \Delta^{\alpha \beta} \partial_{\beta}       \quad ,\\
  \Delta^{\alpha}_{\beta} & = & \Delta^{\alpha \gamma} \Delta_{\gamma \beta} \quad .
\end{array} 
\end{equation}

The transport coefficients provide the link between the kinetic and the
macroscopic layer.  In non-relativistic regimes, the derivation of appropriate
transport coefficients is typically obtained with either Grad's method of
moments \cite{grad-cpam-1949} or the Chapman-Enskog (CE) expansion
\cite{chapman-book-1970}; both techniques provide a consistent connection
between kinetic theory and hydrodynamics, i.e. they provide the same expressions
for the transport coefficients. 
However, it is well known that the two methods give different results
in the relativistic regime.

In recent times, the problem has been extensively studied. 
Theoretical works and numerical investigations seem to converge towards the results provided 
by the CE approach but the question is still open to debate.

Here we consider both the CE and Grad's method of moments
expansion in a general $(d+1)$  space-time coordinate system, deriving all transport coefficients for the relativistic Boltzmann equation in the RTA.
Derivations of some of these coefficients have appeared sparsely in the literature, often for specific quantities and specific space dimensions \cite{denicol-prl-2010, denicol-prd-2012, molnar-prd-2014, 
jaiswal-prc-2013b, tsumura-epj-2012, mendoza-josm-2013, florkowski-prc-2015, tsumura-prd-2015, kikuchi-prc-2015, kikuchi-pla-2016, perciante-josp-2017}. 
For this reason, we consider it useful to gather here for reference the full set of results using both approaches. 
We follow closely the procedure presented in \cite{cercignani-book-2002} for the $(3+1)$-dimensional case.
In the following  we review the procedure used to derive these results, with full details and results collected in Appendix~\ref{sec:appendixC}.

\subsection{Chapman-Enskog expansion}

The Chapman-Enskog expansion consists in splitting the particle distribution function $f$ in two additive terms: 
the equilibrium distribution $f^{\rm eq}$ and a non equilibrium part $f^{\rm neq}$. 
When working in a hydrodynamic regime, it is reasonable to approximate $f^{\rm neq}$ with a small deviation from the equilibrium:
\begin{equation}\label{eq:f-approx-ce-main}
  f = f^{\rm eq} + f^{\rm neq} \sim f^{\rm eq}( 1 + \phi) \quad .
\end{equation}
with $\phi$ of the order of the Knudsen number $\rm Kn$, defined as the ratio between the mean free path 
and a typical macroscopic length scale.
The general idea is to determine an analytical expression for the deviation from the equilibrium $f^{\rm eq} \phi$.
We start from Eq.~\ref{eq:rbe-rta} (let us ignore for the moment the forcing term), 
insert Eq.~\ref{eq:f-approx-ce-main} and retain only terms $\mathcal{O}(\rm{Kn})$, giving:
\begin{equation}
  p^{\alpha} \frac{\partial f^{\rm eq}} {\partial x^{\alpha}}= - \frac{p^{\alpha}U_{\alpha}}{c^{2} \tau} f^{\rm eq} \phi \quad .
\end{equation}
To derive the transport coefficients one then proceeds with the following steps:
\begin{enumerate}
  \item Compute the derivative $p^{\alpha} \partial_{\alpha} f^{\rm eq}$ and derive the constitutive equations of a relativistic Eulerian fluid.
  \item Use the balance equations for energy and momentum to eliminate the convective time derivatives and derive the analytic expression of $\phi$.
  \item Use the now known expression for $f^{\rm eq} \phi$ to compute the first and second order tensors (via their integral definitions), compare against their definition in the Landau frame and work out the expression for the transport coefficients.        
\end{enumerate}

See Appendix~\ref{sec:appendixC} for a full discussion and full analytical expressions in an arbitrary number of space dimensions. 
Here we only mention the ultra-relativistic limit:
\begin{align}\label{eq:ce-ur-limit}
  \lambda_{\rm ur} &= \frac{d+1}{d} c^2 k_{\rm B} n \tau \quad , \\
      \mu_{\rm ur} &= 0                                  \quad , \\
     \eta_{\rm ur} &= \frac{d+1}{d+2} P \tau             \quad .
\end{align}

\subsection{Grad's moments method}

The starting point for the derivation of Grad's method of moments is similar to that
of the Chapman Enskog expansion, with the splitting of the particle distribution function into
two terms, the equilibrium and the non-equilibrium part. 
The way the non-equilibrium part is derived is however significantly different; while CE makes use of
a small parameter of the order of the Knudsen number, Grad's method is based on the expansion 
of the distribution function onto a set of orthonormal basis functions. 
The expansion is then truncated, setting to zero the kinetic moments beyond a prescribed order.
When applying this formalism in the non-relativistic framework, the expansion is based on the
Hermite polynomials, since their projection coefficients deliver the kinetic moments of the distribution function.
In Appendix~\ref{sec:appendixF} we define a set relativistic polynomials having this same property,
which we will use as the expansion basis for Grad's method. 
It is important to stress that this approach, which was also used in the 14-moment approximation of Israel-Stewart, 
presents significant pitfalls which have been identified and corrected by Denicol et al \cite{denicol-prd-2012}.
In particular these authors have shown the importance of using a irreducible set of tensors, such as for example
$\{1, \Delta^{\alpha\beta}p_\beta, p^{<\alpha}p^{\beta>}, \dots\}$. 
For these reasons, we remark that the procedure sketched here (and described in full details in Appendix~\ref{sec:appendixC}) 
should be improved as described in Ref. \cite{denicol-prd-2012}, and extended to an arbitrary number of space dimensions.

We start giving the definition of the entropy density $s$: 
\begin{align}
  s = -\frac{k_{\rm B}}{nc} U^\alpha \int p_\alpha f \ln{f} \frac{d^dp}{p_0} \quad .
\end{align}

The derivation, constrained to the maximization of $s$, can be summarized in the following steps:
\begin{enumerate}
  \item Using Lagrange multipliers method, find an expansion for $f$ that extremizes the entropy density $s$, 
        with the constraints given by definition of $N^\alpha U_\alpha$, $U_\alpha T^{\alpha\beta}$ 
        and $U_\alpha T^{\alpha<\beta\gamma>}$ (see Appendix~\ref{sec:appendixC}). 
  \item Using Grad's ansatz for $f$ we compute the third order moment $T^{\alpha\beta\gamma}$.
  \item The above expression is then plugged into Eq.~\ref{eq:rbe-rta} to determine the non-equilibrium components
        of the energy-momentum tensor.
  \item By applying appropriate projectors it is then possible to derive the constitutive equations for the heat-flux, 
        dynamic pressure and the pressure deviator. The expressions for the transport coefficients are derived 
        by comparison with Eq.~\ref{eq:heat-flux}, Eq.~\ref{eq:stress-tensor}, and Eq.~\ref{eq:dynamic-pressure}.     
\end{enumerate}

Once again, detailed derivations and results are collected in Appendix~\ref{sec:appendixC}; 
in the ultra-relativistic limit we have:
\begin{align}
  \lambda_{\rm ur} &= \frac{d+1}{d+2} c^2 k_{\rm B} n \tau \quad ,\\
      \mu_{\rm ur} &= 0                                    \quad ,\\ 
     \eta_{\rm ur} &= \frac{d+1}{d+3}P\tau                 \quad .
\end{align}

\begin{figure}[H]
\centering
\includegraphics[width=0.85\textwidth]{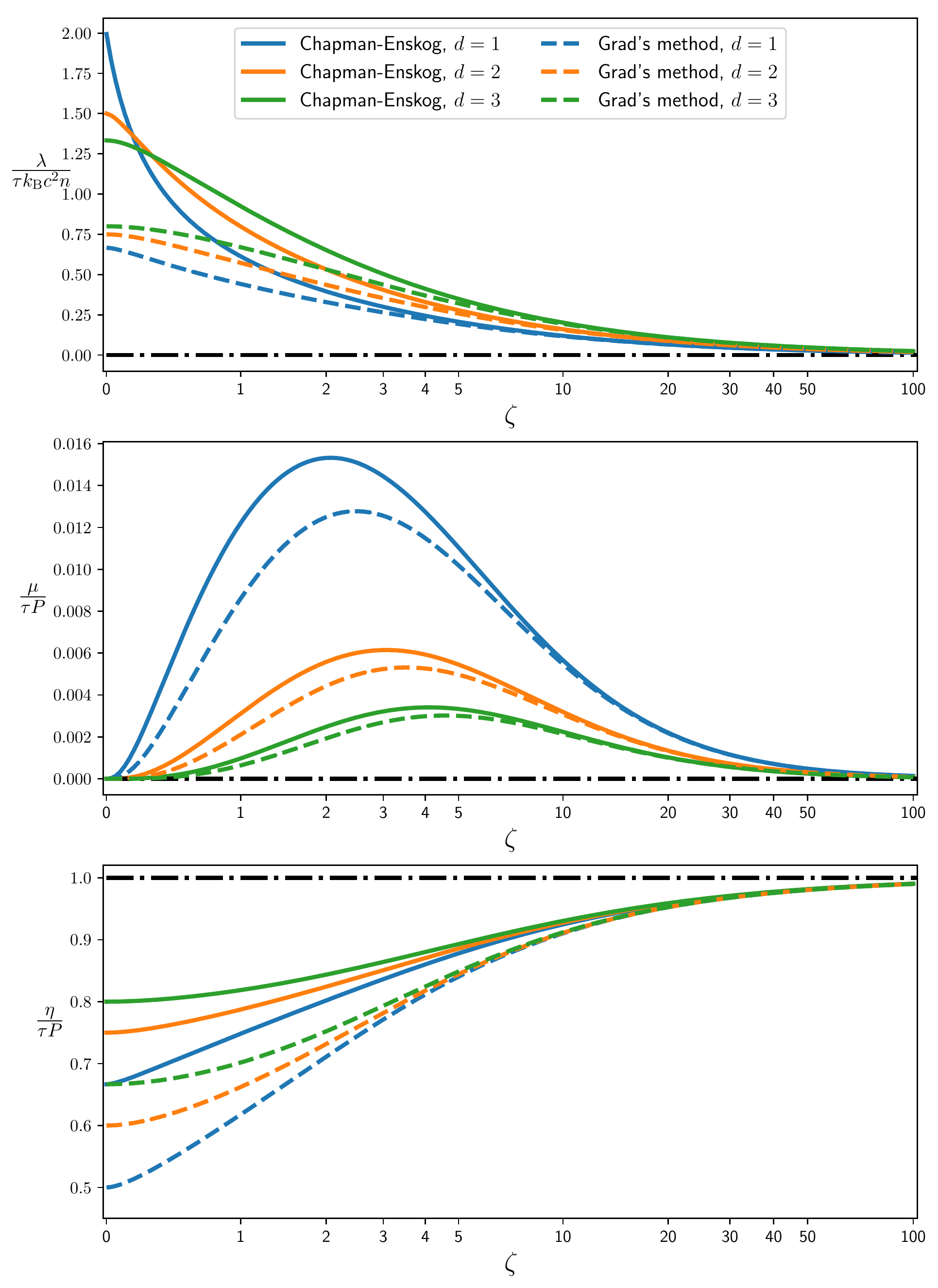}
\caption{Comparison of the non-dimensional thermal conductivity, shear and bulk viscosity in $1,2,3$-dimensions, 
         obtained applying the Chapman-Enskog expansion and Grad's method of moments to the relativistic Boltzmann equation in the relaxation time approximation. 
         The black dotted line represents the limit for $d \rightarrow \infty$; this result is the same using both CE and Grad's method, and shows that in this limit all transport coefficients are classical.
        }\label{fig:transport-coeff-analytic}
\end{figure}

As already remarked, the two methods give different results for the values of all
transport coefficients, even if they tend to agree as one approaches the non
relativistic regime. This is clearly shown 
in Fig.~\ref{fig:transport-coeff-analytic}
where the behavior of $\eta$, $\mu$ and $\lambda$, 
as predicted by the two approaches, is shown as
a function of $\zeta$. Lacking any realistic option for experimental
verification, we will see in later sections that our numerical experiments
strongly point to the CE approach. 
A mathematically nice result (although of little interest for physical purposes) is
that, in the limit of an infinite number of spatial dimensions, all coefficients
remain constant at their non relativistic value over the full kinematic range.

Finally, the behavior of the thermal conductivity needs a further explanation.
It is well known, and it can directly be seen from Eq.~\ref{eq:heat-flux}, 
that the heat flux present a significant difference between the relativistic and the non-relativistic form; indeed for a relativistic iso-thermal fluid there could be a non-zero heat-flux due to a pressure gradient.
Looking at Fig.~\ref{fig:transport-coeff-analytic} one may be puzzled as $\lambda$ seems to go to $0$ 
in the non-relativistic limit. This is so because, for later convenience, we plot $\frac{\lambda}{\tau k_B c^2 n}$. 
If one recasts the expression as $\frac{ \lambda m}{k_{\rm B} P \tau}$ and considers the limit for large $\zeta$, 
one obtains
\begin{equation}
  \lambda = \frac{k_{\rm B}}{m} P \tau \left( \frac{d+2}{2} - \frac{3 (d+2)}{2 \zeta} + \dots \right) \quad ,
\end{equation} 
whose first term is the well-known non-relativistic value.

\section{Relativistic Lattice Boltzmann Methods} \label{sec:rlbm}

In recent times, numerical schemes based on the Lattice Boltzmann Method (LBM) have emerged as
a promising tool for the study of dissipative relativistic hydrodynamics 
\cite{mendoza-prl-2010,romatschke-prc-2011,romatschke-prd-2012,mendoza-prd-2013,gabbana-pre-2017,ambrus-prc-2018,coelho-cf-2018}.
The advantage of this approach is that by working at a mesoscopic level viscous effects 
are naturally included, with relativistic invariance and causality preserved by construction.

In this section we present in full details the algorithmic extension of the LBM to the study of relativistic fluids, 
describing the derivation of a model which allows to cover a wide range of relativistic regimes, 
in principle all the way from fluids of ultra-relativistic massless particles down to non-relativistic fluids.

\subsection{From continuum to the lattice}

We here outline the procedure followed to derive the relativistic lattice Boltzmann equation,
following a procedure similar to the one used with non-relativistic \cite{higuera-epl-1989, xiaoyi-pre-1997, shan-prl-1998, martys-pre-1998} 
and earlier ultra-relativistic LBMs \cite{romatschke-prc-2011, mendoza-prd-2013}. 
In this section, we use natural units, $c = k_{\rm B} = 1$, which helps write many formulas in a more compact form.
\begin{enumerate}
%
\item 
We start by writing Eq.~\ref{eq:rbe-rta} in terms of quantities that can be discretized on a regular lattice,
by dividing the left and right hand sides by $p^0$:
\begin{equation} \label{eq:rbe-rta-lattice-friendly}
  \partial_t f + v^i \nabla_i f 
  = 
  \frac{U^{\alpha} p_{\alpha}}{\tau p^0} (f^{\rm eq} - f) - \frac{m K^{\alpha}}{p^0} \frac{\partial f }{\partial p^{\alpha}}\quad,
\end{equation}
with $v^i = p^i / p^0$ the components of the microscopic velocity. In Eq.~\ref{eq:rbe-rta-lattice-friendly}
the time-derivative and the propagation term are the same as in the non-relativistic regime; the price to pay is 
an additional dependence on $p^0$ of the relaxation (and forcing) term.
%
\item 
%
Next, we expand $f^{\rm eq}$ in an orthogonal basis; we adopt Cartesian coordinates and use 
a basis of polynomials orthonormal with respect to a weight $\omega( p^0)$ given by the Maxwell-J\"uttner 
distribution in the fluid rest frame (where $U^i = 0$). 

Following a Gram-Schmidt procedure one then derives a set of polynomials 
$\{J^{(i)}, i = 1,2\dots \}$, which are used to build the expansion:
\begin{equation}\label{eq:mj-feq-expansion}
  f^{\rm eq}((p^{\mu}), (U^{\mu}), T) 
  = 
  \omega( p^0) \sum_{k = 0}^{\infty} a^{(k)}( (U^{\mu}), T) J^{(k)} ( (p^{\mu}) ) \quad ,
\end{equation}
where $a^{(k)}$ are the projection coefficients defined as
\begin{equation}\label{eq:mj-projection-coefficients}
  a^{(k)}( (U^{\mu}), T) 
  = 
  \int f^{\rm eq}( (p^{\mu}), (U^{\mu}), T)  J^{(k)}( (p^{\mu}) ) \frac{\diff^d p}{p^0}  \quad .
\end{equation}
The polynomials are derived in such a way that the coefficients $a^{(k)}$ 
coincide by construction with the moments of the distribution function; 
as a result the quantity $f_N^{\rm eq}( (p^{\mu}), (U^{\mu}), T)$, 
obtained truncating the summation in Eq.~\ref{eq:mj-feq-expansion} to $N$, 
correctly preserves the moments of the distribution up to the $N$-th order.
Observe that until now the discussion holds its validity in the continuum. 
%
\item 
%
We now find a Gauss-like quadrature on a regular Cartesian grid able to reproduce correctly the moments of the original distribution up 
to order $N$. We proceed in such a way as to preserve exact streaming, meaning that all quadrature points $v^l_i = p^l_i/p^0$ 
must sit on lattice sites.
At this point, the discrete version of the equilibrium function reads as follows:
\begin{equation}\label{eq:mj-discrete-feq}
  f_{i N}^{\rm eq} = w_i \sum_{k} a^{(k)}( (U^{\mu}), T) J^{(k)} ((p^{\mu}_i)) \quad ,
\end{equation}
with $w_i$ appropriate weights, $(p^{\mu}_i)$ the linked abscissae, and the summation 
running on the total number of orthogonal polynomials up to the order $N$.
\item
%
Once a quadrature rule is defined, it is possible to write down the discrete relativistic Boltzmann equation:
\begin{equation}\label{eq:discrete-rbe}
  f_i(\bm{x} + \bm{v}^{i} \Delta t, t + \Delta t) - f_i(\bm{x}, t) 
  = 
  \Delta t~ \frac{p_i^{\alpha} U_{\alpha}}{p^0_i \tau} (f_i^{\rm eq} - f_i) + F^{\rm ext}_i  \quad ,
\end{equation}
where $F^{\rm ext}_i$ is the discretization of the total external forces acting on the system,
more details will be given in Section~\ref{sec:forcing}.

\end{enumerate}
%

\subsection{Polynomial expansion of the equilibrium distribution function}

In this section we define the polynomial expansion of the Maxwell-J\"uttner distribution in $(d+1)$-dimensions. 
It turns out that using non-dimensional quantities is very useful here; to this purpose, we introduce a reference temperature $T_0$ (and a corresponding energy scale $k_B T_0$) and define the following quantities: 
$\tilde{T} = T / T_0$,  $\tilde{m} = mc^2 / (k_B T_0)$, $\tilde{p}^{\alpha} = c p^{\alpha} / (k_B T_0)$. $T_0$ is in principle arbitrary; we will see in the following that $T_0$ is needed to translate between lattice and physical units; for the moment the reader may consider $T_0$ as a typical temperature/energy scale of the system under study.

We start by constructing a set of polynomials in the variables $\tilde{p}^{\alpha}$, orthogonal with respect to a weighting function given by the equilibrium distribution in the co-moving frame:
\begin{equation}
  \omega(p^0,T_0) = C(\tilde{m},T_0) \exp{ \left( - \tilde{p}^0 \right)  } \quad ;
\end{equation}
$C$ is a normalization factor, which deserves a further remark: while the
normalization factor $B(n,T)$ in Eq.~\ref{eq:maxwell-juttner} carries an 
important physical meaning (as discussed in Section \ref{sec:ideal-hydro}),
$C$ can be chosen in the most expedient way.
In most cases we will find it convenient to take the normalization factor 
$C$ such to satisfy the condition
\begin{equation}\label{eq:numerical-normalization-cond}
  \int \omega(p^0,T_0) \frac{\diff^d p}{p^0} 
  = 
  (\frac{k_B T_0}{c})^{d-1} \int \omega(p^0,T_0) \frac{\diff^d \tilde{p}}{\tilde{p}^0} 
  =
  1 \quad ,
\end{equation}
implying
\begin{equation}\label{eq:numerical-normalization}
  C(\tilde{m},T_0) = 
  \frac{1}{2^{\frac{d+1}{2}} \pi^{\frac{d-1}{2}} \tilde{m}^{\frac{d-1}{2}} K_{\frac{d-1}{2}} \left( \tilde{m} \right)}~\frac{c}{k_B T_0^{d-1}} \quad .
\end{equation}

Starting from the set $ \mathcal{V} = \{ 1, \tilde{p}^{\alpha}, \tilde{p}^{\alpha} \tilde{p}^{\beta}, \dots \}$ we apply the Gram-Schmidt 
procedure to derive the polynomials up to a desired order. 
We label the polynomials with the notation $J^{(n)}_{\alpha_1 \cdots \alpha_n}$, where $n$ is the order of the polynomial and the $\alpha$ indexes corresponds to the components of $(\tilde{p}^{\alpha})$ they depend upon. All integrals needed to carry out this procedure are computed in Appendix~\ref{sec:appendixB}.
The first polynomials (up to order $1$) in $(d+1)$ dimensions are easily written:
\begin{align*}
  J^{(0)}   &= 1                                                                                                \quad ,         \\
  J^{(1)}_0 &= \frac{\tilde{p}^0 - \tilde{G}_{d-2}}{\sqrt{\tilde{m}^2 - \tilde{G}_{d-2}^2+ d \tilde{G}_{d-2}}}  \quad ,         \\
  J^{(1)}_i &= \frac{\tilde{p}^i}{\sqrt{\tilde{G}_{d-2}}}                                                       \quad ;         \\
\end{align*}
here we use the shorthand notation $\tilde{G}_{d-2} = G_{d-2}(\tilde{m})$, with $G_d$ defined in Eq. \ref{eq:defG}. 
Their ultra-relativistic limit
\footnote{Some of the expressions that we consider here become singular in the massless limit for $d = 1$; 
          we will return to this point later in this section}
is given by:
\begin{align*}
  J^{(0)}   &= 1                                            \quad , \\
  J^{(1)}_0 &= \frac{\tilde{p}^0}{\sqrt{d-1}} - \sqrt{d-1}  \quad , \\
  J^{(1)}_i &= \frac{\tilde{p}^i}{\sqrt{d-1}}               \quad . \\
\end{align*}
The corresponding projection coefficients (defined through \ref{eq:mj-projection-coefficients} 
and taking into account the normalization of $f^{\rm eq}$ given by Eq. \ref{eq:phys-normalization}) are given by:
\begin{align*}
  a^{(0)}   &= \frac{cn}{k_B T_0}~ \frac{1}{G_{d-2}\tilde{T}}                                   \quad , \\
  a^{(1)}_0 &= \frac{cn}{k_B T_0} ~\left(U^0 - \frac{\tilde{G}_{d-2}}{\tilde{T}G_{d-2}}\right)
               \frac{1}{ \sqrt{\tilde{m}^2 + \tilde{G}_{d-2}( d - \tilde{G}_{d-2})}  }     \quad , \\
  a^{(1)}_i &= \frac{cn}{k_B T_0} ~\frac{U^i}{\sqrt{\tilde{G}_{d-2}}}                           \quad ; \\
\end{align*}
where $G_{d-2}$ (as opposed to $\tilde{G}_{d-2}$) is again a shorthand for
$G_{d-2} = G_{d-2}(m/T) \equiv G_{d-2}(\tilde{m}/\tilde{T})$. 
The ultra-relativistic limit reads:
\begin{align*}
  a^{(0)}   &= \frac{cn}{k_B T_0} ~\frac{1}{(d-1) \tilde{T}}                                 \quad , \\
  a^{(1)}_0 &= \frac{cn}{k_B T_0} ~\left(\frac{U^0}{\sqrt{d-1}} - \frac{1}{\tilde{T} \sqrt{d-1})}\right) \quad , \\
  a^{(1)}_i &= \frac{cn}{k_B T_0} ~\frac{U^i}{\sqrt{d-1}}                                    \quad . \\
\end{align*}
Having derived both the polynomials and the projections we can then write down 
the first order expansion version of the Maxwell-J\"uttner distribution in $(d+1)$-dimension:
\begin{align*}
  f^{\rm eq} &= \frac{cn}{k_B T_0}~\omega( \tilde{p}^0 )
                \left( 
                \frac{1}{G_{d-2} \tilde{T} } +  
                \frac{(\tilde{p}^0 - \tilde{G}_{d-2})} {\tilde{G}_{d-2}(d-\tilde{G}_{d-2})+\tilde{m}^2}
                \left(U^0 - \frac{\tilde{G}_{d-2}}{\tilde{T}G_{d-2}}\right)
                - \frac{1}{\tilde{G}_{d-2}} p^{i} U_{i}  
                \right) \quad . 
\end{align*}
The expression in the ultra-relativistic limit is slightly simpler:
\begin{align*}
  f^{\rm eq} & = \frac{cn}{k_B T_0}~\omega( \tilde{p}^0 )~
                 \frac{d/\tilde{T}+ U^0 \tilde{p}^0 -(d-1) U^0 - \tilde{p}^{i} U_{i}  }{d-1} \quad .
\end{align*}
Expressions at higher orders are rather bulky and are therefore given as supplementary material \cite{SOM}. 
Here we only stress the general structure of the expansion:
\begin{itemize} 
  \item all polynomials are adimensional and written in terms of $\tilde{p}^{\alpha}$ and of $\tilde{m}$;
  \item all expansion coefficients are the product of $\frac{cn}{k_B T_0}$ and again an adimensional expression, that depends on $U^{\alpha}$, $\tilde{T}$ and $\tilde{m}$;
  \item the resulting expressions for $f^{\rm eq}$ are again the product of $\frac{cn}{k_B T_0}$ and an adimensional expression.
\end{itemize} 

This structure will make it very simple to relate lattice-defined quantities with the corresponding physical ones. 
See later on this point.
 
\begin{figure}[htb]
  \centering
  \includegraphics[width=\columnwidth]{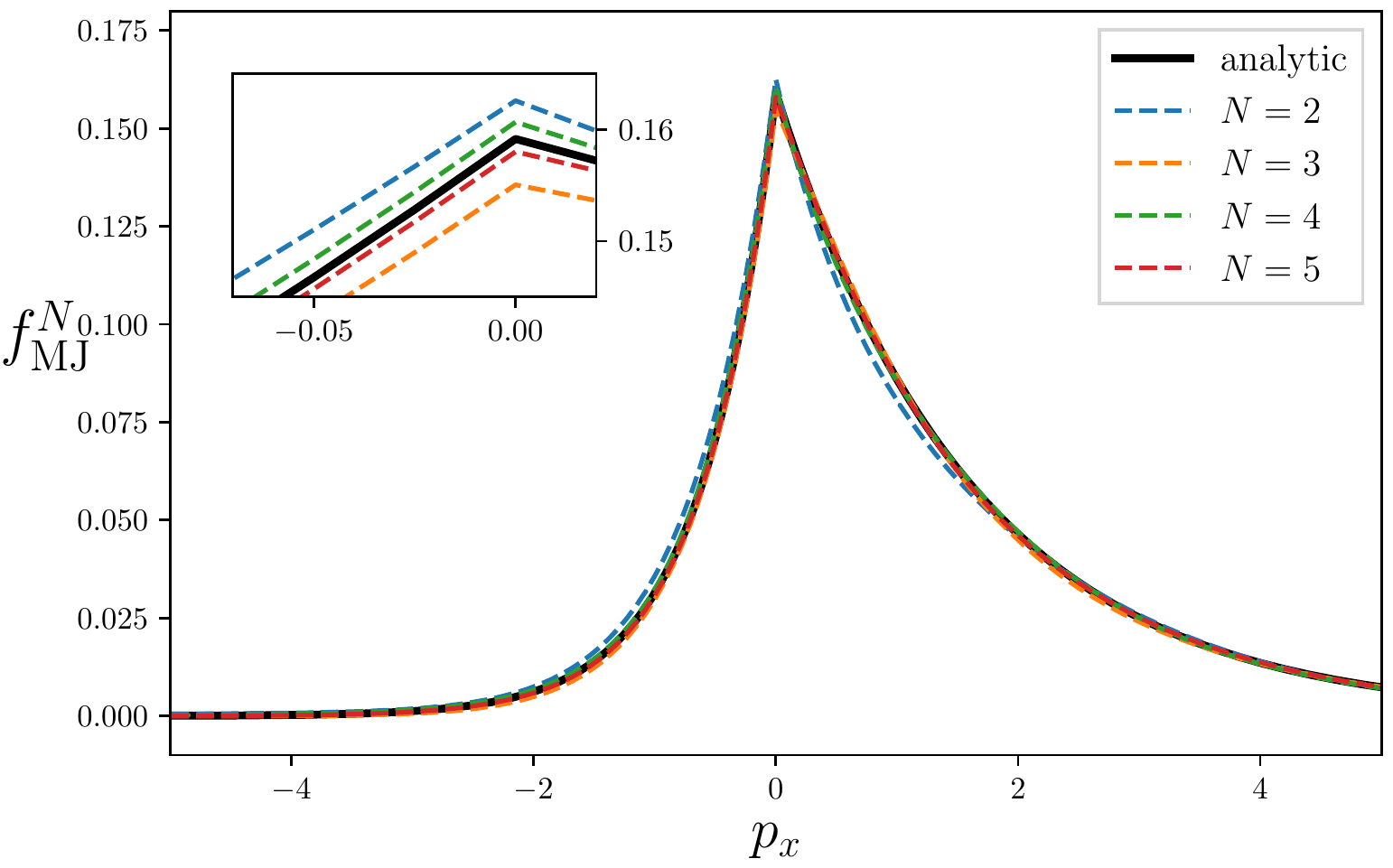}
  \caption{Comparison of the analytic Maxwell J\"uttner distribution in $(2+1)$ dimensions 
           against approximations at various orders $N$, computed using an orthogonal polynomial basis. 
           The distributions are shown as functions of $\bm{p} = (p_x, 0)$, having fixed
           all the other parameters to $\tilde{m} = 0$, $\tilde{T} = 1$, $n = 1$ and $\beta = |U^i|/U^0 = 0.5$.
          }\label{fig:mj2D-m0-beta05}
\end{figure}

Fig.~\ref{fig:mj2D-m0-beta05} shows the expansion of $f^{\rm eq}$ up to the fifth order in 
$(2+1)$ dimensions in the massless limit and compares with the analytical expression. 
$f^{\rm eq}$ is plotted as a function of $\bm{\tilde{p}} = (\tilde{p}_x, 0)$
with $n$, $\tilde{T}$ and $T$ all equal to unity and $\beta = |U^i|/U^0 =  0.5$.
%

The massless limit in $(1+1)$ dimensions needs special care, as the normalization factor of 
$\omega(p^0,T_0)$ defined by Eq. \ref{eq:numerical-normalization} is in this case proportional 
to $1/K_0(\tilde{m})$ and diverges when $\tilde{m} \rightarrow 0$, so the weighting function is ill-defined.  
In this case, in order to define a valid kernel for the Gram-Schmidt procedure, we use for $\omega(p^0,T_0)$
a normalization factor analogous to the one defined in Eq.~\ref{eq:phys-normalization}, that, in this case, writes
\begin{equation}
  C(\tilde{m}, T_0) = \frac{c}{k_B T_0}\frac{1}{2 \tilde{m} K_1(\tilde{m})} \quad .
\end{equation}

We denote with $\mathcal{J}^{(n)}_{\alpha_1 \cdots \alpha_n}$ the polynomials
derived starting from the above defined weighting function, which up to order $1$ take the following form:
\begin{align*}
  \mathcal{J}^{(0)}   &= \frac{1}{\sqrt{\frac{\tilde{G}_1-2}{\tilde{m}^2}}}                                                      \quad , \\
  \mathcal{J}^{(1)}_0 &= \frac{\tilde{p}^0}{\sqrt{-\frac{\tilde{m}^2}{\tilde{G}_1-2}+\tilde{G}_1-1}}
                        -\frac{\tilde{m}^2}{\left(\tilde{G}_1 -2\right) \sqrt{-\frac{\tilde{m}^2}{\tilde{G}_1-2}+\tilde{G}_1-1}} \quad , \\
  \mathcal{J}^{(1)}_i &= \tilde{p}^i                                                                                             \quad . \\
\end{align*}

For the projections, we have:
\begin{align*}
  \mathcal{A}^{(0)}   &= \frac{n}{T_0} \frac{(G_1-2) \tilde{T}}{\tilde{m}^2 \sqrt{\frac{\tilde{G}_1-2}{\tilde{m}^2}}}                 \quad , \\
  \mathcal{A}^{(1)}_0 &= \frac{n}{T_0} \frac{U^0}{\sqrt{-\frac{\tilde{m}^2}{\tilde{G}_1-2}+\tilde{G}_1-1}}
                        -\frac{(G_1-2) \tilde{T}}{\left(\tilde{G}_1-2\right) \sqrt{-\frac{\tilde{m}^2}{\tilde{G}_1-2}+\tilde{G}_1-1}} \quad , \\
  \mathcal{A}^{(1)}_i &= \frac{n}{T_0} U^i                                                                                            \quad . \\
\end{align*}
One may now check that while independent limits of polynomials and projections are still divergent, 
the limit of the product of each polynomial with its corresponding projection is convergent also in the massless limit; 
the limiting value for $f_{N}^{\rm eq}( (\tilde{p}^{\mu}), (U^{\mu}), \tilde{T})$  is then well-behaved:
\begin{equation}\label{eq:mj-expanded-1D-massless}
  f_{N}^{\rm eq}( (\tilde{p}^{\mu}), (U^{\mu}), \tilde{T}) 
  = 
  \omega(p^0,T_0) \sum_{k = 0}^{N} \lim_{\tilde{m} \rightarrow 0} \left( a^{(k)}( (U^{\mu}), \tilde{T})  J^{(k)} ( (\tilde{p}_i^{\mu})) \right) \quad .
\end{equation}
Up to first order, one obtains:
\begin{equation}
f^{\rm eq} = \frac{c n \omega(\tilde{p}_0)}{k_B T_0} \left[ \frac{1 - \tilde{p}^0}{\tilde{T}} + \tilde{p}^0 U^0 + \tilde{p}^x U^x \right]
\end{equation}
This expression has the same structure as for the general case that we have discussed before. Note however, that the fact that polynomials 
and projections do not have independent finite limits in the massless case will require special care for the construction of Gaussian quadratures. 

\subsection{Gauss-type quadratures with prescribed abscissas}\label{sec:gauss-quad}

The discrete formulation of the theory discussed above is based on a Gauss-type quadrature on a Cartesian grid. 
As we move onto this discrete lattice, from now on we use natural units ($ c = 1$ and $k_B = 1$) allowing to write down mathematically slimmer expressions; as ultimately all grid-defined quantities are adimensional, any specific choice on the preferred dimensional units will only affect the conversion factors between physical and numerical units (see later for details). 

In order to ensure that all quadrature points lie on lattice sites,
and to preserve the moments of a distribution up to a desired order $N$, we
need to determine the weights and the abscissas of a quadrature such
to satisfy the orthonormal conditions \cite{philippi-pre-2006}:
\begin{equation}\label{eq:mj-orthogonal-conditions}
  \int \omega(p^0,T_0) J_l^{(m)}( (\tilde{p}^{\mu}) ) J_k^{(n)}( (\tilde{p}^{\mu}) ) \frac{\diff^d \tilde{p}}{\tilde{p}^0} 
  = 
  \sum_i w_i J_l^{(m)}( (\tilde{p}^{\mu}_{i}) )J_k^{(n)}( (\tilde{p}^{\mu}_{i}) ) = \delta_{lk}\delta_{mn} \quad ,
\end{equation}
with $(\tilde{p}^\mu_{i})$ the discrete $(d+1)$ momentum vectors.
A convenient parametrization of $(\tilde{p}^\mu_{i})$ writes as follows:
\begin{equation}\label{eq:discrete-momentum-vectors}
  (\tilde{p}^\mu_{i}) = p^0_i (1, v_0 n_{i}) \quad ,
\end{equation}
where $n_{i} \in \mathbb{Z}^d$ are the vectors forming the stencil $G = \{ n_i ~|~ i = 1,2,\dots,i_{max} \}$ 
defined by the (on-lattice) quadrature points, $v_0$ is a free parameter that can be freely chosen such 
that $v_i = v_0 || n_i || < 1, \forall i$, and $\tilde{p}_i^0$ is defined as
\begin{equation}\label{eq:discrete-p0}
  \tilde{p}_i^0 = \tilde{m} \gamma_i = \tilde{m} \frac{1}{\sqrt{1 - |v_i|^2}} \quad .
\end{equation}
In order to determine a quadrature we proceed as follows: i) select a specific value for $\tilde{m}$, 
ii) choose a set of velocity vectors $G$, containing a sufficient number
of elements such that the left hand side of Eq.~\ref{eq:mj-orthogonal-conditions} is a full ranked matrix,
iii) look for a solution of Eq.~\ref{eq:mj-orthogonal-conditions} formed by non-negative weights 
$(w_i \ge 0, \forall i)$.

\begin{figure}[htb]
  \centering
  \includegraphics[width=.48\columnwidth]{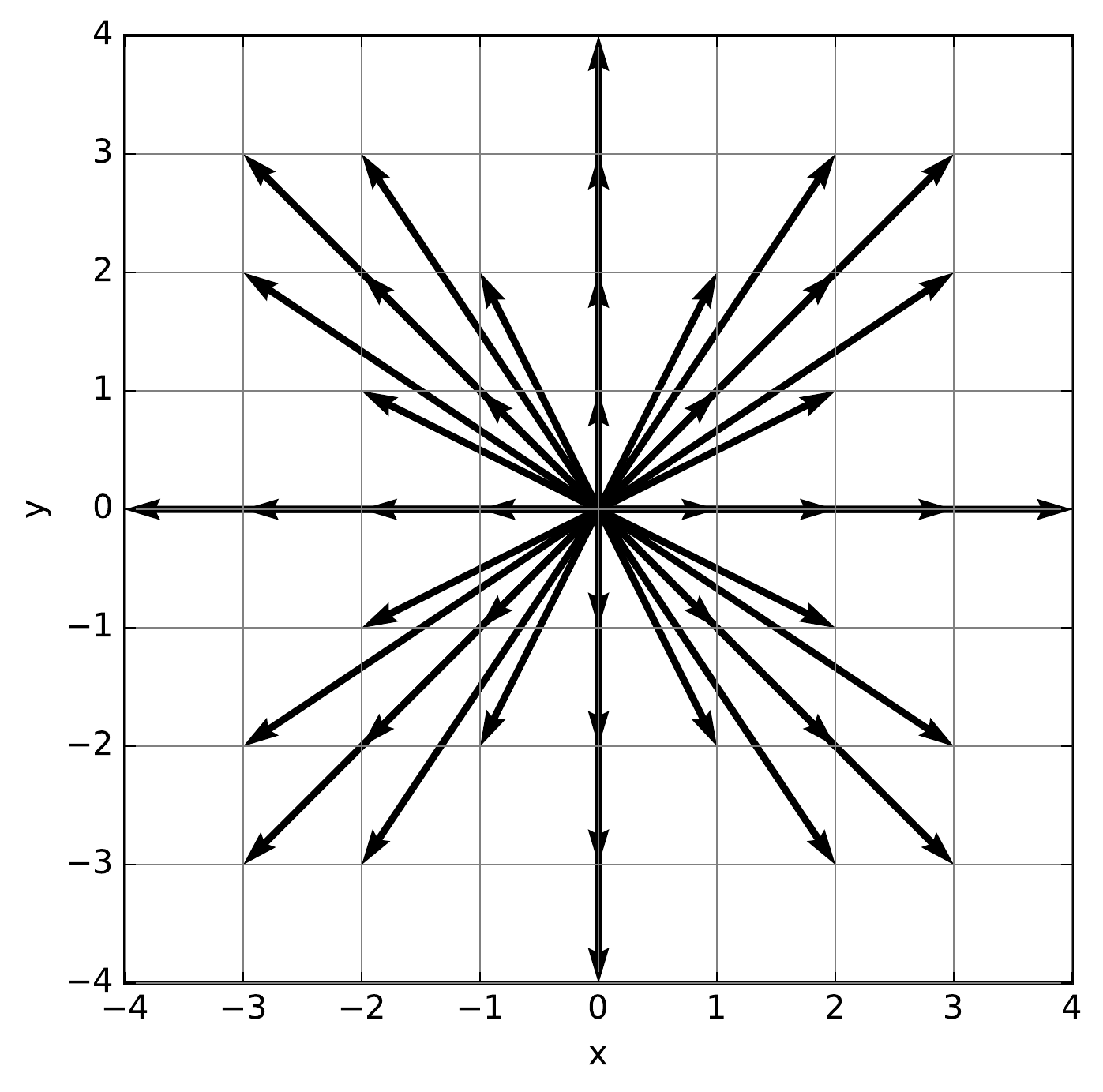} 
  \includegraphics[width=.48\columnwidth]{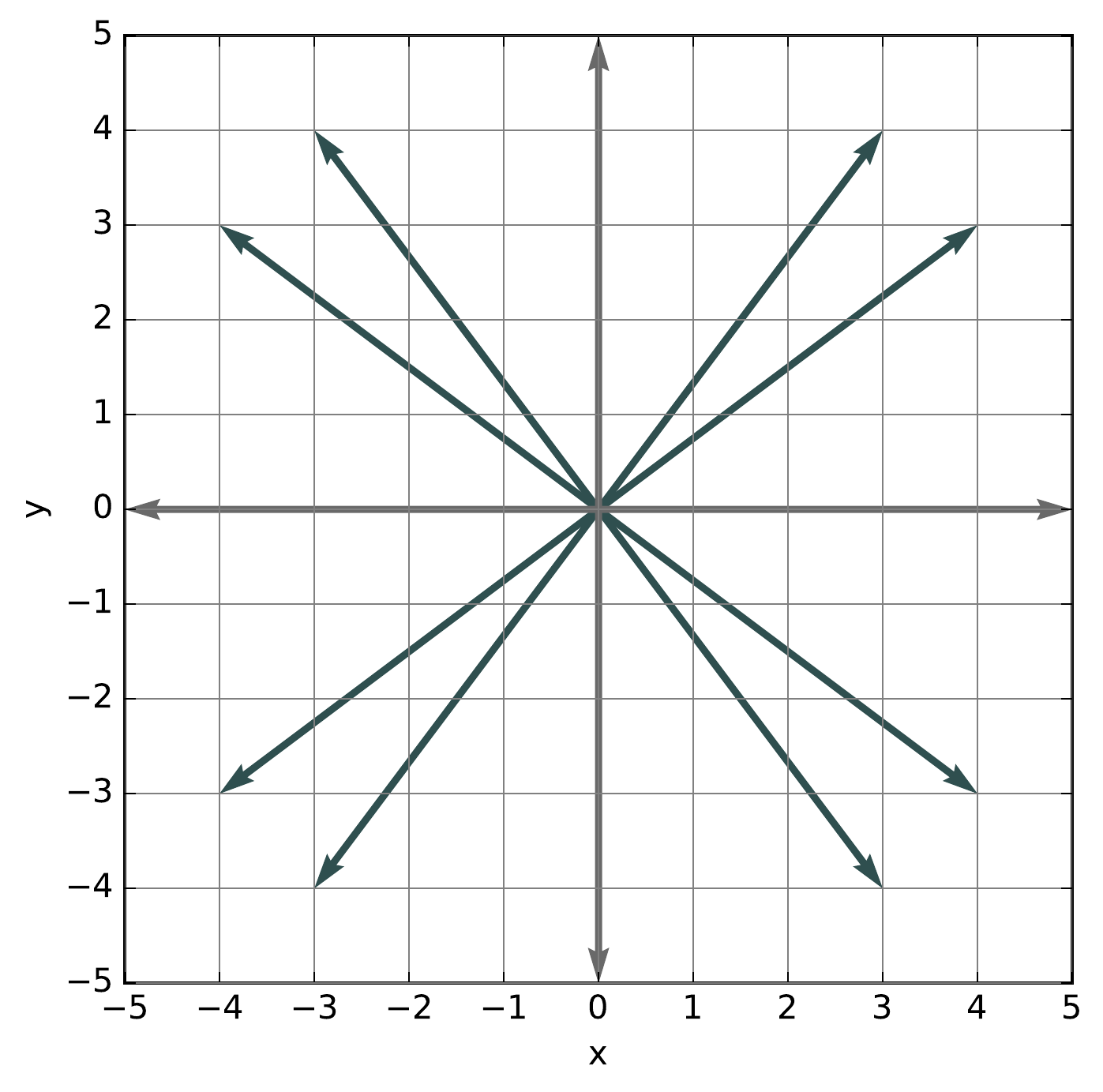} 
  \caption{ Two examples of stencil compatible with a third order quadrature,
            respectively for $\tilde{m} = 5$ (left) and $\tilde{m} = 0$ (right).
            The points forming the stencil for $\tilde{m} = 0$ lie on the intersection 
            between the Cartesian grid and a circle of radius $5$.
          }\label{fig:stencil}
\end{figure}

Observe that while the parametrization in Eq.~\ref{eq:discrete-momentum-vectors} is general 
and can be used to determine quadratures for wide ranges of values of $\tilde{m}$, 
the limit case of massless particles requires a slightly different approach,
as Eq.~\ref{eq:discrete-p0} is not well defined  for $\tilde{m} = 0$; in this case
we let $\tilde{p}^0_i$ be free parameters (as already suggested in \cite{mendoza-prd-2013}) to be
determined such as to satisfy Eq.~\ref{eq:mj-orthogonal-conditions}. 
We can have several energy shells associated to each vector and therefore we add a second index to
Eq.~\ref{eq:discrete-momentum-vectors}:
\begin{equation}\label{eq:ur-discrete-momentum-vectors}
  \left(\tilde{p}^\mu_{i,j}\right) = \tilde{p}^0_j \left(1, \frac{n_i}{||n_i||}\right) \quad ,
\end{equation}
where the index $j$ labels different energy shells, and $||n_i||$
has to be the same for all the stencil vectors since all particles travel
at the same speed $v_i = c = 1, \forall i$.  
Examples of stencils in $2d$ for the massive and massless case are shown in Fig.~\ref{fig:stencil}. 

As a concrete example, we consider the $(2+1)$ dimensional case and solve Eq.~\ref{eq:mj-orthogonal-conditions}
with $\{J^{(i)}, i = 1,2\dots \}$ the orthogonal polynomials in Appendix~\ref{sec:appendixF},
$(\tilde{p}^\mu_i)$ the three-momentum vectors following the parametrization in 
Eq.~\ref{eq:discrete-momentum-vectors} and $w_i$ suitable weights.
We follow the procedure described in \cite{shan-pre-2010,shan-jocs-2016}, 
building a stencil by adding as many symmetric groups as necessary to match the number of linearly independent 
components of Eq.~\ref{eq:mj-orthogonal-conditions}. For example, considering quadratures giving a
second-order approximation, the system of Eqs.~\ref{eq:mj-orthogonal-conditions} has
$6$ linearly independent components, so one needs to build a stencil with (at least) $6$ different symmetric groups.
Likewise, at third order there are $10$ independent components, so we need $10$ groups. Yet higher order approximations
require stencils with even larger numbers of groups.

Having selected a numerical value for the rest mass $\tilde{m}$, and a stencil $G = \{ n_i ~|~ i = 1,2,\dots,i_{max} \}$, 
Eq.~\ref{eq:mj-orthogonal-conditions} leads to a linear system of equations, parametric on $v_0$:
\begin{equation}
  A(v_0) \bm{w} = \bm{b} \quad .
\end{equation}
Here $A$ is a $l \times k$ matrix ($l$ being the number of possible combinations of the orthogonal polynomials, $k$
the number of groups forming the stencil), $\bm{b}$ is a known binary vector, and $\bm{w}$ is the vector of unknowns.
Since the Gaussian quadrature requires strictly positive weights in order to guarantee numerical stability,
we need to select values of $v_0$ (if they exist) such that $w_i > 0 ~ \forall i$. For low-order approximations it is possible 
to compute an analytic solution, writing each weight $w_i$ as an explicit function of the free parameter $v_0$, but this become 
quickly very hard and, already at the second-order, numerical solutions are necessary.
A possible formulation of the problem writes as follows:
\begin{equation}\label{eq:quadrature-problem}
\begin{aligned}
  \min   & \begin{bmatrix} -\bm{c}_1 \\ ~~\bm{c}_2 \end{bmatrix}^T \begin{bmatrix} \bm{w}^{-} \\ \bm{w}^{+} \end{bmatrix}         \quad &, \\
  s.t. ~ & A(v_0) \bm{w}  = \bm{b}                            \quad &, \\
         & 0 < v_0 \leq v_{\rm max}                           \quad &.
\end{aligned}
\end{equation}
The vector of unknowns $\bm{w}$ has been split into two sub vectors, respectively $\bm{w}^{+}$ formed by its 
nonnegative components, and $\bm{w}^{-}$ accounting the negative components. Vectors $\bm{c}_1$ and $\bm{c}_2$
are all-ones vectors matching the dimensions of $\bm{w}^{-}$ and $\bm{w}^{+}$.
We also assume that $A(v_0)$ is a fully-ranked matrix. This can be achieved applying a pre-processing phase 
where redundant rows are removed, for example by applying a $\rm QR$ or $\rm LU$ factorization.
Note that an implicit constraint on $\bm{w}$ is given by normalization factor chosen for 
the weighting function $\omega( \tilde{p}^0 )$.
For example, if the normalization factor is taken such to satisfy Eq.~\ref{eq:numerical-normalization-cond}
it follows directly that the weights will sum to unity:
\begin{equation}
  \sum_i w_i = 1 \quad .
\end{equation}

Observe that in Eq.~\ref{eq:quadrature-problem} we have not constrained $\bm{w}$ to be nonnegative.
By allowing negative values for $\bm{w}$, it is simple to find solutions using, for example,
a line search method to scan the feasible region spanned by the admissible values for $v_0$. Each solution of the minimization
procedure is then accepted only in the case $w_i \ge 0~\forall i$, as this requirement improves numerical stability and is consistent 
with a (pseudo-)particle interpretation of the RLBM. 

In general, many different solutions to the quadrature problem exist. 
We have performed a detailed exploration of the available phase-space, implementing a solver 
for Eq.~\ref{eq:quadrature-problem} based on the \textit{LAPACK} library with several instances 
running in parallel on a cluster of CPUs. The solver takes as
input a stencil $G$ and tries to find a solution for Eq.~\ref{eq:quadrature-problem} 
by scanning several values of $v_0$ with a simple line search strategy.

\begin{figure}[htb]
  \centering
  \includegraphics[width=\columnwidth]{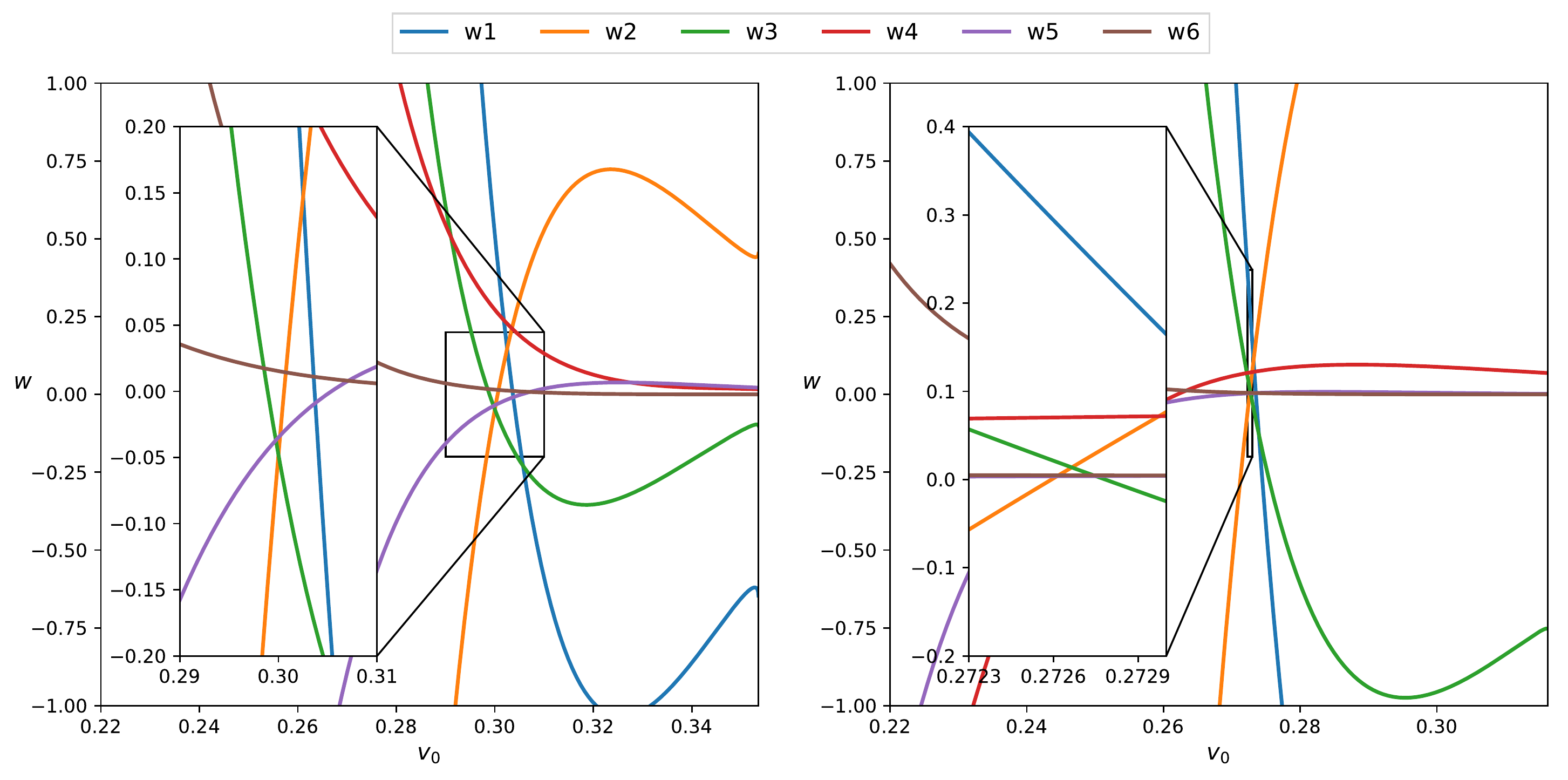} 

  \caption{Visual representation of the parametric solution of  Eq.~\ref{eq:mj-orthogonal-conditions},
           having chosen $\tilde{m} = 5$. The left panel makes use of the stencil
           $G = \{ (0, 0) \bigcup (\pm  1, 0)_{\texttt{FS}} \bigcup (\pm  1, \pm  1)_{\texttt{FS}} \bigcup (\pm  2, 0)_{\texttt{FS}}  \bigcup (\pm  2, \pm  1)_{\texttt{FS}}$ $\bigcup (\pm  2, \pm  2)_{\texttt{FS}} \}$.
           The panel on the right was obtained using the stencil
           $G = \{ (0, 0) \bigcup (\pm  1, 0)_{\texttt{FS}} \bigcup (\pm  1, \pm  1)_{\texttt{FS}} \bigcup (\pm  2, \pm 1)_{\texttt{FS}} \bigcup (\pm  2, \pm  2)_{\texttt{FS}}$ $\bigcup (\pm  3, \pm  1)_{\texttt{FS}} \}$.
           In this second case we can identify a region for which $w_i (v_0) \geq 0 ~ \forall i$  
           (orange colored interval), giving a set of solutions that can be used to build a numerically stable quadrature.
          }\label{fig:v0-vs-w}
\end{figure}

To give an example, we look for a second order quadrature in $2 d$ at $\tilde{m} = 5$ 
using the stencil 
$G = \{  (0, 0) $
                $ \bigcup (\pm  1,      0)_{\texttt{FS}} $
                $ \bigcup (\pm  1, \pm  1)_{\texttt{FS}} $
                $ \bigcup (\pm  2,      0)_{\texttt{FS}} $
                $ \bigcup (\pm  2, \pm  1)_{\texttt{FS}} $
                $ \bigcup (\pm  2, \pm  2)_{\texttt{FS}} $
$\}$, 
where $\rm FS$ stands for full-symmetric.
With this stencil, the longest displacement is given by the set of vectors with length $2 \sqrt{2}$, and therefore
the range of validity of the parameter $v_0$ is $0 < v_0 < 1 / (2 \sqrt{2})$ (this is due to the requirement $v_0 || n_i || \leq 1, \forall i$, 
used in the definition of discrete momentum vectors in Eq.~\ref{eq:discrete-momentum-vectors}). 
A visual representation of the solution for Eq.~\ref{eq:quadrature-problem} is given in Fig.~\ref{fig:v0-vs-w}a, with the
minimum found at $v_0 \sim 0.3005$; in this case we cannot determine a solution for which all the
weights of the quadrature are positive.
We then consider a different stencil
$G = \{  (0, 0) $
                $ \bigcup (\pm  1,      0)_{\texttt{FS}} $
                $ \bigcup (\pm  1, \pm  1)_{\texttt{FS}} $
                $ \bigcup (\pm  2, \pm  1)_{\texttt{FS}} $
                $ \bigcup (\pm  2, \pm  2)_{\texttt{FS}} $
                $ \bigcup (\pm  3, \pm  1)_{\texttt{FS}} $
$\}$,
for which the parameter $v_0$ takes values in $0 < v_0 < 1 / \sqrt{10})$.
From Fig.~\ref{fig:v0-vs-w}b we see that there is a small range of values of $v_0$ where
all the weights take nonnegative values.
%
%
Taking for example $v_0 = 0.2726$, the corresponding weights for the quadrature are:
\begin{align*}
  w_1 = 0.2938928682119484\dots~, \quad w_2 = 0.00136644441345044\dots~, \\
  w_3 = 0.0212650236700010\dots~, \quad w_4 = 0.07032872215612153\dots~, \\
  w_5 = 0.0036974948602444\dots~, \quad w_6 = 0.00477018784553696\dots~.
\end{align*}
Particularly convenient values of $v_0$ are those located at the boundaries of the orange colored interval in Fig.~\ref{fig:v0-vs-w}b, 
since some weights become zero thus allowing the pruning of certain lattice velocities.
In our example one can reduce the full set of $29$ velocities to $25$ by setting either $w_2$ to zero 
(with $v_0 = 0.27259285465\dots$), or $w_3$ to zero ($v_0 =0.27278322823\dots$).

Typically, for a given value of $\tilde{m}$  several different stencils are possible; 
however, each stencil works correctly only in a certain range of $\tilde{m}$.
Still, a reasonably small set of stencils  allows to treat $\tilde{m} \ge 0.5$ at the second 
order and $\tilde{m} \ge 1.2$ at the third order, offering the possibility to cover a very large 
kinematic regime, from almost ultra-relativistic to non-relativistic.

In general, the process of finding quadratures becomes harder and harder as the
order is increased and as $\tilde{m}$ takes smaller and smaller values.  
The reason is that for $\tilde{m} \rightarrow 0$ the pseudo-particles tend to move
all with similar velocities, close to the speed of light, making it difficult
to identify a stencil where all particles travel in one time step at 
different (yet very similar) distances, and still hop from a point of the grid to a neighboring one.

For the limiting case where $\tilde{m} = 0$ this translates in restricting to stencils whose elements sit at the 
intersection between a Cartesian grid and a sphere of given radius. In this case we introduce the parametrization presented in Eq.~\ref{eq:ur-discrete-momentum-vectors},
where following \cite{mendoza-prd-2013} we associate several energy shells to each momentum vector. 
To give an example, we consider a second order quadrature rule for $\tilde{m} = 0$ and solve Eq.~\ref{eq:mj-orthogonal-conditions}
by taking the stencil $G = \{ (\pm  3, \pm  4)_{\texttt{FS}}, (\pm  5, 0)_{\texttt{FS}} \}$ (Fig.~\ref{fig:stencil}b) and the
parametrization in Eq.~\ref{eq:ur-discrete-momentum-vectors} where three different energy shells get associated to each momentum vector.
The solution reads as follows:
\begin{align*}
  \tilde{p}_1^0 = 0.41577455678 \dots &~~~~~~~~~ w_{11} = 0                   & w_{21} = 0.08888662624 \dots \\
  \tilde{p}_2^0 = 2.29428036027 \dots &~~~~~~~~~ w_{12} = 0                   & w_{22} = 0.03481471669 \dots \\
  \tilde{p}_3^0 = 6.28994508293 \dots &~~~~~~~~~ w_{13} = 0.00175356541 \dots & w_{23} = 0.00042187435 \dots   
\end{align*}
The procedure can be iterated at higher orders, although already at order 4 in 2 spatial dimensions one needs to employ stencils with  vectors of length $5 \sqrt{13}$, which is impractical from a computational point of view since implies using very large grids to achieve an adequate spatial resolution; things become even more problematic in $(3+1)$ dimensions. Higher orders would most probably require different strategies, e.g.
off-lattice schemes, which drastically improve the spatial resolution of
the grid, but have as drawbacks the need for interpolation and the introduction
of artificial dissipation effects \cite{coelho-cf-2018,blaga-aip-2017,ambrus-prc-2018}; we do not consider these strategies in this paper.

A special treatment is needed in the $(1+1)$ dimensional case for the massless limit. 
Indeed, as already remarked in the previous section, in this case the massless limit of both polynomials and projections diverges.
It follows that we cannot derive the quadrature through Eq.~\ref{eq:mj-orthogonal-conditions}.
However, we can exploit the fact that it is still possible to obtain an expression for the expansion of the 
equilibrium distribution using Eq.~\ref{eq:mj-expanded-1D-massless}.
We can then express the quadrature problem via the following system of equations:
\begin{equation}
  \int f^{\rm eq}( (\tilde{p}^{\alpha}), (U^{\alpha}), \tilde{T})  \tilde{p}^{\alpha} \dots \tilde{p}^{\omega}  \frac{\diff \tilde{p}^d }{\tilde{p}^0} 
  = 
  \sum_i \sum_j w_{i j} f_{N}^{\rm eq} ( (\tilde{p}_{i j}^{\alpha}), (U^{\alpha}), \tilde{T}) \tilde{p}_{i j}^{\alpha} \dots \tilde{p}_{i j}^{\omega},
\end{equation}
where we explicitly require the preservation of all the moments of the distribution up to a desired order $N$ and use 
all the techniques described before to look for the unknown weights $w_{i j}$.

A graphical view of (a subset) of all stencils that we have found at the $2$-nd and $3$-rd order is
shown in Appendix~\ref{sec:appendixH}, . 
%

\subsection{Forcing Scheme}\label{sec:forcing}

The definition of force in relativity is subject to a certain degree of arbitrariness
due to the lack of certain general properties such as, for example, Newton's third law. 
In the following we will use the definition of Minkowski force \cite{cercignani-book-2002}:
\begin{equation}\label{eq:minkowski-force}
  K^{\alpha} = \frac{\diff p^{\alpha}}{\diff \tau} \quad ,
\end{equation}
where $\diff \tau = \frac{1}{\gamma} \diff t$ is the proper time.
From the definition it follows that the
spatial components of $(K^{\alpha})$ obey
\begin{equation}\label{eq:link-minkowski-classic-force}
  \bm{K} = \gamma ~ \bm{F}  \quad ,
\end{equation}
with $\bm{F}$ the non-relativistic force vector,
whereas the time component is such to satisfy
\begin{equation}
  K^{\alpha} p_{\alpha} = K^0 p^0 - \bm{K} \cdot \bm{p} = 0 \quad .
\end{equation}
Starting from Eq.~\ref{eq:rbe-rta-lattice-friendly}, our task consists in discretizing  the term 
\begin{equation}\label{eq:forcing-term}
  F^{\rm ext} = \frac{m K^{\alpha}}{p^0} \frac{\partial f }{\partial p^{\alpha}}  \quad .
\end{equation}
by taking into consideration the effects of external forces on the (pseudo)-particles used in our description.

Following \cite{shan-jofm-2006}, we assume the distribution function to be not far from equilibrium,
\begin{equation}
  \frac{\partial f}{\partial p^{\alpha}} \approx \frac{\partial f^{\rm eq}}{\partial p^{\alpha}} \quad ;
\end{equation}
at this point, we use the polynomial expansion of the equilibrium distribution 
\begin{equation}
  \frac{\partial f}{\partial p^{\alpha}} \approx \frac{\partial f^{\rm eq}}{\partial p^{\alpha}} 
  = 
  \omega( p^0) \sum_{k = 0}^{\infty} b^{(k)}( (U^{\mu}), T) J^{(k)} ( (p^{\mu}) ) \quad ,
\end{equation}
with the projection coefficients defined as
\begin{equation}
  b^{(k)}( (U^{\mu}), T) = \int \frac{\partial f^{\rm eq}( (p^{\mu}), (U^{\mu}), T) }{\partial p^{\alpha}}
                           J^{(k)}( (p^{\mu})) \frac{\diff^d p}{p^0}  \quad ;
\end{equation}
an even simpler approach starts from the observation that the derivative of
the analytic Maxwell-J\"uttner distribution is given by
\begin{equation}
  \frac{\partial f^{\rm eq}}{\partial p^{\alpha}} = - \frac{U_{\alpha}}{T} f^{\rm eq} \quad ,
\end{equation}
leading to
\begin{equation} 
  F^{\rm ext}
  \approx 
  -\frac{m}{T} \frac{K^{\alpha} U_{\alpha}}{p_0} f^{\rm eq} \quad ,
\end{equation}
which has the clear advantage of requiring one single evaluation of the equilibrium distribution. Both approaches yield consistent results, as we show later on.

As a general remark, the use of the polynomial expansion of the equilibrium distribution is not particularly useful in the relativistic case as it is non trivial to identify the relationship between the coefficients $a^{(k)}$ and $b^{(k)}$ leading to cumbersome analytical form of the resulting expressions and significant computational overheads in the evaluation of the external force; this is even more so, as in our case it is not possible -- as customary in non relativistic LB methods -- to translate the effect of an external force into
a shift in the macroscopic variables of interest \cite{shan-pre-1993, shan-pre-1994, guo-pre-2002, sbragaglia-jofm-2009}.

\section{Numerical recipes} \label{sec:numericalRec}

In this section we provide details on how to implement a RLBM simulation. 
We discuss the conversion from physics to lattice units, 
the numerical scheme and a few practical aspects related
to implementations on modern parallel architectures.

\subsection{From Physical to Lattice units}\label{sec:conversion-phys-lattice}

To relate physical space and time units with the corresponding lattice units, it is convenient
to start by assigning the physical length $\delta x$, corresponding to one lattice spacing. 
Suppose we use $N$ grid points to represent the physical length $L$, the corresponding
lattice spacing $\delta x$ is then:
\begin{equation}
  \delta x = \frac{L}{N} \quad .
\end{equation}

Time and space units are implicitly linked via Eq.~\ref{eq:discrete-rbe}, 
where at each time step pseudo-particles move from position $\bm{x}$ to 
$\bm{x} + \bm{v}^{i} \delta t$. Since we constrain both source and destination 
positions to lie on a Cartesian grid, it follows:
\begin{equation}
v_0 N_{i} \delta t = N_{i} \delta x \quad ,
\end{equation}
with $N_{i} \in \mathbb{Z}^d$. This, in turn, provides the following 
relation between time and space units in the lattice:
\begin{equation}\label{eq:v0dt=dx}
  v_0 = \frac{\delta x}{\delta t} \quad , 
  \frac{v_0}{c} = \frac{\delta x}{c \delta t} 
\end{equation}

The conversion of all mass and energy related quantities is performed by choosing a value for the reference temperature $T_0$ and a corresponding value for the reference energy $k_B T_0$, already encountered in the previous sections in the definition of non-dimensional quantities on the lattice.
While the choice of $T_0$ is in principle arbitrary, a sensible choice can have a major impact on the accuracy of the results.
In fact one can expect better results when $T_0$ is chosen such that the numerical values of the temperature in lattice units are $\sim 1$, since such value was used as expansion origin for the equilibrium distribution function.

At this point we have defined the translation of lengths, time and mass units between physics and lattice.
The conversion of other derived quantities follows straight.
In the following, we provide a few examples, where we distinguish between
physics and lattice units, indicating quantities with a $p$ or $l$ subscript respectively.
The conversion of the particle number density writes as
\begin{equation}
  n_p = n_l \frac{1}{(\delta x)^d} \quad .
\end{equation}
Similarly, a generic velocity can be converted using
\begin{equation}
  v_p = v_l \frac{\delta x}{\delta t} = v_l ~ v_0 \quad .
\end{equation}
As a final example we translate in lattice units the shear viscosity, 
for which we take the general expression
\begin{equation}
  \eta_p = f(\zeta) P_p \tau_p  \quad ,
\end{equation}
with $f(\zeta)$ a function solely depending on the dimensionless relativistic parameter $\zeta$.
Using the EOS of an ideal gas we can write:
\begin{equation}
  \eta_p = f(\zeta) n_p k_B T_p \tau_p 
         = f(\zeta) \frac{n_l}{(\delta x)^d} T_l k_B T_0 \tau_L \delta t
         = \eta_L k_B T_0 \frac{\delta t}{(\delta x)^d} \quad .
\end{equation}

\subsection{Relativistic Lattice Boltzmann Algorithms}

The initial conditions for the RLBM algorithm consist in prescribing
the values of $f_i( \bm{x}, t_0 )$ at the initial time $t_0$. 
A typical choice is to prescribe the equilibrium distribution function
with a given initial profile for temperature, density and velocity, thus setting $f_i( \bm{x}, t_0 ) = f_i^{\rm eq}$.

For each time step, the following operations are performed to evolve 
the distribution function at each single grid point:

\begin{enumerate} 
  \item We start by computing the first and second moment of the distribution:
        $$ N^{\alpha}       = \sum_i f_i \tilde{p}^{\alpha}_i \quad ,$$
        $$ T^{\alpha \beta} = \sum_i f_i \tilde{p}^{\alpha}_i \tilde{p}^{\beta}_i \quad .$$
  \item The energy density $\epsilon$ and the four velocity $U^{\alpha}$ are obtained solving the eigenvalue problem:
        $$ \epsilon U^{\alpha} = T^{\alpha \beta} U_{\beta} \quad , $$
        with $\epsilon$ corresponding to the largest eigenvalue of $T^{\alpha \beta}$, 
        and $U^{\alpha}$ being the correspondent eigenvector.
  \item Next, we compute the particle density from
        $$ n = U_{\alpha} N^{\alpha} \quad .$$
  \item We then compute the temperature from the EOS (see Section~\ref{sec:ideal-hydro}).

  \item We now have all the fields required to compute the equilibrium distribution function:
        $$ f_{i N}^{\rm eq}( (\tilde{p}^{\mu}), (U^{\mu}), \tilde{T}) 
           = 
           w_i \sum_{k = 0}^{N} a^{(k)}( (U^{\mu}), \tilde{T}) J^{(k)} ( (\tilde{p}_i^{\mu})) \quad . 
        $$

  \item If present, we compute the Minkowski forcing term (see Section~\ref{sec:forcing}).

  \item We determine the local value of the relaxation time $\tau$ (which typically is either 
        a constant or determined in such a way that the ratio of shear viscosity and entropy density $\eta / s$ is constant).

  \item Finally, we evolve the system over one time step, via the discrete Boltzmann equation:    
        $$
        f_i(\bm{x} + \bm{v}^{i} \Delta t, t + \Delta t)
        = 
        \left( 1 - \Omega \right) f_i(\bm{x}, t) + \Omega f_i^{\rm eq}(\bm{x}, t) + F_i^{\rm ext} 
        $$
        where
        $$
        \Omega = \Delta t~ \frac{\tilde{p}_i^{\mu} U_{\mu}}{\tilde{p}^0 \tau} \quad .
        $$
        is the dimensionless relaxation parameter controlling the transport coefficients.
\end{enumerate}
%
\subsection{Parallel implementation on GPUs.}

One of the main reasons for the widespread resort to LBM algorithms is computational efficiency \cite{succi-cf-2019}. 
The strength of LBM is embedded in the stream-collide paradigm, which, being numerically -exact- (zero roundoff), 
stands in contrast with the advection-diffusion scheme used in a macroscopic fluid-flow representation. 

The streaming phase consists in moving particles according to the discrete velocities 
defined by the stencil, and thus, unlike advection, following a regular pattern
regardless of the complexity of the fluid flow.
Moreover, streaming is exact in the sense that there is no round-off error
since it consists only of memory shifts, with no floating-point operations involved.
We remark that efficient memory access has become a main point of optimization 
in the modern large-scale implementations \cite{shet-pre-2013,shet-ijmp-2013,calore-ijfpca-2019,succi-cf-2019}.

Instead, the collide step performs all the floating-point operations required 
to implement the collisional operator. The locality of the collisional operator
makes it possible to update each grid point in parallel, making LBM an excellent
target for highly scalable implementations on modern HPC architectures.

The relativistic formulation presented in the previous sections preserves all the 
computational virtues of the classical algorithm.
The complexities in the analytic expressions of the polynomial expansion of the 
distribution, in the EOS (etc..), reflects in a significantly higher demand of floating point
operations required to update a single grid point, easily one or two orders of 
magnitude more with respect to the classical LBM.

In Tab~\ref{tab:comp-figures}, we collect a few figures of merit regarding 
the performances of RLBM codes in 2 and 3 dimensions on a recent NVIDIA Pascal GPU.
The simulation parameters are the same used in the Green-Taylor vortex benchmark
described in Section~\ref{sec:calibration-shear}. We simulate the specific case $\zeta = 5$
on a periodic square grid of $512^2$ points in 2d, and $192^3$ points in 3d, using 
the following stencils: 
$  G = \{ (0, 0)$  $\bigcup (\pm  1,      0)_{\texttt{FS}} $ $ \bigcup (\pm  1, \pm  1)_{\texttt{FS}} $ $ \bigcup (\pm  2,      0)_{\texttt{FS}}$,
                   $\bigcup (\pm  2, \pm  1)_{\texttt{FS}} $ $ \bigcup (\pm  2, \pm  2)_{\texttt{FS}} $ $ \bigcup (\pm  3,      0)_{\texttt{FS}}$,
                   $\bigcup (\pm  3, \pm  2)_{\texttt{FS}} $ $ \bigcup (\pm  3, \pm  3)_{\texttt{FS}} $ $ \bigcup (\pm  4,      0)_{\texttt{FS}}$ 
      $\}$,
for the 2d case, and
$G  = \{ (0, 0, 0)$
                    $\bigcup (\pm 1,     0,     0)_{\texttt{FS}} $ $\bigcup (\pm 1, \pm 1, \pm 1)_{\texttt{FS}}$
                    $\bigcup (\pm 2,     0,     0)_{\texttt{FS}} $ $\bigcup (\pm 2, \pm 1, \pm 1)_{\texttt{FS}}$
                    $\bigcup (\pm 2, \pm 2,     0)_{\texttt{FS}} $ $\bigcup (\pm 2, \pm 2, \pm 1)_{\texttt{FS}}$
                    $\bigcup (\pm 2, \pm 2, \pm 2)_{\texttt{FS}} $ $\bigcup (\pm 3,     0,     0)_{\texttt{FS}}$
                    $\bigcup (\pm 3, \pm 1, \pm 1)_{\texttt{FS}} $ $\bigcup (\pm 3, \pm 2,     0)_{\texttt{FS}}$
$\}$
for the 3d case, where we recall $\texttt{FS}$ stands for full-symmetric.

Thanks to the high arithmetic intensity of the algorithm 
(defined as the ratio of total floating-point operations to total data movement),
it is comparatively simple to sustain a large fraction of the performance peak of the target architecture.
A detailed analysis on the GPU-porting and optimization of RLBM will be reported elsewhere.

\begin{table}[tbh]
%
%
\centering
\begin{tabular}{lrr}
                                      &              &                 \\
\toprule 
                                      & 2D RLBM ($\zeta = 5$) & 3D RLBM ($\zeta = 5$) \\
\midrule                                    
Stencil vectors                       & 45           &  143            \\
FLOP/site                             & $\sim 66000$ &$  \sim 210000$  \\
Arithmetic intensity                  & 92           &  92             \\
Collide MLUPS                         & 55           &  14             \\
Collide TFLOPS (${\cal E}_c$ \%)      & 3.7 (70 \%)  &  3.1 (60 \%)    \\
\bottomrule 
\end{tabular}
%
\caption{
Overview of the performances of a NVIDIA Pascal P100 GPU (1792 cores for a performance peak of 
5.3 TFLOPS in double precision arithmetics) running our RLBM codes in 2 and 3 space dimensions. 
We list the number of floating point operations (FLOP) required to update one 
grid point, the arithmetic intensity and the performance of the collide kernel,
expressed both in MLUPS (Million Lattice Updates per Second) and TFLOP per second.
${\cal E}_c$ is an estimate of the GPU sustained performance with respect to peak performance.}\label{tab:comp-figures}
\end{table}

\section{Numerical Results I: Calibration of Transport Coefficients} \label{sec:calibration}

In section \ref{sec:dissipative-hydro} we have summarized the steps needed to derive 
the analytical expressions for the transport coefficients 
of an ideal relativistic gas in $(d+1)$ dimensions, using both Chapman-Enskog and Grad's moments method. 
Unfortunately, to the best of our knowledge, no experimental setup is available to discern which (if any) of the 
two methods gives the correct results. For this reason, the lattice kinetic scheme developed in the 
previous pages can be used to tell the two methods apart. 

Here we summarize and extend the results presented in \cite{gabbana-pre-2017b,coelho-cf-2018,gabbana-pre-2019,gabbana-arx-2019},
which show that the transport coefficients calculated following Chapman-Enskog's approach are in 
better agreement with numerical results than those obtained via using Grad's method.
We present numerical results for the shear viscosity, thermal conductivity and of the bulk viscosity as well. 
In addition, we extend previous results to $1,2,3$ space dimensions.

The numerical results presented in this section base on schemes using third order quadratures, 
which are made available as supplemental information \cite{SOM}.

\subsection{Shear viscosity}\label{sec:calibration-shear}
%
As discussed in Section~\ref{sec:dissipative-hydro}, the analytic form for the shear viscosity
predicted by the Chapman-Enskog expansion and Grad's method of moments is different.
Both methods provide results in the form
\begin{equation}\label{eq:shear-viscosity-general-form}
  \eta = f(\zeta) P \tau  \quad ,
\end{equation} 
but with a different dependence on the relativistic parameter $\zeta$, as expressed in the 
above equation by the function $f(\zeta)$ (see Appendix~\ref{sec:appendixC} for
the analytic expression of $f(\zeta)$ in the two cases and for a full comparison).

Here, we describe the procedure followed to measure $f(\zeta)$ from simulations.
We first consider an almost divergence-free flow, which allows to neglect compressible effects, and to simplify 
the energy-momentum tensor to
\begin{equation}\label{eq:2nd-order-tensor-simplified}
  T^{\alpha \beta} = (\epsilon + P) U^{\alpha} U^{\beta} - P \eta^{\alpha \beta} + \pi^{< \alpha \beta >}  \quad .
\end{equation} 
As a benchmark, we take the Taylor-Green vortex \cite{taylor-prsl-1937}, a well known example 
of a decaying flow, exhibiting an exact solution for the classical Navier-Stokes equations, 
and for which we can derive an approximate solution in the relativistic regime.
We start from the following initial conditions, in a $d > 1$ periodic domain, 
\begin{align}\label{eq:tg-initial-conditions}
  u_x &= \phantom{-} u_0 \cos (x) \sin (y) \quad x, y \in [0, 2\pi] \quad , \\
  u_y &= - u_0 \cos (y) \sin (x) \quad ,   \notag                 \\
  u_z &= u_w =  \dots = 0        \quad .   \notag
\end{align}
with $u_0$ a given value for the initial velocity,
and assume that the time dependent solution
takes the same form as in the classical case:
\begin{align}\label{eq:tg-general-solution}
  u_x &= \phantom{-} u_0 \cos(x) \sin(y) F(t) \quad x, y \in \left[0, 2 \pi\right] \quad , \\
  u_y &= - u_0 \cos(y) \sin(x) F(t) \quad , \notag \\ 
  u_z &= u_w =  \dots = 0           \quad . \notag
\end{align}
In order to determine the analytic expression of $F(t)$, we solve the conservation equations by
inserting, starting from Eq.~\ref{eq:2nd-order-tensor-simplified}:
\begin{equation}\label{eq:sys-eqs}
  \partial_{\beta} T^{\alpha \beta} =  (\epsilon + P) U^{\alpha} \partial_{\beta} U^{\alpha} + \partial_{\beta} \pi^{<\alpha \beta>} = 0
\end{equation}
Next, we plug Eq.~\ref{eq:tg-general-solution} into Eq.~\ref{eq:sys-eqs}
and perform a first order expansion of the resulting expression, giving
\begin{equation}
  \begin{bmatrix}
    0 \\
    \phantom{-} U^0 (\epsilon + P ) \sin{(x)} \cos{(y)} F'(t) \\
             -  U^0 (\epsilon + P ) \cos{(x)} \sin{(y)} F'(t) \\
    0
  \end{bmatrix}
+
  \begin{bmatrix}
    0 \\
    \phantom{-}2 \eta U^0 F(t) \sin{(x)} \cos{(y)} \\
             - 2 \eta U^0 F(t) \cos{(x)} \sin{(y)} \\
    0
  \end{bmatrix}
=
  \begin{bmatrix}
    0 \\
    0 \\
    0 \\
    0
  \end{bmatrix}
\end{equation}
where for improved readability we have kept separated the two additive terms of Eq.~\ref{eq:sys-eqs}.

From the above we directly get the following differential equation:
\begin{equation}
  2 \eta F(t) + (P + \epsilon) F^{'}(t) = 0 \quad ,
\end{equation}
which can be solved under the assumption of an (approximately) constant value of $P + \epsilon$:
\begin{equation}\label{eq:tg-relativistic-approx}
  F(t)  = \exp{ \left( - \frac{2 \eta}{P + \epsilon} t  \right) } F(0) \quad .
\end{equation}
%
\begin{figure}[htb]
  \centering
  \includegraphics[width=.99\textwidth]{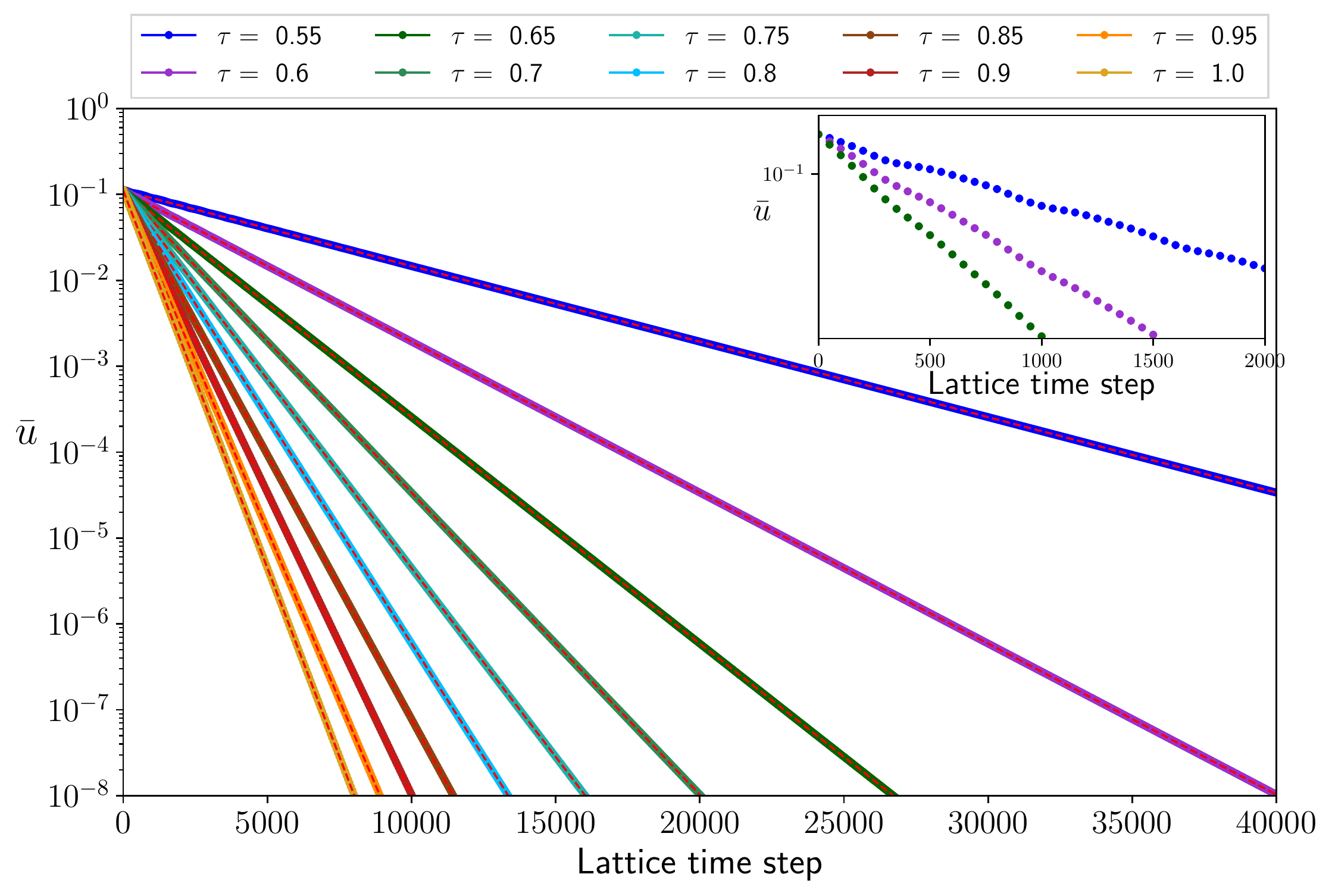}
  \caption{Simulated time evolution of $\bar{u}$ for selected $\tau$ values
           on a $L = 400$ square lattice. Simulation are performed using 
           a $(3+1)$-dimensional solver, with initial numerical parameters $\zeta = 0$, $u_0 = 0.2$, $n = 1$, $T = 1$. 
           Dashed lines are fits to the exponential decay predicted by Eq.~\ref{eq:tg-relativistic-approx}. 
           The inset shows non-linear effects in the early phases of the flow.
          }\label{fig:taylor-green-decay}
\end{figure} 

Next, it is expedient to introduce an observable $\bar{u}(t)$,
\begin{equation}
  \bar{u}^2(t) = \int_0^{2 \pi} \int_0^{2 \pi} \left(u_x^2(t,x,y) + u_y^2(t,x,y) \right) \diff x \diff y \quad ,
\end{equation}
that is directly proportional to $F(t)$, as easily seen from Eqs~\ref{eq:tg-general-solution}. 
We perform several simulations of the Taylor-Green vortex, on periodic boxes of side $L = 400$.
Temperature and density are set to unity on each grid point, while the velocity fields is 
initialized using Eq.~\ref{eq:tg-initial-conditions} and with $u_0 = 0.2$.
In order to better characterise the numerical fit of $f(\zeta)$, we consider a broad range 
of $\zeta$ values, smoothly bridging between ultra-relativistic to near non-relativistic regimes.
The relaxation of time is set to a constant value throughout each simulation, with numerical
values spanning between $0.6$ and $1.0$; at fixed time intervals we then get an 
estimate for $\bar{u}^2(t)$, calculated as:
\begin{equation}
  \bar{u}^2(t) = \frac{1}{L^2} \sum_{i=1}^L \sum_{j=1}^L u_x(t,i,j)^2 + u_y(t,i,j)^2 \quad .
\end{equation}

In Fig.~\ref{fig:taylor-green-decay}, we show a few example of simulations, 
featuring the time evolution of $\bar{u}$, with $\zeta = 0$ and for several 
different values of the relaxation time, clearly exhibiting an exponential decay. 

For each set of mesoscopic values, we perform a linear fit of $\log( \bar{u} )$
extracting a corresponding value of $\eta$ via Eq.~\ref{eq:tg-relativistic-approx}.
Finally, by comparison with Eq.~\ref{eq:shear-viscosity-general-form}, we estimate
the value of $f(\zeta)$ at different values of $\zeta$.
In Table~\ref{tab:shear-from-numerics}, we show a few results obtained following 
this procedure in the $(3+1)$-dimensional case. One appreciates that, for each different 
value of $\zeta$, measurements of $\eta(\tau)$ yield a constant value of $f(\zeta)$. 

\begin{table}
\centering 
\begin{tabular}{|c|c|c|c|c|c|c|c|}
\hline 
        & \multicolumn{7}{c|}{$f(\zeta)$} \\
\hline 
$\tau$  & $\zeta$ = 0 & $\zeta$ = 1.6 & $\zeta$ = 2 & $\zeta$ = 3 & $\zeta$ = 4 & $\zeta$ = 5 & $\zeta$ = 10 \\
\hline
  0.600 & 0.8003 & 0.8319 & 0.8448 & 0.8587 & 0.8892 & 0.8994 & 0.9311  \\
  0.700 & 0.8002 & 0.8318 & 0.8447 & 0.8584 & 0.8888 & 0.8990 & 0.9302  \\
  0.800 & 0.8002 & 0.8318 & 0.8447 & 0.8583 & 0.8887 & 0.8989 & 0.9300  \\
  0.900 & 0.8002 & 0.8318 & 0.8447 & 0.8583 & 0.8887 & 0.8988 & 0.9299  \\
  1.000 & 0.8002 & 0.8317 & 0.8446 & 0.8582 & 0.8887 & 0.8988 & 0.9299  \\
\hline 
\end{tabular}
\caption{Selected sample values for the estimate of the parameter $f(\zeta)$ 
         for several values of $\tau$ and $\zeta$. Statistical errors for all 
         entries are smaller than $1$ in the last displayed digit.
        }\label{tab:shear-from-numerics}
\end{table}

\begin{figure}[htb]
\centering
\includegraphics[width=0.99\textwidth]{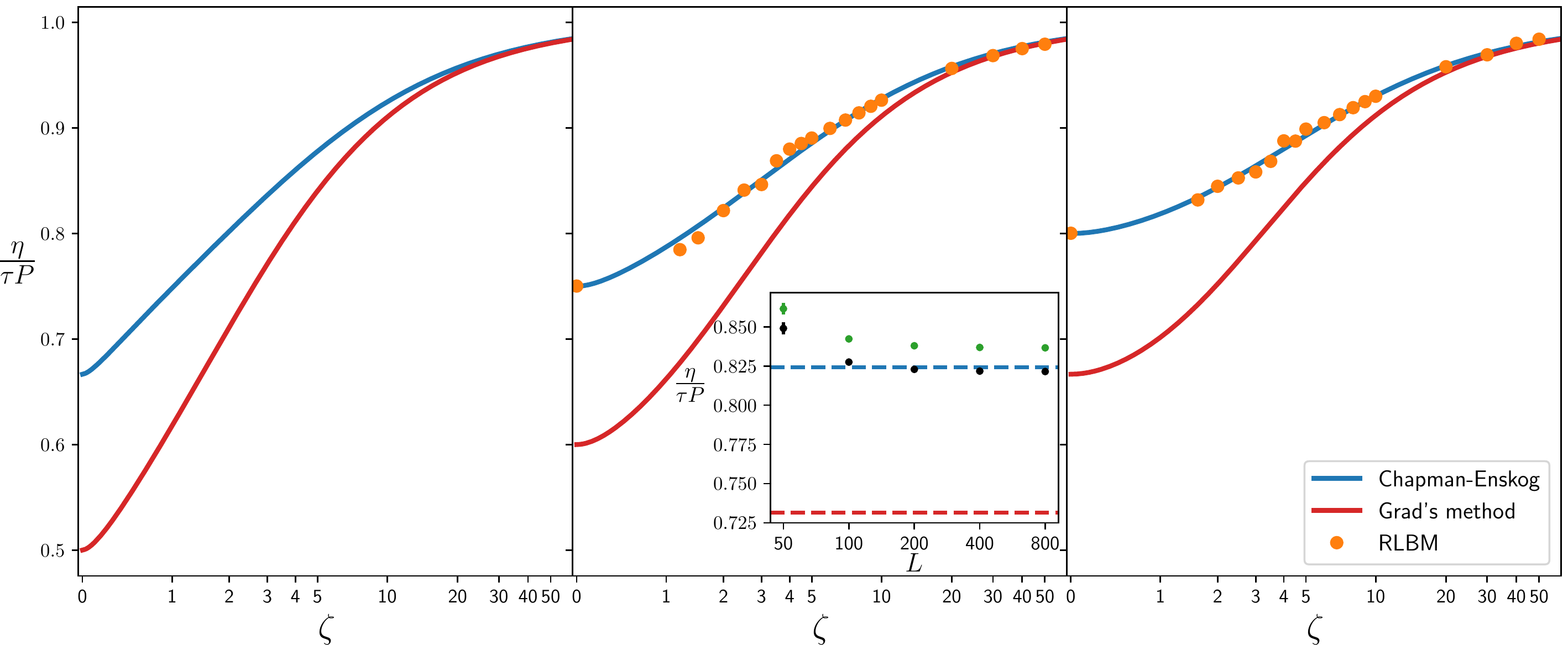}
\caption{ Comparison of the non-dimensional shear viscosity for a relativistic gas in 
          (1+1), (2+1) and (3+1) dimensions, obtained using Chapman-Enskog and Grad's methods. 
          Dots represent numerical measurements of the shear viscosity simulating the time
          decay of Taylor-Green vortex using RLBM. The inset shows a grid-convergence test 
          for the estimated value of the shear in viscosity in the bidimensional case at $\zeta = 2$;
          black and green dots represent the estimates obtained using two different third order
          quadratures (see the main text for their definition), showing small discrepancies of
          about $2 \%$.
        }\label{fig:eta-cmp}
\end{figure}

%
Moreover, from the second column of Tab.~\ref{tab:shear-from-numerics}, we obtain $f(0) = 4/5$ 
to very high accuracy, which is consistent with the result of the Chapman-Enskog expansion in 
the ultra-relativistic limit (see Eq.~\ref{eq:ce-ur-limit}).
In Fig.~\ref{fig:eta-cmp}, we show that the CE prediction almost perfectly 
matches the results of the simulations  
(and we remark that {\em no} free parameters are involved in this comparison) 
over a broad range of values of $\zeta$, in both $(2+1)$ and $(3+1)$-dimensions.
For the $(1+1)$-dimensional case, we only show the analytic results since in this
case we do not have a suitable benchmark with an approximate analytic solution
that can be used to numerically fit the curve.

To conclude, in the inset in Fig.~\ref{fig:eta-cmp} we show two effects:
i) the impact of the grid resolution on the quality of the estimate
ii) the way different quadratures provide slightly different results.
In the example shown we have focused in the case $d = 2$, $\zeta = 2$.
We perform simulations using two different quadratures, with the green
dots obtained using the stencil 
$ QA = \{ (0, 0)$, $ (\pm  1, \pm  1)_{\texttt{FS}} $, $(\pm  2, \pm  2)_{\texttt{FS}} $, $(\pm  3, \pm  4)_{\texttt{FS}}$,
                   $ (\pm  5,      0)_{\texttt{FS}} $, $(\pm  5, \pm  5)_{\texttt{FS}} $, $(\pm  6, \pm  2)_{\texttt{FS}}$,
                   $ (\pm  6, \pm  3)_{\texttt{FS}} $, $(\pm  6, \pm  4)_{\texttt{FS}} $, $(\pm  7, \pm  1)_{\texttt{FS}}$ 
      $\}$,
while the black ones base on the stencil 
$QB = \{ (0, 0)$,  $ (\pm  1, \pm  2)_{\texttt{FS}} $, $(\pm  3, \pm  4)_{\texttt{FS}} $, $(\pm  4, \pm  4)_{\texttt{FS}}$,
                   $ (\pm  5, \pm  5)_{\texttt{FS}} $, $(\pm  6,      0)_{\texttt{FS}} $, $(\pm  6, \pm  3)_{\texttt{FS}}$,
                   $ (\pm  6, \pm  4)_{\texttt{FS}} $, $(\pm  7, \pm  1)_{\texttt{FS}} $, $(\pm  7, \pm  2)_{\texttt{FS}}$ 
      $\}$.
For each quadrature we perform a convergence test with respect to the grid size, 
showing that the estimates tend to stabilize to a constant values as soon as $L \ge 200$.
We attribute the differences observed in the results provided by the
two quadratures in the different error committed in the approximation 
of the higher order tensors; note that the corresponding results 
differ from each other by  approximately 1-2\%, which we can
consider an estimate of our systematic error.

%

\subsection{Thermal conductivity}

The numerical measurements of the thermal conductivity follow the same general steps described
in detail in the previous section. We consider a numerical setup for simulating two 
parallel plates, kept at different constant temperatures $T_{\rm hi}$ and $T_{\rm lo}$,
with $\Delta T = T_{\rm hi} - T_{\rm lo}$. For sufficiently small values of $\Delta T$,
the flow can be approximated to be non-relativistic and as a consequence Eq.~\ref{eq:heat-flux}
reduces to Fourier's law:
\begin{equation}\label{eq:fourier-law}
  q^{\alpha} =   \lambda \nabla^{\alpha} T \quad .
\end{equation}
Under these settings, simulations reach a steady state with an approximately constant temperature
gradient, and a constant heat flux $q^\alpha$ which can be calculated in simulations using
Eq.~\ref{eq:first-moment-neq}: 
\begin{equation}\label{eq:heat-flux-from-first-moment}
  q^{\alpha} =  h_e (N_E^{\alpha} - N^{\alpha}) \quad .
\end{equation}
Combining the two equations above, it is possible to estimate the 
thermal conductivity $\lambda$ and discern between the expressions predicted respectively by 
Chapman-Enskog and Grad's methods (once again, refer to Appendix~\ref{sec:appendixC} for the d-dimensional
analytical form of $\lambda$ for the two cases).

Similarly to the case of shear viscosity $\eta$, we perform several simulations
varying the mesoscopic parameters $\tau$ and $\zeta$.
The simulations are performed on a rod represented using $1600$ points along the $x$ axis,
with all the other dimensions represented by $1$ single point, with periodic boundary conditions.
The leftmost and rightmost points are kept stationary by imposing the equilibrium distribution
calculated with a temperature in numerical units of respectively $T_{\rm hi} = 1.005$ and $T_{\rm lo} = 0.995$, 
zero velocity, while the density is obtained by linear interpolation from the neighboring interior points.
In order to calculate an estimate for the thermal conductivity the system is evolved until a steady 
state is reached. From the final configuration we then compute an approximation of the temperature gradient 
using central finite difference. Next, we compute spatial averages of the quantities 
$< \nabla^{\alpha} T >_x$ and $<q^{x}>_x$ which we use in combination with Eq.~\ref{eq:fourier-law} 
to get an estimate for $\lambda$.

\begin{figure}[htb]
\centering
\includegraphics[width=0.99\textwidth]{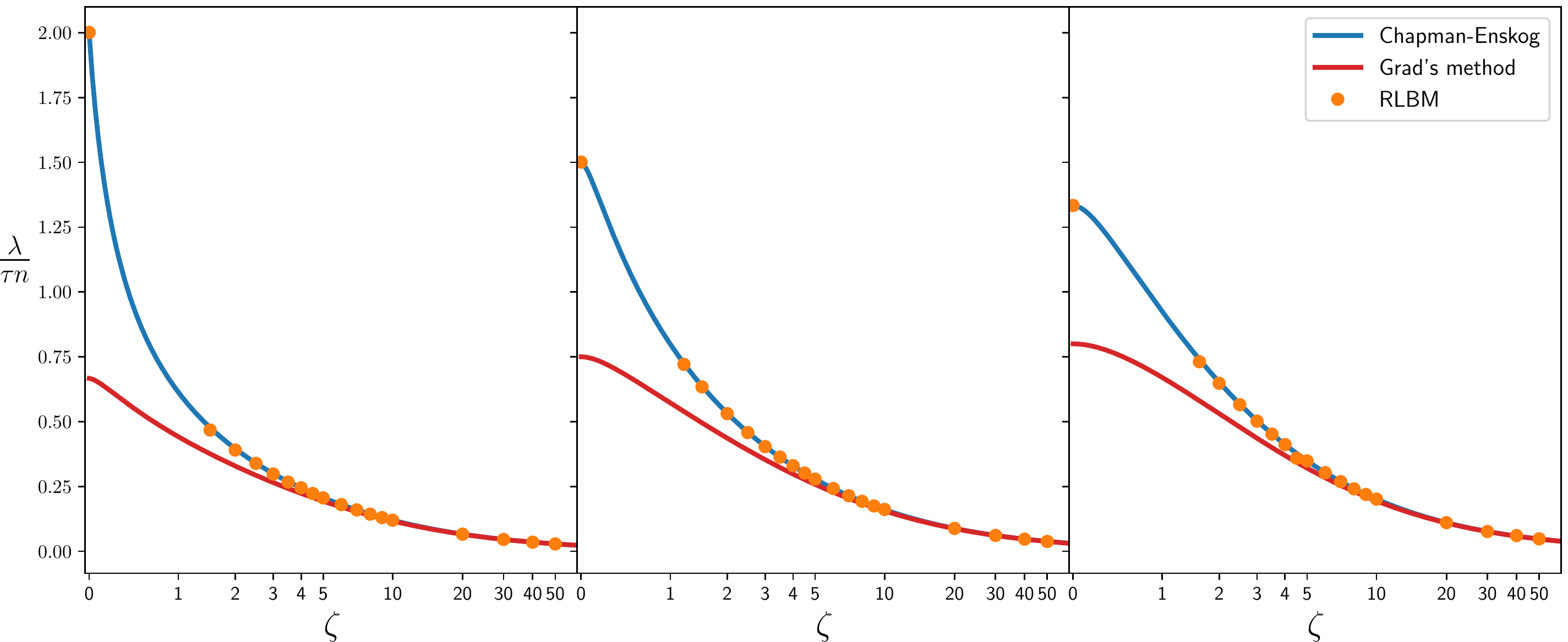}
\caption{ Comparison of the non-dimensional thermal conductivity for a relativistic gas in 
          (1+1), (2+1) and (3+1) dimensions, obtained using Chapman-Enskog and Grad's methods. 
          Dots represent numerical measurements from RLBM simulations.
        }\label{fig:lambda-cmp}
\end{figure}

The results obtained are summarized in the plots in Fig.~\ref{fig:lambda-cmp}, showing that 
in 1, 2 and 3 spatial dimensions, the numerics are in excellent agreement with the predictions 
of the Chapman-Enskog expansion.

%
%
\subsection{Bulk viscosity}
%
\begin{figure}[htb]
\centering
\includegraphics[width=0.99\textwidth]{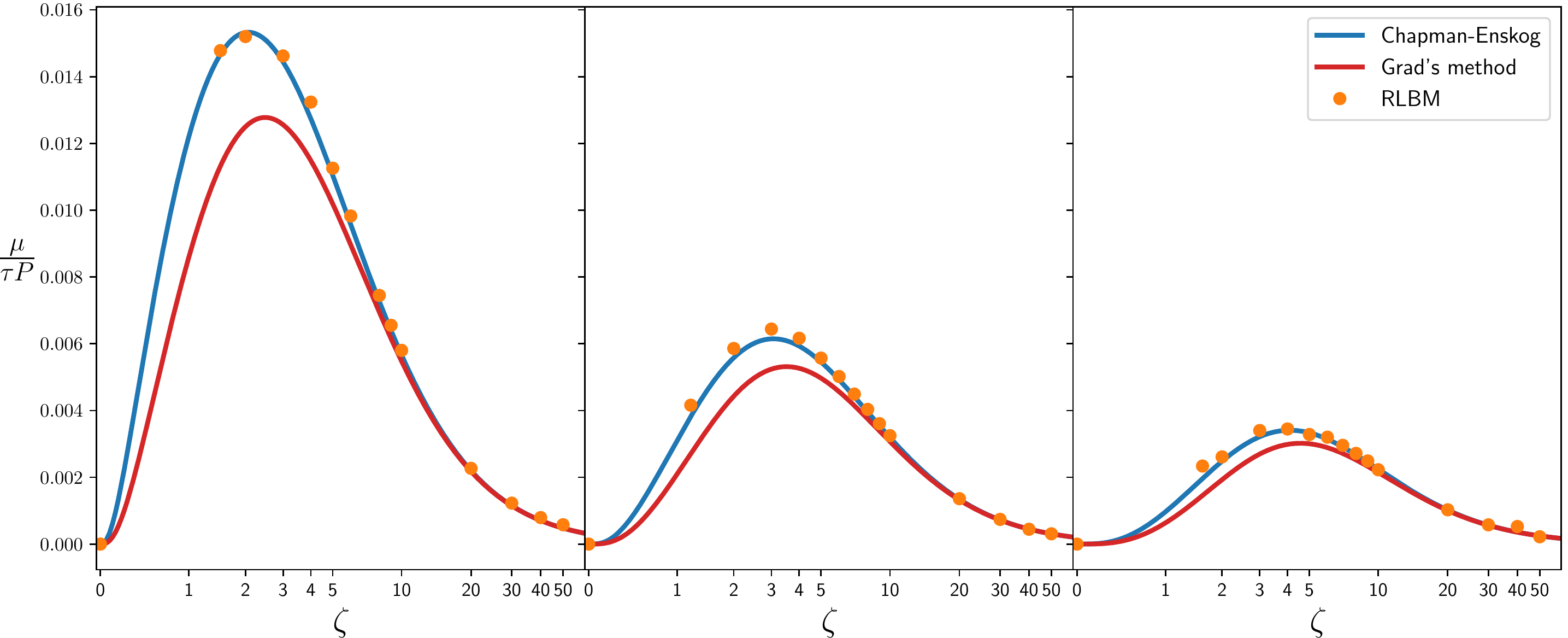}
\caption{ Comparison of the non-dimensional bulk viscosity for a relativistic gas in 
          (1+1), (2+1) and (3+1) dimensions, obtained using Chapman-Enskog and Grad's methods. 
          Dots represent numerical measurements from RLBM simulations.
        }\label{fig:mu-cmp}
\end{figure}
%
The measurement of the bulk viscosity requires the analysis of a flow with sizable compressibility effects.
A popular example that serves our purpose is the Riemann problem and
the generation of shock waves (which will be studied in detail in the forthcoming sections). 
However, the presence of strong discontinuities makes the numerical analysis challenging 
and the error committed in our data-fits too large to discriminate between the predictions 
of the analytic form of the bulk viscosity $\mu$ due to Chapman-Enskog and Grad's method.
We therefore turn to the analysis of a time decaying sinusoidal wave, which still gives the possibility 
to observe compressible effects, yet with the advantage of well behaved derivatives.

We consider a periodic domain with the following initial condition:
\begin{align}
  u_x &= u_0 \sin(x)       \quad x \in [0, 2\pi] \quad , \\
  u_y &= u_z =  \dots = 0  \quad . \notag
\end{align}
with $u_0$ a given initial velocity. 
Particle density and temperature are initially set to a constant value.
In the simulations we measure the dynamic pressure $\varpi$ from the trace 
of the energy momentum tensor (Eq.~\ref{eq:second-moment-neq}):
\begin{equation}\label{eq:varpi-from-tensor}
  \varpi = -\frac{1}{d}( T^{\mu}_{\mu}-{T_{E}}_{\mu}^{\mu})  \quad .
\end{equation}
By combining the above equation with Eq.~\ref{eq:dynamic-pressure} it is 
then possible to perform numerical measurements of the bulk viscosity through
\begin{align}\label{eq:mu-def}
  \mu = -\frac{\varpi}{\nabla_\alpha U^\alpha} \quad .
\end{align}

We perform simulations on a mono-dimensional system with $1600$ used to represent 
the $x$ axis, and $1$ single point for all the other components and periodic boundary 
conditions. Density and temperature are set to unit, while the initial amplitude 
of the wave is set to $u_0 = 10^{-5}$.
Like in the previous cases discussed in this section, we perform several simulations 
with different values of the mesoscopic parameters $\tau$ and $\zeta$, and perform
the following steps to obtain accurate measures of the bulk viscosity:
i) we follow the time evolution of the system for half a period of the sine wave
and at each time step $t$ we compute $\varpi$ via Eq.~\ref{eq:varpi-from-tensor} and
calculate an estimate of $\nabla_\alpha U^\alpha$ using a central finite difference scheme. 
ii) we calculate the spatial average $< \mu (t) >_x$ from Eq.~\ref{eq:mu-def}, ignoring points
where $| \nabla_\alpha U^\alpha | < 10^{-10}$.
iii) we improve our estimate for $\mu$ by averaging on several time steps.

The results results presented in
Fig.~\ref{fig:mu-cmp} lead to the same conclusions as for $\eta$ and $\lambda$, with 
clear evidence that the Chapman-Enskog procedure is in excellent agreement with the numerical results.

We conclude this section with a remark: As shown in Fig~\ref{fig:eta-cmp}-~\ref{fig:mu-cmp} our 
simulations do not cover the domain $0 < \zeta \lesssim 1$. The reason, as it has been 
explained in Section~\ref{sec:gauss-quad}, is that in this parameter range the definition 
of quadratures lying on Cartesian grid becomes a hard task since the physical constrains 
that one needs to satisfy are not suitable for a lattice formulation. 
While off-lattice quadratures would allow the description of wider a kinematic range,
the numerical evaluation of the transport coefficients would have been weakened 
by the numerical artifacts introduced by an interpolation scheme.

\section{Numerical Results II: Benchmark and Validation} \label{sec:bench}

In this section we provide a few validation tests together with example of applications
of the RLBM. We start with the validation of the forcing scheme which we use to 
reproduce the results of a simple non-relativistic Poiseuille-flow.
We then consider the Riemann problem, a benchmark commonly used in both non-relativistic  and relativistic numerical 
hydrodynamics, in order to assess the stability and the 
accuracy of numerical solvers. We validate the code in a nearly inviscid regime,
for which analytic solutions are available, then explore viscous regimes 
for fluid of both mass-less and massive particles, comparing with other numerical
solvers available in the literature.

Next, we give an example of simulation in three spatial dimensions, 
relevant for the study of the early stage formation of the quark-gluon plasma.

We conclude by presenting  simulations of the electrons flow in graphene,
studying realistic setups which have been recently used in actual experiments.

\subsection{Validation of the forcing scheme}\label{sec:forcing-validation}

In Section~\ref{sec:forcing} we have presented two approaches to introduce a generic 
Minkowski force $K^{\alpha}$ in the numerical scheme. 
Benchmark that can be adopted for testing the correctess 
of the forcing scheme are not abundant: among them we list Ref. \cite{greif-prd-2014}, in which a computation of electric conductivity in a QGP \cite{greif-prd-2014} is performed.
Here we follow an even simpler approach and take into consideration the effect of applying 
a weak gravitational field to the (pseudo)-particles forming a relativistic fluid.
In the non-relativistic case, the most standard benchmark is given by the Poiseuille-flow,
describing the motion of a fluid between two parallel plates under the effect of gravity
(or of a pressure gradient).

In the following, we will then directly compare with the analytic solution of the classical
Poiseuille-flow, assuming a sufficiently small gravity acceleration $\bm{g}$.
Starting from Eq.~\ref{eq:link-minkowski-classic-force}, we can define the Minkowski
force in terms of $\bm{g}$ as
\begin{equation}
  \bm{K} = \gamma ~ \bm{F}  
         = \gamma ~   m \bm{g}
         = \frac{p^0}{c} \bm{g} \quad .
\end{equation}

\begin{figure}[t]
\centering
\includegraphics[width=0.99\textwidth]{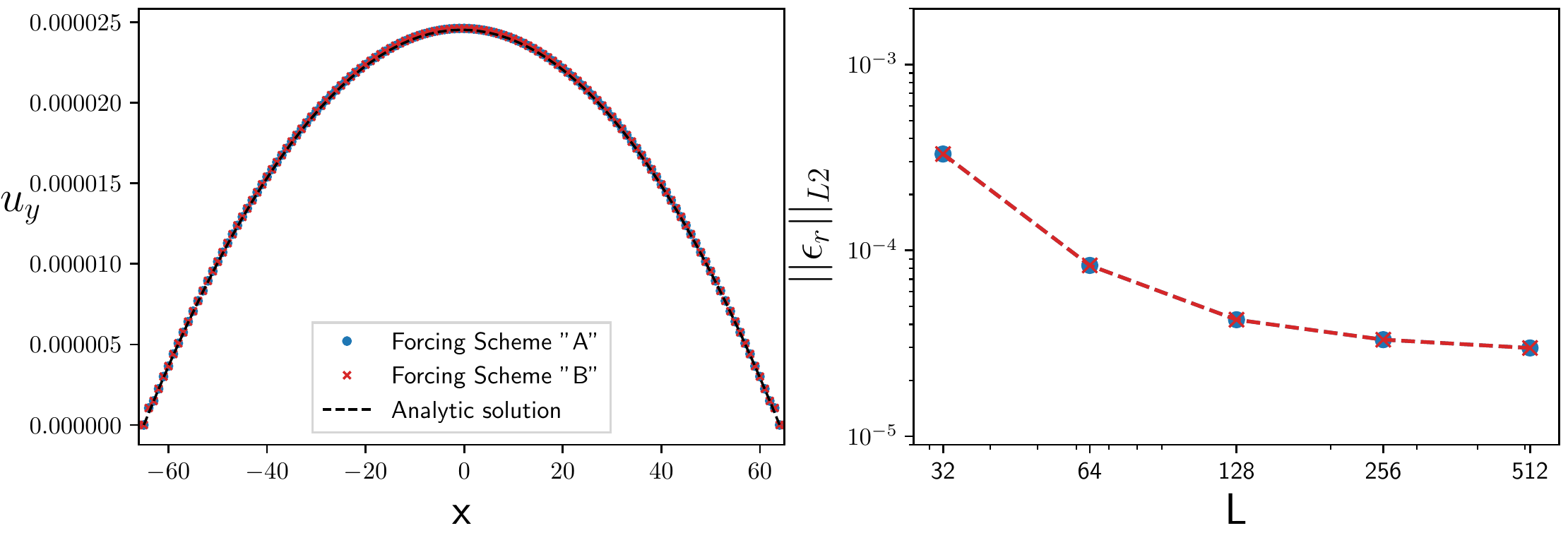}
\caption{Validation of the forcing scheme for the RLBM algorithm by solving a
         non-relativistic Poiseuille flow. 
         In the left panel we show the parabolic velocity profile obtained 
         using a RLBM scheme in two spatial dimensions, working
         on a $L \times 64$ grid, and considering two different implementations of
         the forcing scheme: 
         In scheme "A" the forcing term is discretized using a polynomial expansion,
         while for scheme "B" we compute explicitly the derivative of the 
         Maxwell-J\"uttner distribution (see Section~\ref{sec:forcing} for details).
         The panel on the right shows the relative error, computed with respect 
         to the analytic solution. Both solution provide the same level of accuracy,
         with the error decreasing as the grid size is increased.
        }\label{fig:force-validation}
\end{figure}

In Fig.~\ref{fig:force-validation} we validate the two different implementations of
the forcing scheme described in section~\ref{sec:forcing}: 
We call scheme "A" the forcing term discretized using a polynomial expansion,
while scheme "B" refers to the case where we compute explicitly the derivative 
of the Maxwell-J\"uttner distribution.
We work in two spatial dimensions and perform simulations on grids of size $L \times 64$. 
Periodic boundary conditions are applied along the $y$-axis, while at $x = 0$ and $x = L-1$
no-slip boundary conditions are used to simulate two parallel plates.
We apply a gravity-like force acting parallel to the wall boundaries, of magnitude
$|\bm{g}| = 10^{-8}$ in numerical units.

In the left panel in Fig.~\ref{fig:force-validation} we can see that both
implementations of the forcing term correctly reproduce the parabolic profile of
the fluid flow. On the right panel we also show a more quantitative comparison,
with the relative error as a function of the number of points $L$ used to
represent the box width. 
The right panel in Fig.~\ref{fig:force-validation} shows the relative error
computed with respect to the analytic solution as:
\begin{equation}
  \epsilon_r = \sqrt{ \frac{ \sum_x \left( u_y(x) - u^{\rm ref}_y(x) \right)^2 }{ \sum_x \left( u^{\rm ref}_y(x) \right)^2 } } \quad ,
\end{equation}
where we consider the analytic solution of the classic Poiseuille flow:
\begin{equation}
  u^{\rm ref}_y(x) = \frac{| \bm{g} |}{2 \nu} x(x - L)    \quad .
\end{equation}
The relative error shows saturation at a plateau value of about $10^{-5}$. 
This is due to the specific procedure adopted to apply the non-slip 
boundary conditions at the solid walls, which consists of imposing a local equilibrium 
at zero flow speed. This provides a dramatic gain in simplicity at the cost of ignoring 
non-equilibrium effects due to the near-wall velocity gradient. 
The standard bounce-back technique, ensuring second
order accuracy is conceptually straightforward but extremely laborious in the
presence of solid boundaries, due to the large number of discrete velocities
implied. For the purpose of the present benchmark, we have opted for simplicity,
also on account of the fact that, consistently with previous work \cite{mohamad-epjsp-2009},
for slow flows such as those of Fig.~\ref{fig:force-validation}, the relative
error appears to be quite negligible. Should the physical problem require the
handling of fast flows with strong near-wall gradients, a more accurate
treatment of boundary conditions would certainly be needed.

While the differences between scheme "A" and "B" are negligible in terms of 
precision, they are instead relevant in terms of computational requirements.
Comparing the execution time for the simulations used to produce the results in
Fig.~\ref{fig:force-validation} we observe that scheme "A" is $1.5-1.7$ times
more expensive than "B", due to the necessity to compute  the extra terms in the
polynomial expansion of the force term. These overheads can be even larger in
3-dimensions, where the coefficients of the expansions depend on Bessel
functions.

\subsection{Relativistic Sod's Shock tube}

The 1-d Riemann shock tube test is a widely adopted benchmark for the validation of numerical 
hydrodynamics methods. 
This benchmark has an exact time-dependent solution, both in the non-relativistic \cite{sod-jocp-1978} 
and in the relativistic \cite{thompson-jofm-1986,rezzolla-book-2013} regimes, and can be used to test the ability 
of a numerical solver to evolve flows in the presence of strong discontinuities and large gradients.

From a physical point of view, the problem consists of a tube filled with a
gas which initially is in two different thermodynamical states on 
either side of a membrane placed at $x = 0$.
As a result, the macroscopic quantities describing the fluid present a discontinuity at the membrane.
Once the membrane is removed the discontinuities decay producing shock/rarefaction waves,
depending on the initial configuration chosen for the two different chambers.
%
\begin{figure}[h!]
\centering
\includegraphics[scale=0.93]{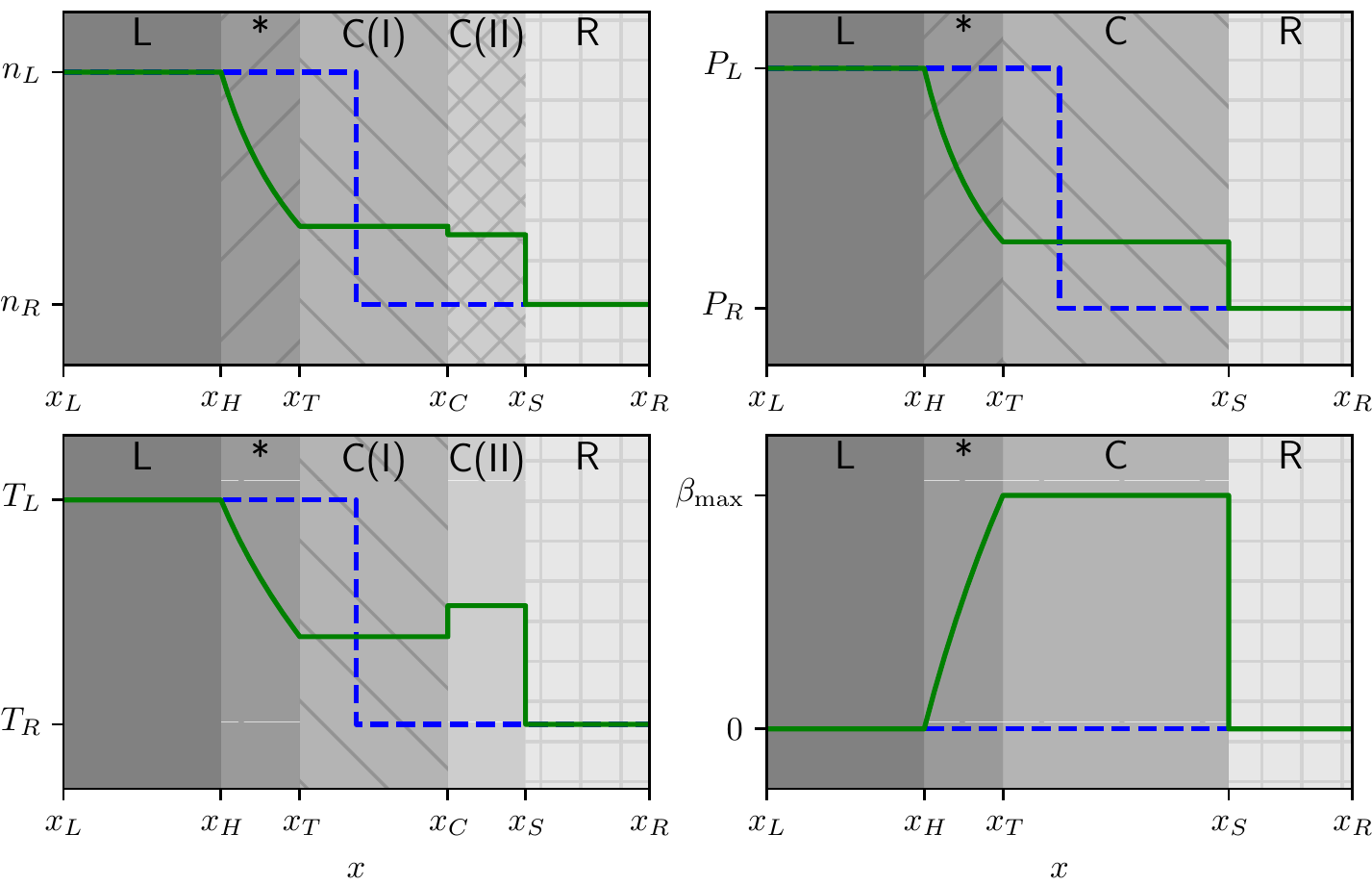}
\caption{Example of analytic solution of the Sod's shock tube problem
         in the inviscid limit, for an ultra-relativistic equation of state in $(3+1)$ dimensions. 
         Region $L$ and $R$ have not been reached yet by the rarefaction wave and preserve 
         therefore their initial state; region $*$ contains the rarefaction wave. 
         The shock wave is present in regions $C(I)$ and $C(II)$ and the interface between these regions 
         is associated to discontinuities in density and temperature.
         The blue dotted lines represent the initial conditions while the green lines represent the solution at a certain time $t > 0$.
        }\label{fig:analytic-inviscid-sod}
\end{figure}

The Sod's shock tube problem is a particular instance of the Riemann problem, with 
the following initial conditions:
\begin{equation}
  (P, n, \beta) = 
    \begin{cases}
      (P_{\rm L}, n_{\rm L}, 0) & x < 0 \\
      (P_{\rm R}, n_{\rm R}, 0) & x > 0
    \end{cases}
  \quad .
\end{equation} 
Let us assume $n_{\rm L} > n_{\rm R}$ and $P_{\rm L} > P_{\rm R}$, where the subscript $\rm L$ and $\rm R$ 
refer respectively to the left and right sides of the membrane.
With these initial conditions, the time evolution of the flow is characterized 
by two distinct components: a rarefaction wave traveling from the initial field 
discontinuity to the left, and a shock wave traveling from the initial 
field discontinuity to the right. 

If we consider an inviscid fluid, it is possible to describe the time evolution of the system 
analytically by solving the conservation equations.
However, the derivation of a solution in regimes other than the classical and 
ultra-relativistic one is a hard task which necessarily requires numerical integrations \cite{thompson-jofm-1986}.
For this reason we restrict the first part of our analysis to the ultra-relativistic regime.
%
\begin{figure}[H]
\centering
\includegraphics[width=.93\columnwidth]{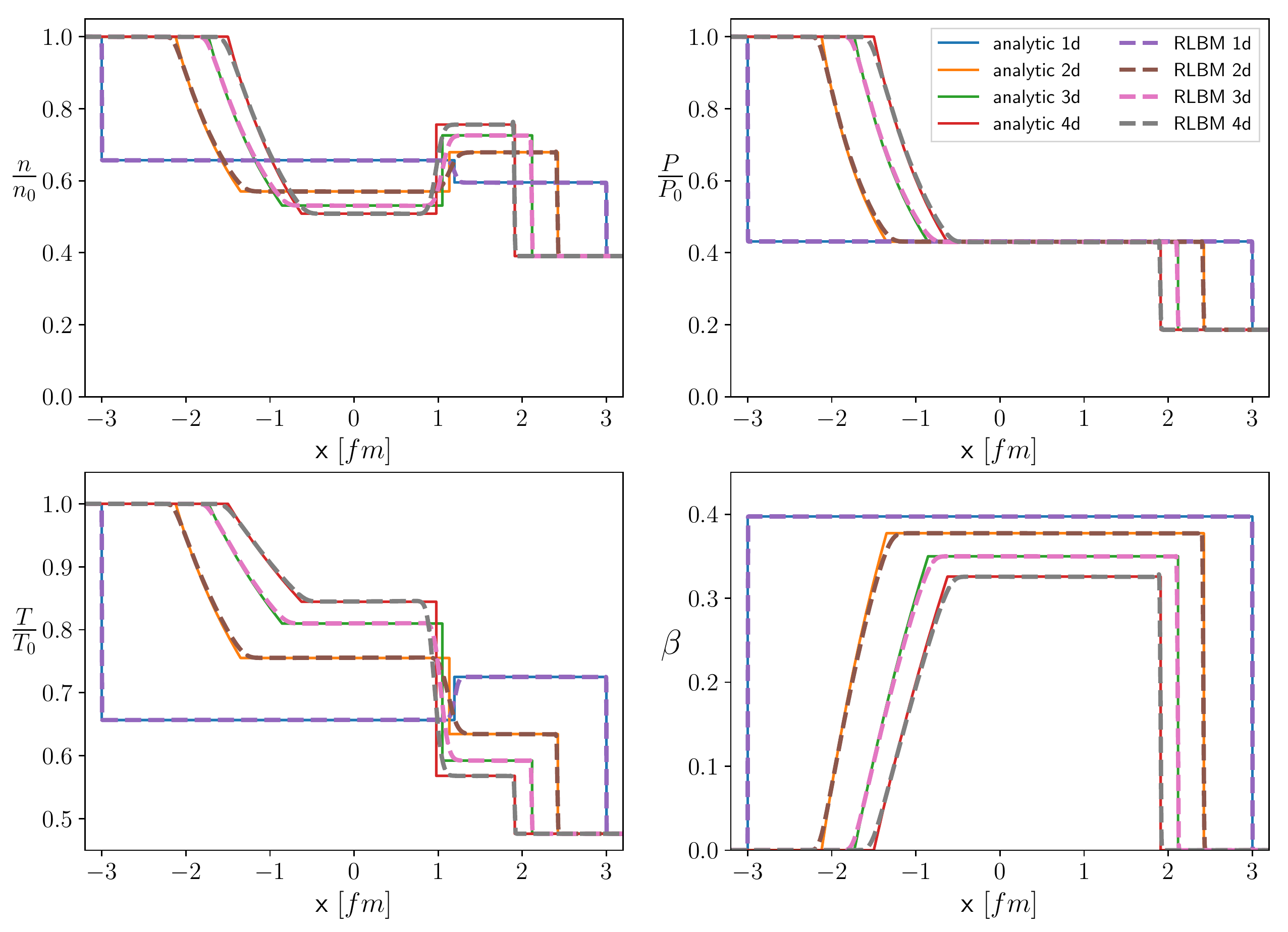}
\caption{  Sod's shock tube for a gas of massless particles in
           (d+1)-dimensions (d=1,2,3,4), at $t = 3.0~\rm{fm/c}$.
           Lines represent the analytic solution, while dotted 
           lines are result of RLBM simulations at $\eta / s = 0.002$.
           Top: left) temperature profile right) pressure profile.
           Bottom: left) density profile right) $\beta = |U^i|/U^0$.
           All macroscopic quantities are plotted in non-dimensional units by 
           dividing for their correspondent initial values at $x = -3.0~\rm{fm}$.
           All numerical results are obtained using third order quadratures 
           (see Appendix~\ref{sec:ur-quadratures} for details on the velocity sets),
           using $6400\times1\times1$ grids.
        }\label{fig:sod-inviscid-rlbm-Ddim}
\end{figure}

At a given time $t > 0$, the flow domain can be characterized by defining the 
different macroscopic quantities in the five regions shown in Fig.~\ref{fig:analytic-inviscid-sod}. 
Their definition, together with the general d-dimensional analytical solution, is reported in Appendix~\ref{sec:appendixD}.

For testing the inviscid regime we consider the following initial setup:
$P_{\rm L} = 5.43~\rm{GeV/fm^3}$ and $P_{\rm R} = 1~\rm{GeV/fm^3}$, 
with corresponding initial temperatures $T_{\rm L} = 400~\rm{MeV}$ and $T_{\rm R} = 190~\rm{MeV}$.
In order to convert from physics to lattice units we follow the discussion
in section~\ref{sec:conversion-phys-lattice}.
We start by setting our reference temperature $T_0$ equal to $T_{\rm L}$, 
thus $T_0 = T_{\rm L} = 400~\rm{MeV}$, which translates the initial temperatures 
on the lattice to $T_{\rm L} = 1$ and $T_{\rm R} = 0.475$. We also choose the initial values for the 
particle number density to be $n_{\rm L} = 1$ and $n_{\rm R} = 0.39$, which correctly reproduce the ratio $P_{\rm L} / P_{\rm R}$.  
%
\begin{figure}[H]
  \centering
  \includegraphics[width=.92\columnwidth]{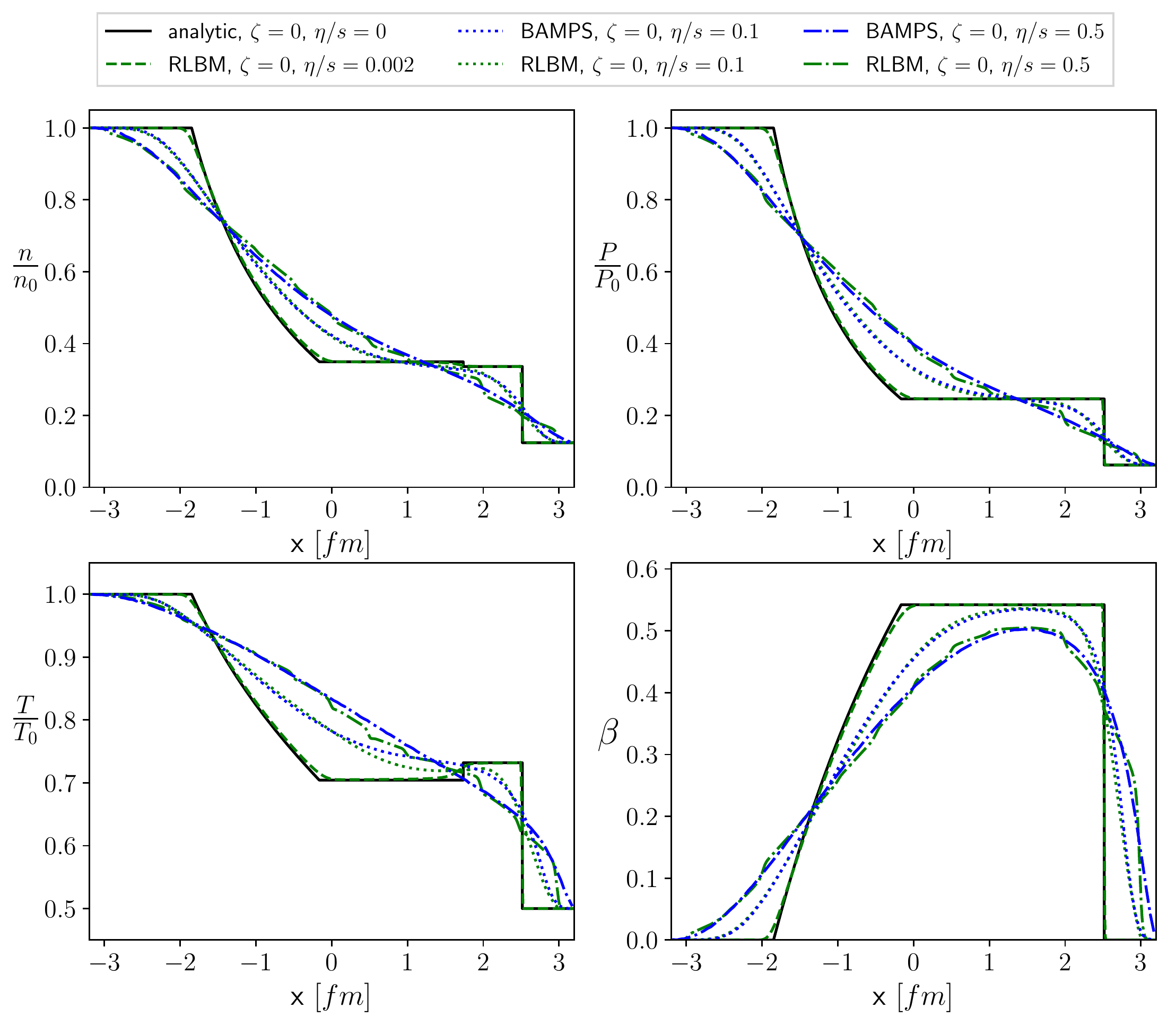}
  \caption{Sod's shock tube for a gas of massless particles 
           at $t = 3.2~\rm{fm/c}$, for several values of $\eta / s$.
           The results of simulations of the RLBM (green lines) at $\eta / s = 0.1$
           and $\eta /s = 0.5$ are compared against the results provided by BAMPS (blue lines).
           Top: left) density profile right) pressure profile.
           Bottom: left) temperature profile right) $\beta = |U^i|/U^0$.
           Density, temperature and pressure are plotted in non-dimensional units by 
           dividing for their correspondent initial values at $x = -3.2~\rm{fm}$.
           The simulations have been performed on grids of size $6400\times1\times1$,
           using the ultra-relativistic third order quadrature in Appendix~\ref{sec:ur-quadratures}.
          }\label{fig:bamps-rlbm-m0}
\end{figure}
%
We perform our tests on a grid of size $L_x \times 1 \times 1$, half of which represents 
the physical domain defined in the interval $(-3.2~\rm{fm}, 3.2~\rm{fm})$, while the other half forms
a mirror image that allows using periodic boundary conditions. Taking for example $L_x = 6400$,
it follows that on our grid $6.4 ~\rm{fm}$ corresponds to  $3200$ grid points, 
that is $\delta x = 0.002~\rm{fm}$. The corresponding value of $\Delta t $ is quadrature 
dependent; considering for example the third order quadrature for $\zeta = 0$ in (3+1)-dimensions
given in Appendix~\ref{sec:ur-quadratures} and having $v_0 = 1 / \sqrt{41}$, 
we obtain $\Delta t \approx 0.013~\rm{fm / c}$.
Since RLBM algorithms cannot handle zero-viscosity flows, we approximate
the inviscid regime using the lowest sustainable ratio between the shear viscosity 
and the entropy density ($\eta / s$).
In Fig.~\ref{fig:sod-inviscid-rlbm-Ddim} we show a validation of the code in 1,2,3 and 4 spatial dimensions
at $t = 3.0~\rm{fm/c}$, where in all simulations we have used $\eta / s = 0.002$.
The macroscopic profiles compare well with the analytical solution, 
and indeed we can clearly recognise the five different regions characterizing the flow.

\begin{figure}[htb]
\centering
\includegraphics[width=0.99\textwidth]{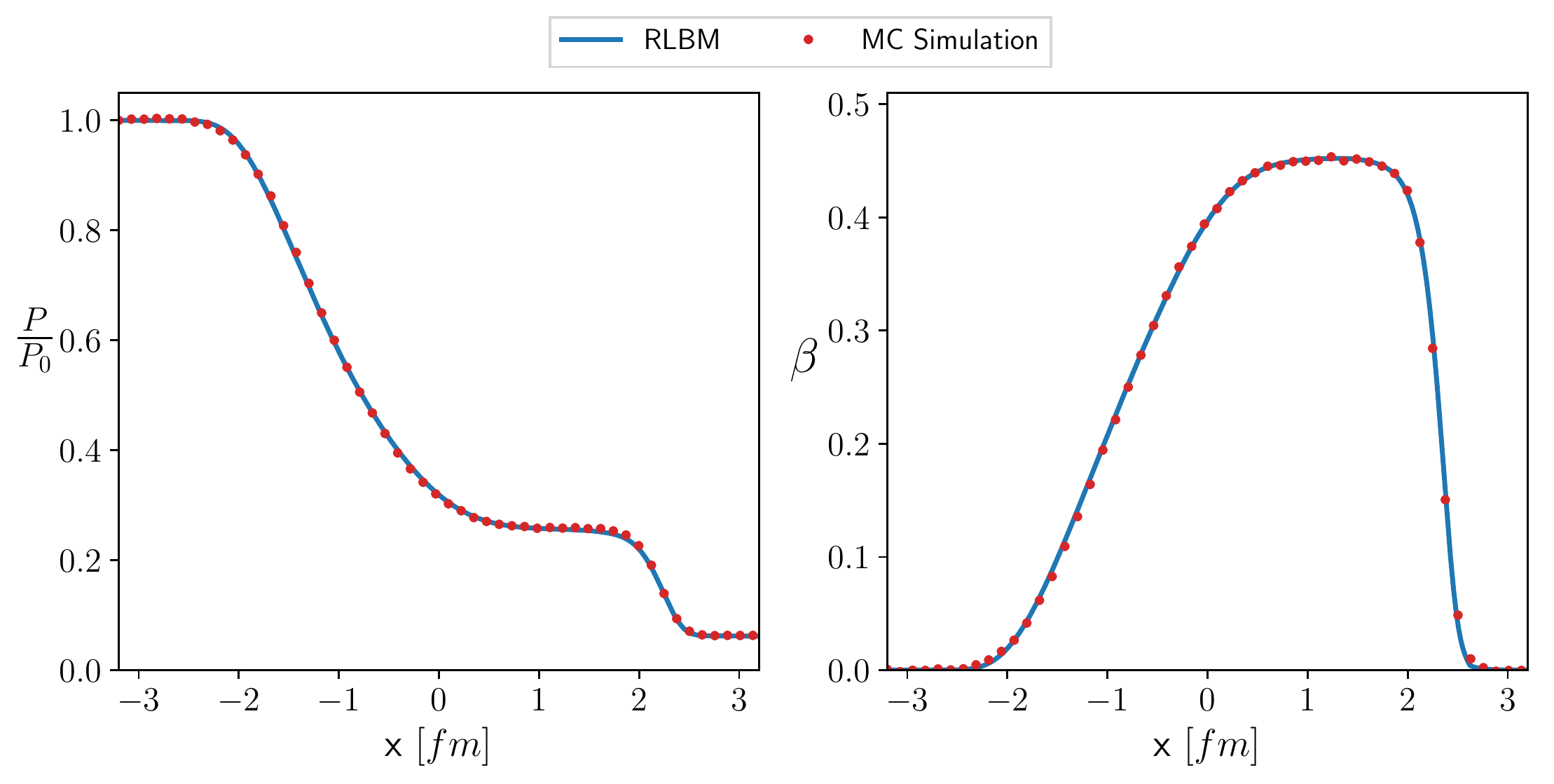}
\caption{  Sod's shock tube for a gas of massive particles $( m = 800~\rm{MeV})$
           at $t = 3.2~\rm{fm/c}$, at $\eta / s = 1 / 2 \pi$.
           The results of simulations of the RLBM (lines) are compared with
           those obtained solving the RBE with a Monte Carlo approach. 
           Left: pressure profile. Right) $\beta = |U^i|/U^0$.
           RLBM simulations have been performed on grids of size $6400\times1\times1$,
           using a third order quadrature identified by the velocity set $G_{\rm C}^{(\rm o3)}$ 
           in Table~\ref{fig:quadrature-3D}.
        }\label{fig:cmp-massive}
\end{figure}

When a non-zero viscosity is introduced, dissipation smoothens 
the interfaces between the different regions. Since in the 
viscous regime it is not possible to provide an exact solution, 
we compare with the results of other numerical solvers, such as 
the Boltzmann approach multi-parton scattering program (BAMPS) \cite{xu-prc-2007,bouras-prl-2009}.

The initial conditions in this case are: $P_{\rm L} = 5.43~\rm{GeV/fm^3}$, $P_{\rm R} = 0.339~\rm{GeV/fm^3}$, 
$T_{\rm L} = 400~\rm{MeV}$ and $T_{\rm R} = 200~\rm{MeV}$.  
We use a $(3+1)$-dimensional EOS, and the same relation for the entropy density used in BAMPS:
$ s = 4 n - n \ln{( n / n^{\rm eq} )}$, where $n^{\rm eq}$ comes from the equilibrium function, 
$n^{\rm eq} = d_{\rm G} T^3 / \pi^2$, with $d_{\rm G} = 16 $ the degeneracy of the gluons \cite{el-prc-2009}.

In Fig.~\ref{fig:bamps-rlbm-m0} we present the results of simulations for a few selected values 
of $\eta / s$, corresponding to different viscous regimes:
$\eta /s = 0.002$ is the nearly inviscid hydrodynamic regime discussed above, 
a highly viscous flow at $\eta /s = 0.1$ where an hydrodynamic approach is still justified, 
and finally $\eta / s = 0.5$ where we enter a transition towards a ballistic regime (thus going beyond hydrodynamics). 
For $\eta / s = 0.1$ the RLBM simulation is in excellent agreement with the results provided by BAMPS. Here we can 
observe that as the viscosity is increased, the interfaces between the different regions becomes smoother,
and eventually cannot be distinguished anymore when we move to $\eta / s = 0.5$: in this last example we are transitioning 
towards a ballistic regime, where the hydrodynamic approach becomes questionable.

We conclude this section taking into consideration relativistic shock waves for fluids of 
massive particles, a setup which has been rarely addressed in previous studies \cite{rezzolla-book-2013}.
Lacking an analytical solution to the problem, we compare the results of RLBM simulations
against the results produced by another numerical solver, which solves the full relativistic
Boltzmann equation using a Monte-Carlo approach based on the test-particles method \cite{plumari-prc-2012,ruggieri-plb-2013,plumari-prc-2015,plumari-epc-2019}. 
In Fig.~\ref{fig:cmp-massive} we show an example for a gas of particles of mass $m = 800~\rm{MeV}$
and the same initial conditions used in the previous case: $T_{\rm L} = 400~\rm{MeV}$, $T_{\rm R} = 200~\rm{MeV}$,
$P_{\rm L} = 5.43~\rm{GeV/fm^3}$ and $P_{\rm R} = 0.339~\rm{GeV/fm^3}$. The ratio $\eta / s$ is kept fixed to
the value $1 / 2 \pi$. We appreciate an excellent agreement in both the pressure and velocity profiles.
A more detailed analysis will be reported elsewhere.

\subsection{Quark-gluon plasma}

The study of quark-gluon plasma is the most natural application ground for the RLBM.
In this section, we provide a very preliminary example in which we simulate the evolution 
of the initial stages of heavy-ion collisions.
We consider the same numerical setup used in \cite{florkowski-arxiv-2019}, with an 
initial Gaussian distribution for the temperature profile
\begin{equation}
  T = T_{\rm r}~g(x,y,z), 
  \quad 
  g(x,y,z) = \exp{(-\frac{x^2}{2 \sigma_x^2} -\frac{y^2}{2 \sigma_y^2} -\frac{z^2}{2 \sigma_z^2})} \quad ,
\end{equation}
and likewise for the particle density $n = n_{\rm r}~g(x,y,z)$.
%
\begin{figure}[htb]
\centering
\includegraphics[width=0.99\textwidth]{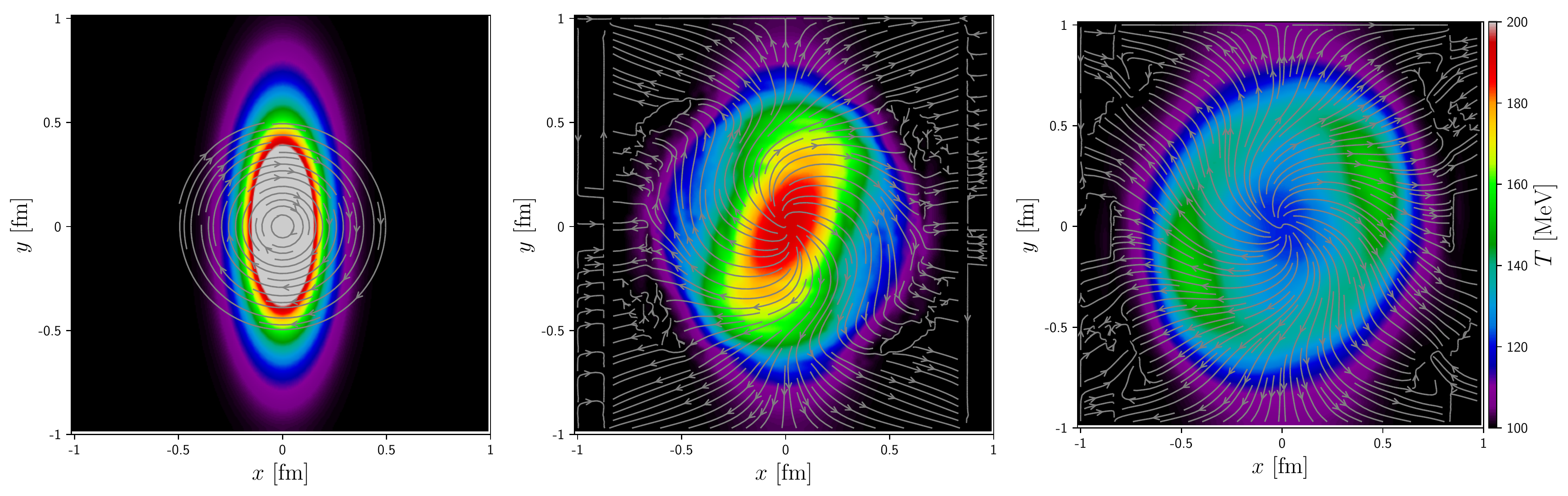}
\caption{ 
          Temperature profile of the system in the $x-y$ plane at $z = 0~{\rm fm}$, 
          at three different time steps, from left to right $t=0~{\rm fm /c}, t=0.35~{\rm fm /c}, t=0.7~{\rm fm /c}$.
          White lines represent the velocity streamlines of the fluid.
        }\label{fig:qgp-example}
\end{figure}
%
The initial temperature at the center of the fireball is $T_{\rm r} = 200~{\rm MeV}$, with
$n_{\rm r} = 1.25~{\rm fm^{-3}}$, and the particles mass $m = 1~{\rm GeV}$.
We also add a background temperature of $T = 100~{\rm MeV}$ and density of $n_{\rm r} = 0.625~{\rm fm^{-3}}$
to avoid numerical instabilities far from the center of the fireball, where we would have a zero temperature
which cannot be sustained by the RLBM solver.

The initial velocity profile is given by
\begin{equation}
  U^{\alpha} = \gamma \left( 1, -h(r) y, h(r) x, 0 \right)  ,
  \quad 
  h(r) = \frac{1}{r} \tanh{(\frac{r}{r_0})}
\end{equation}
with $r$ the distance from the center of the fireball, and $r_0$ a parameter describing the 
strength of the flow. We take $r_0 = 1~{\rm fm}$.

Under these initial conditions, the evolution of the system is triggered by an initial rotation around the $z$ axis.
In Fig.~\ref{fig:qgp-example} we show the temperature profile of the system in the x-y plane at $z = 0~{\rm fm}$,
at three different time steps. We can see that the symmetry of the rotating ellipsoid is quickly broken, 
with the single source splitting into two separate hot spots. The evolution is qualitatively similar to 
the one portrayed in \cite{florkowski-arxiv-2019}, with a more quantitative analysis left to future detailed work.

\subsection{Hydrodynamic flow of electrons in graphene}

We now take into consideration an ``exotic'' application of
dissipative relativistic hydrodynamics and apply the RLBM to the study 
of the dynamics of the electrons in graphene sheets.
The motivation for adopting these type of solvers comes from the fact that
electrons in graphene sheets follow an ultra-relativistic dispersion relation, 
so they can be regarded as a fluid of massless (quasi-) particles 
whose energy depends on momentum as $\epsilon_{\rm F} = v_{\rm F} | \bm{p} |$, with 
the Fermi speed $v_{\rm F} \sim 10^6 m/s$ mimicking the role of the speed of light in true relativistic systems.

\begin{figure}[H]
{
\begin{overpic}[width=.925\columnwidth]{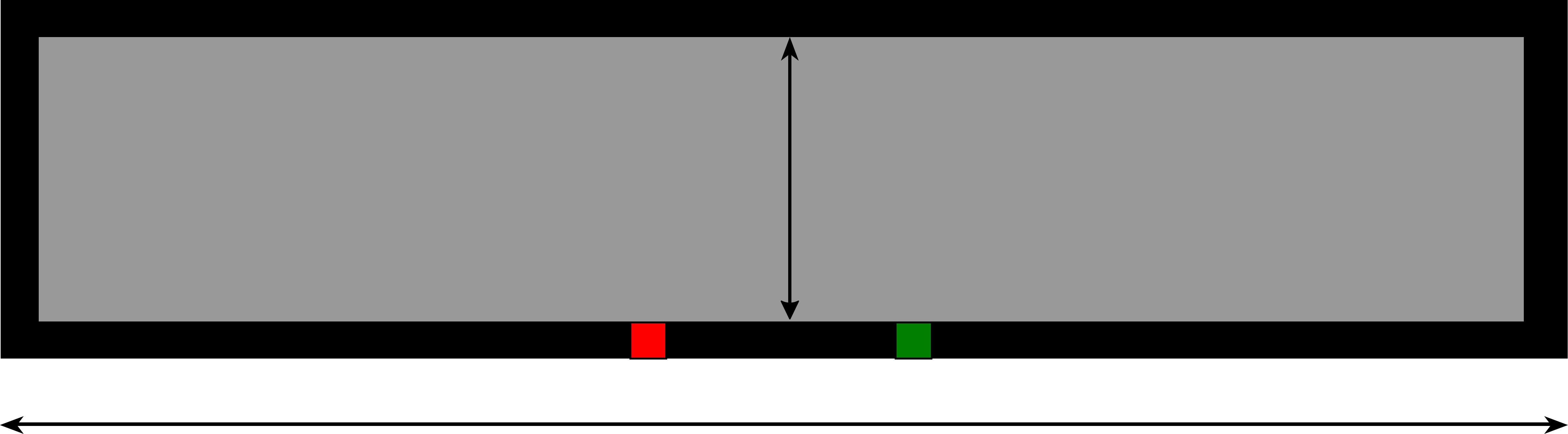}
      \put(46,15){$W$}
      \put(49,1.5){$L$}
\end{overpic}  
}
{
\centering
\includegraphics[width=0.99\textwidth]{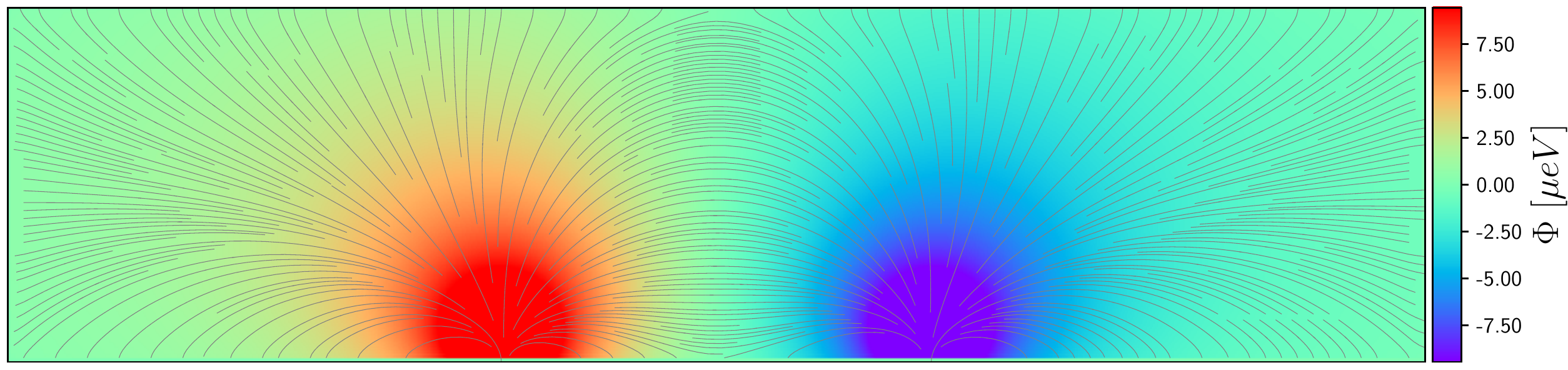}
\includegraphics[width=0.99\textwidth]{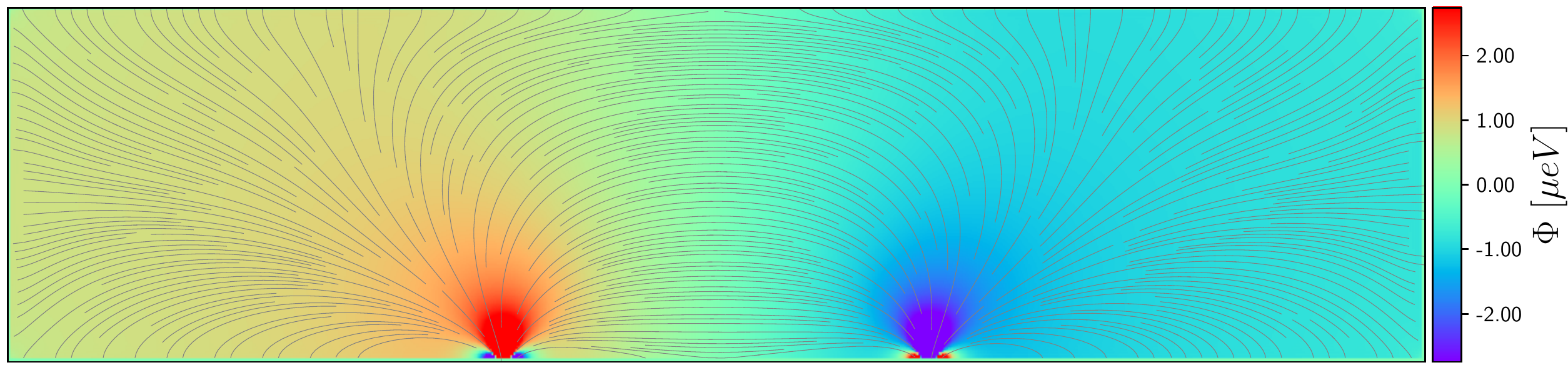} 
\includegraphics[width=0.99\textwidth]{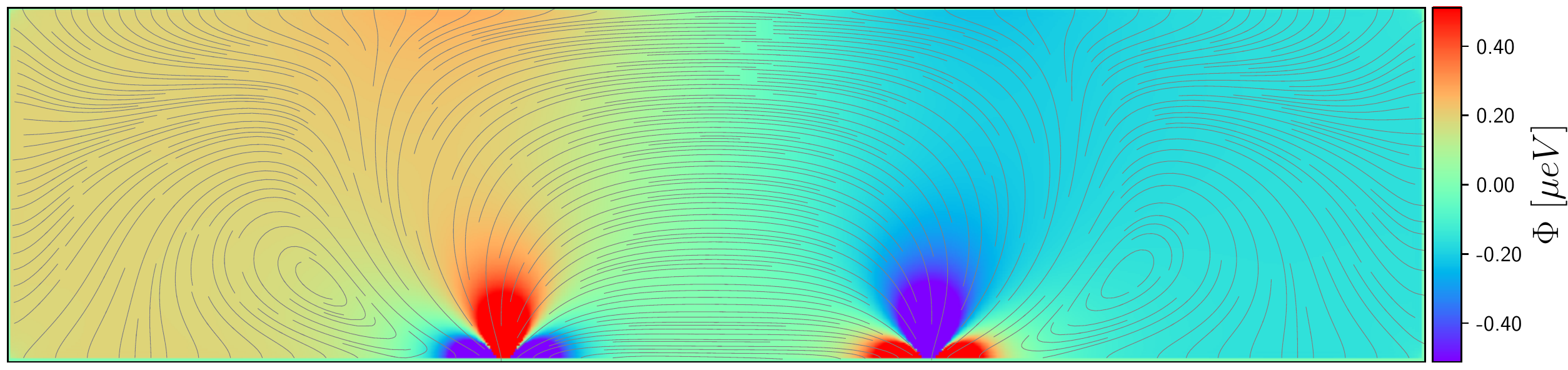}
}
\caption{ a) Sketch of the "vicinity-geometry" used in simulations:
          Two contacts are used to inject (red area) and drain (green area) 
          a current from a graphene sheet of width $W$. The current is zero
          at the boundaries of the sample (black areas).
          b-d) Snapshot of three different simulations at the steady-state, 
               all using $W=2 \mu m$, $\bar{n} = 2\cdot10^{-12} ~ \rm cm^{-2}$, $I = 10^{-7} ~ \rm A$, 
               $\nu = 0.005~ m^2/s$ and $C_g / e^2 = 3.03 \cdot 10^{35}~J^{-1} m^{-2}$,
               and for different values of $\tau_{\rm D}$.
               In b) $\tau_{\rm D} = 1~\rm{ps}$, in c)  $\tau_{\rm D} = 20~\rm{ps}$, 
               in d) $\tau_{\rm D} = 200~\rm{ps}$
               The color map describes the electrochemical potential. 
               Lines represent the electrons velocity streamlines.
        }\label{fig:whirlpool-snapshots}
\end{figure}

In this section we provide a validation of this numerical approach by simulating
steady-state flows in the so-called "vicinity-geometry", which has been 
subject of several theoretical and experimental studies \cite{torre-prb-2015, pellegrino-prb-2016, berdyugin-science-2019} 
to outline phenomena such as negative nonlocal resistance, current whirlpools, and measuring the Hall viscosity
of graphene's electrons fluid.
The geometrical setup is sketched in Fig.~\ref{fig:whirlpool-snapshots}, which represents
a single-layer graphene sheet of dimension $L \times W$ 
(usually encapsulated between one or more layers of dielectric materials, such as boron nitride), 
in which two electrodes are used to inject and drain a constant current within the device.
%
\begin{figure}[htb]
  \centering
  \includegraphics[width=0.99\textwidth]{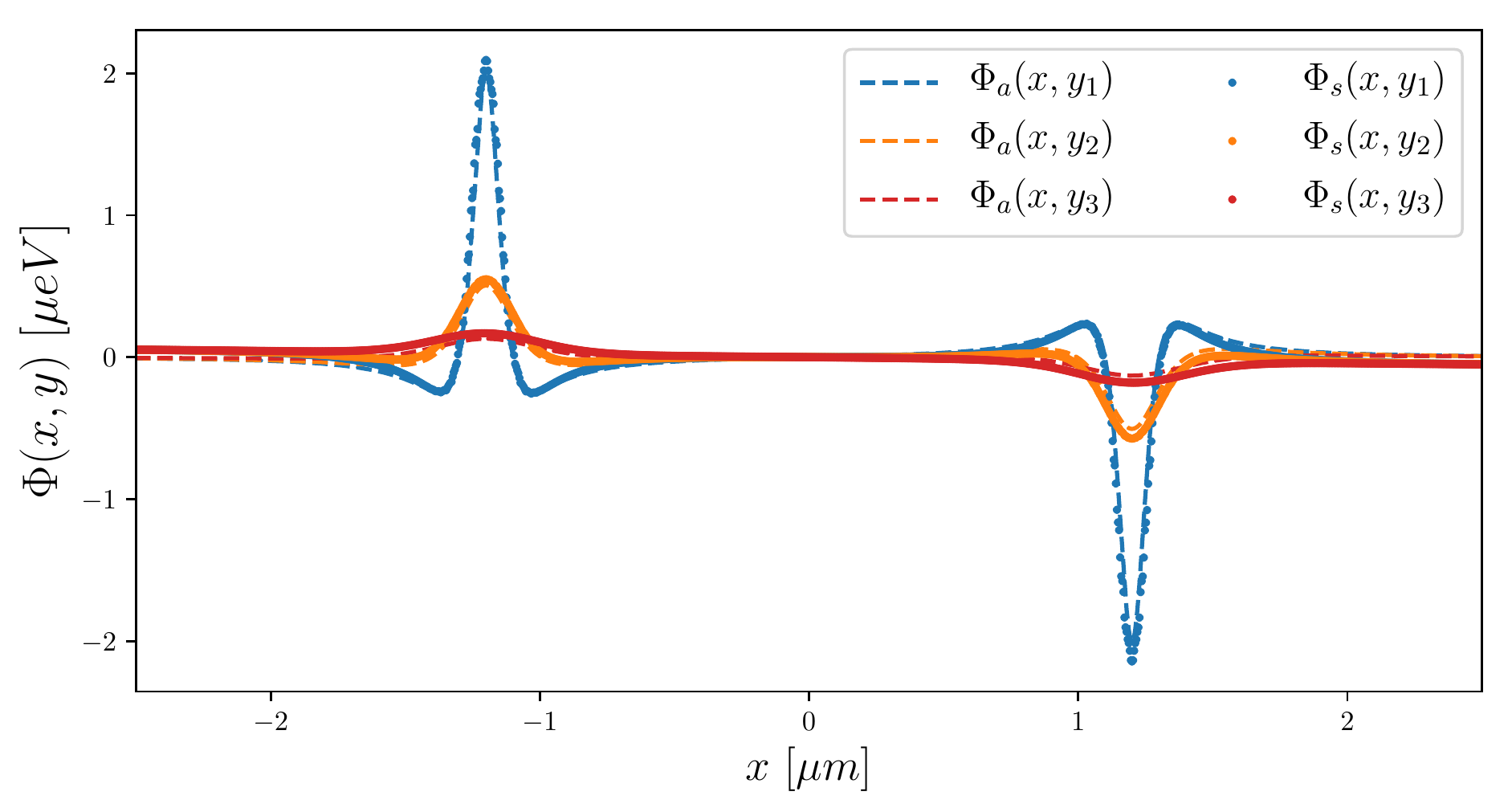}
  \caption{ Electrochemical potential measured at several fixed distances from
            the side of the graphene sample, respectively
            $y_1 = 0.1~\rm{\mu m}$, $y_2 = 0.2~\rm{\mu m}$, $y_3 = 0.4~\rm{\mu m}$.
            With $\Phi_s(x,y)$ we indicate the results of a numerical simulation,
            which are compared with the analytic solution $\Phi_a(x,y)$ for the benchmark
            (Eq.~\ref{eq:phi-analytic}).
          }\label{fig:cmp-electrochemical-potential-rlbm}
\end{figure}
%

Experimental measurements have shown voltage drops in the proximity of the injectors, which 
are found to be dependent on the viscosity of the electron fluid \cite{torre-prb-2015}. 
An accurate analytic approximation of the experimental results for the electrochemical potential is derived in the same paper:
\begin{equation}\label{eq:phi-analytic}
  \Phi(x,y) = \frac{m \nu I}{\bar{n} e^2 W^2} \left[ F_2(x-x_0,y) - F_2(x+x_0,y) \right] \quad ,
\end{equation}
where for single layer graphene the effective mass $m$ is defined as $m = \epsilon_{\rm F} / v_{\rm F}^2$, 
$I$ is the magnitude of the injected current, $\bar{n}$ the equilibrium density, $W$ the width of the device, 
$e$ the elementary charge, $\nu$ the kinematic viscosity, and
\begin{equation}
  F_2(x,y) = \pi\frac{ 1 + \cosh(\pi x/W) \cos(\pi y/W) }{ \left[ \cosh(\pi x/W) \cos(\pi y/W) \right]^2 } \quad .
\end{equation}

The expression above has been obtained by assuming an infinitely long sample.
In our simulations, we use a grid with aspect ratio $L / W = 4$. Regarding boundary conditions,
current is set to zero at the boundaries of the sample, while the equilibrium value of the 
distribution is imposed at the grid points used to represent the contacts \cite{succi-book-2018}.
 
Since $\Phi$ is not a direct observable of our lattice formulation, we need to perform a parameter 
matching procedure (see \cite{gabbana-prl-2018} for details) in order to compare with Eq.~\ref{eq:phi-analytic}:
\begin{equation}
  \Phi(x,y) = \varphi(x,y) - \frac{\delta P(x,y)}{e \bar{n}} \quad ,
\end{equation}
where $\varphi({\bm r},t)$ is the electrical potential self-induced by the electrons motion withing 
the $2d$ graphene layer, and $\delta P(x,y) \approx T(x,y) ( n(x,y) - \bar{n}(x,y) )$ is the local pressure difference.  
Fig.~\ref{fig:cmp-electrochemical-potential-rlbm} compares our simulation against the analytical 
prediction of Eq.~\ref{eq:phi-analytic}, showing very good agreement at several distances from the boundary layer.

For a treatment of the problem closer to the experimental setup, it is important to include not only effects  
due to electric forces but also interactions with phonons and impurities.
To this end, we have included an external forcing term:
\begin{equation}\label{eq:forcing-for-graphene}
  \bm{F} = - e \nabla \varphi - \frac{1}{\tau_{\rm D} } n \bm{u} \quad ;
\end{equation}
the first term is the contribution due to the electric field, which
in principle would require the solution of the full Poisson equation;
since this approach is computationally expensive, we  compute $\varphi(x,y)$ by employing a local 
capacitance approximation \cite{tomadin-prb-2013}:
\begin{equation}
  \varphi(x,y) = - \frac{e}{C_{\rm g}} \bar{n}(x,y) \quad ,
\end{equation}
with $C_g$ the geometrical capacitance per unit area, depending on both the geometrical 
and permittivity properties of the dielectric layer.
The second term in Eq.~\ref{eq:forcing-for-graphene} is used to parametrize
phonon-electron scattering as a friction term, with $\tau_{\rm D}$ the single scattering time. 
Albeit very simple, this parametrization has proven successful in describing experiments
in the linear-response regime \cite{bandurin-science-2016, kumar-natphys-2017, bandurin-natcomm-2018}.
%

In Fig.~\ref{fig:whirlpool-snapshots} we show a few examples which qualitatively summarize the results 
of the simulations. The simulations are conducted using the following physical parameter:
$W=2 \mu m$, $\bar{n} = 2\cdot10^{-12} ~ \rm cm^{-2}$, $I = 10^{-7} ~ \rm A$, 
$\nu = 0.005~ m^2/s$ and $C_g / e^2 = 3.03 \cdot 10^{35}~J^{-1} m^{-2}$.
In panel b), we show a simulation with $\tau_D = 1~\rm ps$, in which the interaction with phonons 
is sufficiently strong to prevent the onset of a hydrodynamic regime.
In panel c), we show the result of a simulation with $\tau_D = 20~\rm ps$, where we observe a weak signature 
of a hydrodynamic behavior in the proximity on the current injector
(note the formation of a negative potential). Finally in panel d) we show that for 
$\tau_D = 200~\rm ps$ we observe the formation of current whirlpools.

For the geometric setup here analyzed the authors in \cite{pellegrino-prb-2016}
predicted the formation of backflow near the injectors for arbitrarily small
values of the vorticity diffusion parameter $D_{\nu} = \sqrt{ \nu \tau_{\rm D}}$.
However, for similar geometries they also found that the transition occurs when
a specific threshold is overrun, i.e. $D_{\nu} > W / (\sqrt{2} \pi)$, which 
can be expressed in a more explicit form as 
\begin{equation}
  \nu > \nu_{*} = \frac{W^2}{2 \pi^2~\tau_D} \quad .
\end{equation}
The above expression is in good agreement with our simulations
since for the parameters used in panel b) we get $\nu_{*} \approx 0.2~\rm m^2/s$,
for panel c), where we observe the first signatures of the hydrodynamic regime
we have $\nu_{*} \approx 0.01~\rm m^2/s$, thus a figure about the same order 
of magnitude of the actual viscosity, and finally $\nu_{*} \approx 0.001~\rm m^2/s$
for the parameters used in the simulation corresponding to panel d).

\section{Conclusions}

This paper has presented an exhaustive account of the conceptual and mathematical
development of a suite of Lattice Boltzmann methods capable of describing 
relativistic dissipative fluid flows over a wide kinematic range, ranging from
ultra-relativistic ($\zeta \to 0$) to near non-relativistic ($\zeta
\rightarrow \infty$), and -- in principle -- in any number of space
dimensions. All these methods are based on the expansion of the J\"uttner
distribution onto an appropriate basis of orthogonal polynomials and on its
discretization, via Gauss-like quadrature, on a regular and uniform lattice. 
Quadratures have been studied up to the fifth order, even though -- on
computational efficiency grounds --  only the third order can be tackled in
practice. While building largely on previous developments, the
present work considerably streamlines previous formulations
with a uniform treatment of all spatial dimensions.

The main advantage of the kinetic approach to relativistic hydrodynamics
is to ensure built-in relativistic covariance and causality also in the presence of dissipative effects. 
This is welcome but still leaves an open question as to the connection between
the meso-scale  relaxation time and the macroscopic transport coefficients.
This problem is as practical as it is conceptual, since different theoretical treatments yield different
results in the relativistic regime; several recent developments suggest
that the Chapman Enskog expansion as the correct route.  
It would be obviously interesting to have this result confirmed by a
carefully planned experimental setup, hopefully to be available in the
coming years. 
Meanwhile, this work provides a neat numerical answer: we have compared in detail the predictions of the CE
approach and of Grad's moment and gathered convincing numerical
evidence that, in $1$, $2$ and $3$ spatial dimensions, the CE approach is the
one correctly linking all macroscopic transport coefficients, $\lambda$,
$\eta$ and $\mu$ to the mesoscopic relaxation time $\tau$. This solves
the problem in the relaxation-time approximation but, given the universal
nature of the macroscopic coefficients, strongly suggests that this is likely 
to offer the correct path also for more complex formulations of the Boltzmann equation.

In any case, the analysis described above provides a neat and accurate
calibration procedure that enables the present RLB scheme for the 
simulation of a variety of relativistic hydrodynamic problems.

The computational framework derived from the present RLB method has been extensively
validated in several contexts where analytical results or
semi-analytical approximations are available, and over a broad range of
kinematic regimes. Our tests also include an "exotic" application to the
electron dynamics in graphene sheets, opening the way to accurate
and realistic numerical simulations of these materials.

We have made a specific effort to describe in detail all steps
of our derivations and results, so as to allow the interested reader to
reproduce them.

The RLB method can hopefully find potential use for a wealth of different
relativistic fluid problems across scales, from astrophysics and cosmology to 
high-energy physics and material science. 

Much is left for the future.
At the methodological level, new lines of development may include -- in the short
time frame -- comparison with different solvers, with a close eye on the
trade-off between computational efficiency and physical accuracy, development
of accurate boundary conditions, and -- on a longer time frame 
-- an analysis of the relevance of quantum effects, as described for instance by 
quantum versions of the Maxwell-J\"uttner distribution.
Moreover, off-lattice scheme could represent an efficient solution to 
further extend the parameter regimes supported by the method, allowing covering
BH-regimes, and fluids consisting of particles of vanishingly-small non-zero mass.

As to applications, the customization of the RLB scheme to the detailed study of quark-gluon plasmas
dynamics in current and future high-energy experiments, appears a very appealing topic for future research.

\section*{Acknowledgments}

AG has been supported by the European Union's Horizon 2020 research and
innovation programme under the Marie Sklodowska-Curie grant agreement No. 642069.
DS has been supported by the European Union's Horizon 2020 research and
innovation programme under the Marie Sklodowska-Curie grant agreement No. 765048.
SS acknowledges funding from the European Research Council under the European
Union's Horizon 2020 framework programme (No. P/2014-2020)/ERC Grant Agreement No. 739964 (COPMAT).
All numerical work has been performed on the COKA computing cluster at Universit\`a di Ferrara. 


\appendix
\renewcommand*{\thesection}{\Alph{section}}
\section{Notation}\label{sec:appendixA}

In this appendix we summarize the notation used throughout this work. 
We also introduce useful projector operators, including some of their relevant properties, 
which have been often used in the main text and on which we extensively rely 
for the calculation of the transport coefficients.

We work in a flat (d+1) dimensional space-time, parametrized by the coordinates
\begin{equation}
  \left( x^{\alpha} \right) = \left( c t, \bm{x} \right) \quad ,
\end{equation}
with the first component giving the time coordinate, while the second term represents the  
usual spatial coordinates in a $d$ dimensional Euclidean space.  

The notation $\left( A^{\alpha} \right) $ is used to represent a vector $\in \mathbb{R}^{d+1}$ 
(for simplicity, we refer to these vectors as \emph{four vectors} in the text, even when $d \neq 3$)
while the boldface notation $\bm{A}$ is used for vectors $\in \mathbb{R}^{d}$.
Greek indexes are used to denote $(d+1)$ space-time coordinates 
and Latin indexes for $d$ dimensional spatial coordinates. 
Einstein's summation convention over repeated indexes is adopted.

We use the metric tensor $\eta^{\alpha \beta} = diag(1, -\mathds{1})$,
with $\mathds{1} = \left( 1, \dots, 1 \right) \in \mathbb{N}^d $.

The Minkowski-orthogonal projector to the fluid velocity $U^{\alpha}$ is defined as 
\begin{align}
  \Delta^{\alpha\beta} & = \eta^{\alpha\beta} - \frac{1}{c^2} U^\alpha U^\beta \quad , \\
  \Delta_{\alpha\beta} & = \eta_{\alpha\beta} - \frac{1}{c^2} U_\alpha U_\beta \quad .
\end{align}

By construction, the product of $\Delta_{\alpha\beta}$ with the quadri-velocity $U^\beta$ 
is therefore equal to zero, and its trace equals the number of spatial dimensions:
\begin{align}
  \Delta^{\alpha\beta} U_\beta & = \Delta_{\alpha\beta} U^\beta = 0 \quad , \\
  \Delta^{\alpha}_{\alpha}     & = d                                \quad . 
\end{align}

It  is useful to introduce the short-hand notation
\begin{equation}
  \Delta^{\alpha}_{\gamma}  
  = 
  \Delta^{\alpha\beta} \Delta_{\beta\gamma} 
  = 
  \delta^{\alpha}_{\gamma} - \frac{1}{c^2} U^{\alpha} U_{\gamma} \quad ,
\end{equation}

together with the following easily verifiable properties:
\begin{align}
  \Delta_{\alpha}^{\gamma} \Delta_{\gamma\beta}     &= \Delta_{\alpha\beta}       \quad ,\\ 
  \Delta^{\alpha}_{\gamma} \Delta^{\gamma}_{\beta}  &= \Delta^{\alpha}_{\beta}    \quad .
\end{align}

These projectors can be applied to express a generic quadri-vector $A^{\alpha}$ as
the sum of two terms, respectively orthogonal and parallel to $U^{\alpha}$:
\begin{equation}
  A^\alpha = \Delta^{\alpha\beta} A_\beta + \frac{1}{c^2} U^\alpha U^\beta A_\beta \quad .
\end{equation}

Likewise, we can decompose a generic tensor $T^{\alpha\beta}$ in order to identify 
the symmetric $U^\alpha$-orthogonal components $T^{(\alpha\beta)}$:
\begin{equation}
  T^{(\alpha\beta)} 
  = 
  \frac{1}{2} \left( \Delta^\alpha_\gamma \Delta^\beta_\delta + \Delta^\beta_\gamma \Delta^\alpha_\delta \right) T^{\gamma \delta} 
  = 
  \frac{1}{2} \left( \Delta^{\alpha\gamma} \Delta^{\beta\delta} + \Delta^{\beta\gamma} \Delta^{\alpha\delta} \right) T_{\gamma \delta} \quad ,
\end{equation}
the antisymmetric $U^\alpha$-orthogonal components $T^{[\alpha\beta]}$:
\begin{equation}
  T^{[\alpha\beta]} 
  = 
  \frac{1}{2} \left( \Delta^\alpha_\gamma \Delta^\beta_\delta - \Delta^\beta_\gamma \Delta^\alpha_\delta \right) T^{\gamma \delta} 
  = 
  \frac{1}{2} \left( \Delta^{\alpha\gamma} \Delta^{\beta\delta} - \Delta^{\beta\gamma} \Delta^{\alpha\delta} \right) T_{\gamma \delta} \quad ,
\end{equation}
and the symmetric, traceless, $U^\alpha$-orthogonal components $T^{<\alpha\beta>}$:
\begin{align}                                              
  T^{<\alpha\beta>} 
  = T^{(\alpha\beta)} - \frac{1}{d} \Delta^{\alpha\beta}\Delta_{\gamma\delta} T^{(\gamma\delta)} \quad .
\end{align}
The following properties naturally follow from the definitions above:
\begin{align}  
  U_{\alpha}T^{(\alpha\beta)}   &= 0    &     \eta_{\alpha\beta}T^{(\alpha\beta)}   &= \Delta_{\mu\nu}T^{\mu\nu}    &\Delta_{\alpha\beta}T^{(\alpha\beta)}  &= \Delta_{\mu\nu}T^{\mu\nu}     \quad ,\\
  U_{\alpha}T^{[\alpha\beta]}   &= 0    &     \eta_{\alpha\beta}T^{[\alpha\beta]}   &= 0    &\Delta_{\alpha\beta}T^{[\alpha\beta]}  &= 0                                                    \quad ,\\
  U_{\alpha}T^{<\alpha\beta>}   &= 0    &     \eta_{\alpha\beta}T^{<\alpha\beta>}   &= 0    &\Delta_{\alpha\beta}T^{<\alpha\beta>}  &= 0                                                    \quad .        
\end{align}

It is at times useful to introduce a decomposition also for the quadri-gradient $\partial_\alpha$, 
by defining the convective time derivative $D$ and its orthogonal component $\nabla^\alpha$:
\begin{align}
  \partial^\alpha = \Delta^{\alpha\beta} \partial_\beta + \frac{1}{c^2} U^\alpha U^\beta \partial_\beta 
                  = \nabla^{\alpha} + \frac{1}{c^2} U^\alpha D \quad ,
\end{align}
with the following useful properties:
\begin{align}
  U_\alpha \nabla^{\alpha} &= 0                             \quad , \\
  \nabla^{\alpha} U_\alpha &= \partial^{\alpha} U_{\alpha}  \quad .
\end{align}

\section{Integrals of the Maxwell-J\"uttner distribution}\label{sec:appendixB}

We compute here some integrals often used in the development
of the numerical methods presented in this work and in the definition of the transport coefficients.

\subsection{Integrals $Z^{\alpha_1 \dots \alpha_n}$}

Let
\begin{align}
Z = & \int e^{-\frac{p_\mu U^\mu}{k_{\rm B} T}} \frac{\diff^d p}{p_0} \\
Z^{\alpha_1 \dots \alpha_n} = &\int e^{-\frac{p_\mu U^\mu}{k_{\rm B} T}} p^{\alpha_1} \dots p^{\alpha_n} \frac{\diff^d p}{p_0} \quad .
\end{align}
$Z^{\alpha_1 \dots \alpha_n}$ is obtained via differentiation of $Z$ with respect to $U_{\alpha_1} \dots U_{\alpha_n}$:
\begin{equation} \label{eq:Z-derivative}
  Z^{\alpha_1 \dots \alpha_n} = (-k_{\rm B} T)^n \frac{\partial Z}{\partial U_{\alpha_1} \dots\partial U_{\alpha_n}} \quad .
\end{equation}
$Z$ is a Lorentz-invariant quantity which depends only on $U_{\alpha} U^{\alpha} = c^2$; however, 
as we need to compute derivatives with respect to $U^{\alpha}$, we first derive the result 
for an unconstrained $U_{\alpha} U^{\alpha}$ and, after performing the derivatives, 
evaluate the result for $U_{\alpha} U^{\alpha} = c^2$. 

We write:
\begin{equation}
  Z = \int e^{-\frac{p_0 \sqrt{ U^{\mu} U_{\mu}} }{k_{\rm B} T}} \frac{d^dp}{p_0} 
    = \int e^{-\frac{\sqrt{U^{\mu}U_{\mu}}}{k_{\rm B} T}\sqrt{m^2 c^2 + p^2}}\frac{d^dp}{\sqrt{m^2 c^2 + p^2}} \quad ,
\end{equation}
and then switch to spherical coordinates:
\begin{align*}
  d^dp = p^{d-1} dp~d\Omega  && \text{with} && \int d\Omega = \frac{d\pi^{\frac{d}{2}}}{\Gamma\left(1+\frac{d}{2}\right)} \quad ,
\end{align*}
giving 
\begin{align*}
  Z = \frac{d\pi^{\frac{d}{2}}}{\Gamma\left(1+\frac{d}{2}\right)} 
      \int_0^{+\infty} e^{-\frac{\sqrt{U^{\mu}U_{\mu}}}{k_{\rm B} T}\sqrt{m^2 c^2 + p^2}}\frac{p^{d-1}dp}{\sqrt{m^2 c^2 + p^2}} \quad .
\end{align*}
Changing integration variable $p = mc \sqrt{t^2-1}$ and defining
$\zeta = mc^2/k_BT$, we have
\begin{align*}
  Z = \frac{d\pi^{\frac{d}{2}}}{\Gamma\left(1+\frac{d}{2}\right)}(mc)^{d-1} 
      \int_1^{+\infty} (t^2-1)^{\frac{d-1}{2}-\frac{1}{2}} e^{-\frac{\zeta t}{c}\sqrt{U^{\mu}U_{\mu}}}dt \quad .
\end{align*}
Recalling one useful definition of the modified Bessel function of the second kind \cite{abramowitz-book-1965}:
\begin{align*}
  K_\nu (z) = \frac{\pi^{1/2}(z/2)^{\nu}}{\Gamma\left(\nu+1/2\right)}\int_1^{+\infty} e^{-zt}(t^2-1)^{\nu-1/2}dt \quad ,
\end{align*}
we finally obtain:
\begin{align*}
  Z = \pi^{\frac{d-1}{2}}2^{\frac{d+1}{2}}\zeta^{\frac{d-1}{2}}\left(\frac{k_{\rm B} T}{\sqrt{c}}\right)^{d-1} \frac{K_{\frac{d-1}{2}} \left(\frac{\zeta}{c}\sqrt{U^{\mu}U_{\mu}}\right)}{(U^{\mu}U_{\mu})^{\frac{d-1}{4}}} \quad .
\end{align*}

All integrals can now be obtained using Eq.~\ref{eq:Z-derivative}; tedious but straightforward manipulations 
yield a nice and regular structure. Indeed, defining  the coefficients $A_n$:
\begin{align}
  A_n = 2^{\frac{d+1}{2}} \pi^{\frac{d-1}{2}} \zeta^{n+\frac{d-1}{2}} K_{n+\frac{d-1}{2}}(\zeta) \quad ,
\end{align}
one obtains: 
\begin{align*}
 Z =& \left(\frac{k_{\rm B} T}{c}\right)^{d-1} A_0 \\
 Z^{\alpha} =& \left(\frac{k_{\rm B} T}{c}\right)^{d} A_1 \frac{U^\alpha}{c}  \\
 Z^{\alpha\beta} =& \left(\frac{k_{\rm B} T}{c}\right)^{d+1} \left[A_2\frac{U^{\alpha}U^{\beta}}{c^2}-A_1\eta^{\alpha\beta}\right] \\
 Z^{\alpha\beta\gamma} =& \left(\frac{k_{\rm B} T}{c}\right)^{d+2}\left[A_3\frac{U^{\alpha}U^{\beta}U^{\gamma}}{c^3}-A_2\left(\frac{\eta^{\alpha\beta}U^{\gamma}+\eta^{\gamma\beta}U^{\alpha}+\eta^{\alpha\gamma}U^{\beta}}{c}\right)\right] \\
\dots \\
Z^{\alpha_1 \alpha_2 \dots \alpha_n} =& \left(\frac{k_{\rm B} T}{c}\right)^{d+n-1} \sum_{k=0}^{\lfloor \frac{n}{2} \rfloor} (-1)^{k} A_{n-k} \frac{ \prec U_{n-2k} \eta_k \succ }{c^{n-2k}}
 \end{align*}
Where 
\begin{align*}
  \prec U_{n-2k} \eta_k \succ 
  = \underbrace{\eta^{\alpha_1\alpha_2}\dots\eta^{\alpha_{2k-1}\alpha_{2k}}}_{\tiny \text{k terms}}
    \underbrace{U^{\alpha_{2k+1}}U^{\alpha_{2k+2}}\dots U^{\alpha_n}}_{\tiny \text{(n-2k) terms}} 
    + \mbox{permutations of indexes}.
\end{align*}
In the above, terms like $\eta^{\alpha \beta}$, $\eta^{\beta \alpha}$ and $U^{\alpha}U^{\beta}$, $U^{\beta}U^{\alpha}$ are counted only once.
To give an example, for $n=4$, $k=1$, we have:
\begin{align*}
\prec U_{2} \eta_1 \succ &= \eta^{\alpha_1\alpha_2}U^{\alpha_3}U^{\alpha_4} 
                          + \eta^{\alpha_1\alpha_3}U^{\alpha_2}U^{\alpha_4}
                          + \eta^{\alpha_2\alpha_3}U^{\alpha_1}U^{\alpha_4} \\
                         &+ \eta^{\alpha_1\alpha_4}U^{\alpha_3}U^{\alpha_2}
                          + \eta^{\alpha_4\alpha_3}U^{\alpha_2}U^{\alpha_1}
                          + \eta^{\alpha_2\alpha_4}U^{\alpha_1}U^{\alpha_3} \quad .
\end{align*}

\subsection{Integrals $K^{\alpha_1 \dots \alpha_n}$}
Let
\begin{equation}
  K^{\alpha_1 \dots \alpha_n} =     \int f^{\rm eq}                     \frac{p^{\alpha_1} \dots p^{\alpha_s}}{p_\mu U^\mu} \frac{\diff^d p}{p_0} 
                              = B(n,T) \int e^{-\frac{p_\mu U^\mu}{k_{\rm B} T}} \frac{p^{\alpha_1} \dots p^{\alpha_s}}{p_\mu U^\mu} \frac{\diff^d p}{p_0} \quad ,
\end{equation}
with $B(n,T)$ defined in Eq.~\ref{eq:phys-normalization}. We include this normalization factor as these integrals are only encountered in the derivation of the transport coefficients following the Chapman-Enskog expansion.
It immediately follows from the definition that:
\begin{equation}
  U_{\alpha_n} K^{\alpha_1 \dots \alpha_n} = B(n,T) Z^{\alpha_1 \dots \alpha_{n-1} } \quad .
\end{equation}
The tensorial structure of $K^{\alpha_1 \dots \alpha_n}$ is similar to
that of $Z^{\alpha_1 \dots \alpha_n}$:
\begin{align*}
  K^{\alpha}                 & = \frac{n}{c k_{\rm B} T} a_{11} U^\alpha\\
  K^{\alpha\beta}            & = \frac{n}{c}         \left[a_{21} \frac{U^{\alpha}U^{\beta}}{c^2} - a_{22} \eta^{\alpha\beta}\right] \\
  K^{\alpha\beta\gamma}      & = \frac{n k_{\rm B} T}{c^2} \left[a_{31}\frac{U^{\alpha}U^{\beta}U^{\gamma}}{c^3} 
                                 - a_{32}\left(\frac{\eta^{\alpha\beta}U^{\gamma}+\eta^{\gamma\beta}U{\alpha}+\eta^{\alpha\gamma}U^{\beta}}{c}\right)\right] \\
  \dots \\
  K^{\alpha_1 \dots \alpha_n}& = \frac{n (k_{\rm B} T)^{n-2}}{c^{n-3}} \sum_{k=0}^{\lfloor \frac{n}{2} \rfloor} (-1)^{k} a_{n(k+1)} \frac{ \prec U_{n-2k} \eta_k \succ}{c^{n-2k}} 
\end{align*}
with (limiting ourselves to up to 4 Lorentz indexes):
\begin{align*}
  a_{11}  & = \frac{1}{\zeta^2} \left( G_{d} - (d+1) \right)                     \\
  a_{21}  & = 1 + \chi  &&    a_{22}   =     \chi                            \\
  a_{31}  & = G_{d} + 2   &&    a_{32}   =     1                                 \\
  a_{41}  & = \frac{3 G_{d} - 3 \zeta^2 \chi}{d+2} + (d+6) G_{d} + \zeta^2 &&  a_{42} = \frac{(3 d+7) G_{d} - \zeta^2 \chi}{d+2} && a_{43}  = \frac{G_{d} - \zeta^2 \chi}{d+2} \quad ;
\end{align*}
$G_{d}$ is defined in Eq. \ref{eq:defG} and $\chi$ is:
\begin{equation*}
\chi =
\left\{
  \begin{array}{ll}
    \frac{1}{d!!}  \sum_{k=0}^{\lfloor \frac{d-1}{4} \rfloor} \zeta^{2 k} 
           \left( g^{(k+1)}_{\frac{d+1}{2} - 2 k} - G_{d}~g^{(k+1)}_{\frac{d+1}{2} - 2 k - 1} + (-1)^{\frac{d+1}{2}}
           \zeta^{\frac{d-1}{2}} \frac{Ki_{1}(\zeta)}{K_{\frac{d+1}{2}}(\zeta)}
           \right)   
       & \mbox{for odd  d} \\
    \frac{1}{d!!}  \sum_{k=0}^{\lfloor \frac{d-1}{4} \rfloor} \zeta^{2 k} 
           \left( h^{(k+1)}_{\frac{d}{2} - 2 k} - G_{d}~h^{(k+1)}_{\frac{d}{2} - 2 k - 1} + (-1)^{\frac{d}{2}}
           \zeta^{\frac{d-2}{2}} \sqrt{\frac{\pi }{2}} \frac{\Gamma(0,\zeta)}{K_{\frac{d+1}{2}}(\zeta)}
           \right)          
       & \mbox{for even d} 
  \end{array}
\right.
\end{equation*}
Finally, the coefficients $g^{(i)}_j$ and $h^{(i)}_j$ are defined as:
\begin{equation*}
g^{(i)}_j = 
\begin{cases}
   0                                              \\
   1                                              \\
   2 j~g^{(i)}_{j-1} + (2 j - 3)!!                \\
   2 (j + 2i - 2)~g^{(i)}_{j-1} + ~g^{(i-1)}_{j}
\end{cases}
h^{(i)}_j =
\begin{cases}
   0                                                       & \mbox{for } j = 0 \\
   1                                                       & \mbox{for } j = 1 \\
   2 j~h^{(i)}_{j-1} + (2 j - 1)!!                         & \mbox{for } i = 1 \\
   2 (j + 2i - \frac{3}{2})~h^{(i)}_{j-1} + ~h^{(i-1)}_{j} & \mbox{else} \\
\end{cases}
\end{equation*}

We write explicitly a few cases:
\begin{align*}
  \chi &= 1 - \frac{Ki_{1}}{K_1}                                                                              & d= 1 \\
  \chi &= \frac{1}{2} \left( 1 -     \zeta^2 \frac{e^{\zeta} \Gamma(0, \zeta)}{1 + \zeta}             \right) & d= 2 \\
  \chi &= \frac{1}{3} \left( 5 - G_{d} + \zeta   \frac{Ki_{1}}{K_2}                                       \right) & d= 3 \\
  \chi &= \frac{1}{8} \left( 5 - G_{d} + \zeta^4 \frac{e^{\zeta } \Gamma (0,\zeta )}{\zeta ^2+3 \zeta +3} \right) & d= 4
\end{align*}

In the above expressions
\begin{equation*}
  Ki_{\alpha} =  \int_0^{\infty} e^{-\zeta \cosh (t)} \left( \cosh (t) \right)^{- \alpha} \diff t \quad ,
\end{equation*}
is the Bickley-Naylor function and 
\begin{equation*}
  \Gamma(\alpha, x) = \int_x^{\infty} y^{\alpha - 1} e^{-y} \diff y \quad ,
\end{equation*}
is the upper incomplete gamma function.

\section{Derivation of transport coefficients}\label{sec:appendixC}
%
In this appendix section we present the detailed derivation of the transport coefficients
for the relativistic Boltzmann equation in the relaxation time approximation. We present
the analytical derivation in a general $(d+1)$-dimensional Minkowski space-time following
both the Chapman Enskog expansion and Grad's method of moments. Useful for both cases 
are the constitutive equations for the heat flux, the dynamic pressure and the pressure 
deviator, which can be identified by applying suitable projectors, defined in 
Appendix~\ref{sec:appendixA}, to the non equilibrium components of the first and second
order tensors of the particles distribution function (Eq.~\ref{eq:first-moment-neq} and Eq.~\ref{eq:second-moment-neq}).
The resulting expressions write as follows:
\begin{align}
  q^{\alpha}           &= -h_e\Delta^{\alpha\beta}N_{\alpha}                     \quad , \label{eq:qfromn}   \\
  \varpi               &= -P -\frac{1}{d}\Delta_{\alpha\beta}T^{\alpha \beta}    \quad , \label{eq:omfromt}  \\
  \pi^{<\alpha \beta>} &= \Delta^{<\alpha}_\mu \Delta^{\beta>}_\nu T^{\mu\nu} 
                        = \left(\Delta^{\alpha}_\mu \Delta^{\beta}_\nu - \frac{1}{d} \Delta^{\alpha\beta} \Delta_{\mu\nu} \right)T^{\mu\nu}                                      \quad . \label{eq:pifromt}
\end{align}
%

\subsection{Chapman-Enskog expansion}

The Chapman-Enskog expansion consists in splitting the particle distribution function $f$ in two 
additive terms: the equilibrium distribution $f^{\rm eq}$ and a non equilibrium part $f^{\rm neq}$. 
When working in a hydrodynamic regime, it is reasonable to approximate $f^{\rm neq}$ with a small 
deviation from the equilibrium:
\begin{equation}\label{eq:f-approx-ce}
  f = f^{\rm eq} + f^{\rm neq} \sim f^{\rm eq}( 1 + \phi) \quad .
\end{equation}
with $\phi$ of the order of the Knudsen number $\rm Kn$, defined as the ratio between the mean 
free path and a typical macroscopic length scale.
Before going in details, we briefly outline the conceptual steps of the expansion.
The general idea is to determine an analytical expression for the deviation from the equilibrium $f^{\rm eq} \phi$.
We start from Eq.~\ref{eq:rbe-rta} (let us ignore for the moment the forcing term), insert Eq.~\ref{eq:f-approx-ce} and retain only terms $\mathcal{O}(\rm{Kn})$, giving:
\begin{equation}\label{eq:rta-rbe-ce}
  p^{\alpha} \partial_{\alpha} f^{\rm eq} = - \frac{p^{\alpha}U_{\alpha}}{c^{2} \tau} f^{\rm eq} \phi \quad .
\end{equation}
To derive the transport coefficients we will then proceed with the following steps:
\begin{enumerate}
  \item Compute the derivative $p^{\alpha} \partial_{\alpha} f^{\rm eq}$ and derive the constitutive equations of a relativistic Eulerian fluid.
  \item Use the balance equations for energy and momentum to eliminate the convective time derivatives and derive the analytic expression of $\phi$.
  \item Use the now known expression for $f^{\rm eq} \phi$ to compute the first and second order tensors (via their integral definitions), compare against their definition in the Landau frame and work out the expression for the transport coefficients.        
\end{enumerate}

\subsubsection{Constitutive equations of a relativistic Eulerian fluid}

From Eq.~\ref{eq:first-moment-neq} and Eq.~\ref{eq:second-moment-neq} we infer the following
constraints on the particle distribution function:
\begin{align}
  n c^2  &= U^{\alpha} N_{\alpha} = c U^{\alpha} \int f^{\rm eq} p_{\alpha} \frac{\diff^d p}{p_0} 
  = c U^{\alpha} \int f          p_{\alpha} \frac{\diff^d p}{p_0} \quad , \label{eq:landau-constraint1}\\
  \epsilon c^2 &= U^{\alpha} U^{\beta} T_{\alpha\beta}=U^{\alpha} U^{\beta} c \int f^{\rm eq} p_{\alpha} p_{\beta} \frac{\diff^d p}{p_0}
  = U^{\alpha} U^{\beta} c \int f          p_{\alpha} p_{\beta} \frac{\diff^d p}{p_0} \quad . \label{eq:landau-constraint2}
\end{align}   
These conditions together with Eq.~\ref{eq:f-approx-ce} lead to the following constraints for the non-equilibrium part of the distribution:
\begin{align}
  U_{\alpha}           \int p^{\alpha}           \phi f^{\rm eq} \frac{d^{d}p}{p_{0}} & = 0 \quad , \label{eq:f-neq-cond1} \\
  U_{\alpha} U_{\beta} \int p^{\alpha} p^{\beta} \phi f^{\rm eq} \frac{d^{d}p}{p_{0}} & = 0 \quad . \label{eq:f-neq-cond2}       
\end{align}
Furthermore, we multiply Eq.~\ref{eq:rta-rbe-ce} by $\{1, p^{\beta} \}$ and integrate in momentum space, getting:
\begin{align}
 \int            p^{\alpha} \partial_{\alpha} f^{\rm eq}                         \frac{d^{d}p}{p_{0}}
      &= 
      - \frac{1}{c^{2}\tau} U_{\alpha} \int           p^{\alpha} \phi f^{\rm eq} \frac{d^{d}p}{p_{0}} = 0  \label{eq:collisional-cond1} \quad , \\
 \int p^{\beta } p^{\alpha} \partial_{\alpha} f^{\rm eq}                         \frac{d^{d}p}{p_{0}}
      &= 
      - \frac{1}{c^{2}\tau} U_{\alpha} \int p^{\beta} p^{\alpha} \phi f^{\rm eq} \frac{d^{d}p}{p_{0}} = 0  \quad . \label{eq:collisional-cond2}
\end{align}

We now use the definition of the Maxwell-J\"uttner distribution (Eq.~\ref{eq:maxwell-juttner}) 
to calculate $p^{\alpha} \partial_{\alpha} f^{\rm eq}$:
\begin{equation}\label{eq:feq-derivative}
  p^{\alpha} \partial_{\alpha} f^{\rm eq} 
  = 
  f^{\rm eq} p^{\alpha} \left[ \frac{\partial_\alpha n}{n} + \left(1 - G_{d} \right) \frac{\partial_\alpha T}{T} \right]
  +
  f^{\rm eq} \frac{p^{\alpha}p^{\beta}}{k_{\rm B} T} \left[ \frac{U_{\beta}\partial_{\alpha}T}{T} - \partial_{\alpha}U_{\beta} \right] \quad .
\end{equation}

Plugging Eq.~\ref{eq:feq-derivative} into Eq.~\ref{eq:collisional-cond1} and Eq.~\ref{eq:collisional-cond2} gives
\begin{align}
  \left[ \frac{\partial_\alpha n}{n} + \left(1 - G_{d} \right) \frac{\partial_\alpha T}{T} \right] Z^{\alpha}
  +
  \frac{1}{k_{\rm B} T} \left[ \frac{U_{\beta}\partial_{\alpha}T}{T} - \partial_{\alpha}U_{\beta} \right] Z^{\alpha \beta}
  & = 0 \quad , \label{eq:mass-conservation-cond} \\
  \left[ \frac{\partial_\alpha n}{n} + \left(1 - G_{d} \right) \frac{\partial_\alpha T}{T} \right] Z^{\alpha \beta}
  +
  \frac{1}{k_{\rm B} T} \left[ \frac{U_{\beta}\partial_{\alpha}T}{T} - \partial_{\alpha}U_{\beta} \right] Z^{\alpha\beta\gamma}         
  & = 0 \quad , \label{eq:energy-momentum-conservation-cond}
\end{align}
with details on the calculation of integrals $Z^{\alpha_1 \dots\alpha_n}$ given in Appendix~\ref{sec:appendixB}.

It is possible to rearrange the above equations as
\begin{align}
  0 & = D n + n \nabla^{\alpha} U_{\alpha}                                  
  \quad , \label{eq:mass-conservation-cond2} \\
  0 & = -\partial^{\beta}(nT) + n \frac{U^{\beta} DT}{c^2} \left[(2+d)G_{d} + \zeta^2 - G_{d}^2\right] + G_{d} n T \frac{DU^{\beta}}{c^2} 
  \quad . \label{eq:energy-momentum-conservation-cond2}
\end{align}
The first equation above is immediately recognized as the relativistic counterpart of the continuity equation, while the balance equations for energy
and momentum stem from the second equation. To summarize we get:
\begin{align}
        Dn + n \nabla^{\alpha} U_{\alpha}                                & = 0 \quad ,  \label{eq:conservation-particle-number} \\
  n c_v DT + P \nabla_{\alpha} U^{\alpha}                                & = 0 \quad ,  \label{eq:conservation-momentum}        \\
  \nabla^\gamma P - \left( P + \epsilon \right) \frac{DU^{\gamma} }{c^2} & = 0 \quad ,  \label{eq:conservation-energy}         
\end{align}
where Eq.~\ref{eq:conservation-momentum} and Eq.~\ref{eq:conservation-energy} have been derived by 
multiplying Eq.~\ref{eq:energy-momentum-conservation-cond2} respectively by $U_{\beta}$ and $\Delta^{\gamma}_{\beta}$,
using the EOS in Eq.~\ref{eq:eos-ddim}, and having identified the general expression for the heat capacity $c_v$ \cite{cercignani-book-2002}:
\begin{equation}\label{eq:heat-capacity}
  c_v = k_{\rm B} \left[ (2 + d)G_{d} + \zeta^2 -G_{d}^2 -1 \right] \quad .
\end{equation}

\subsubsection{Approximation of the non-equilibrium term of the particle distribution function} 

Combining Eq.~\ref{eq:rta-rbe-ce} with Eq.~\ref{eq:feq-derivative} it is possible to define an analytic expression for $\phi$. 
The final result reads as
\begin{equation}\label{eq:analytic-phi}
 \phi = - \frac{c^2 \tau}{p^\mu U_\mu} p^\alpha
        \left[ \frac{\partial_\alpha n}{n} + \left(1 - G_{d}\right) \frac{\partial_\alpha T}{T}
               +       p^{\beta} \frac{ U_{\beta} \partial_{\alpha} T}{k_{\rm B} T^2}
               - \frac{p^{\beta} \partial_{\alpha} U_{\beta}}{k_{\rm B} T}
        \right] \quad .
\end{equation}
By knowing the deviation of the Maxwell-J\"uttner distribution function it is now possible to 
determine the transport coefficients from the integral definition of the moments of the distribution.
We start by taking into consideration the thermal conductivity.

\subsubsection{Thermal Conductivity}
%
In order to determine the analytic expression for the thermal conductivity we consider the integral
definition of the particle flow tensor together with Eq.~\ref{eq:f-approx-ce}:
\begin{align}
  N^{\beta} &= c \int p^{\beta} f \frac{\diff^d p}{p_0} \notag \\
            &= c \int p^{\beta} f^{\rm eq} \frac{\diff^d p}{p_0} + c \int p^{\beta} \phi f^{\rm eq} \frac{\diff^d p}{p_0} \quad .
\end{align}
We then insert Eq.~\ref{eq:analytic-phi}, giving 
\begin{equation}\label{eq:expanded-four-flow}
  N^{\beta} = c B Z^{\beta} 
             -c^3 \tau  \left[   \frac{\partial_\nu n}{n} K^{\nu \beta}
                               + \left(1 - G_{d} \right) \frac{\partial_\nu T}{T} K^{\nu \beta}
                               + \frac{\partial_{\nu}T}{k_{\rm B} T^2} B Z^{\beta \nu} 
                               - \frac{\partial_{\nu}U_{\delta}}{k_{\rm B} T} K^{\delta \nu \beta}
                        \right]   \quad .
\end{equation}

Next, we apply to the above equation the projector $\Delta^{\alpha}_{\beta}$, giving
\begin{equation}\label{eq:projected-flow-tensor-tmp}
  \Delta^{\alpha}_{\beta} N^\beta 
  =
  -c^2 \tau \left[ -  a_{22} \nabla^{\alpha} n    - a_{22} \left(1 - G_{d} \right) \frac{n\nabla^{\alpha} T}{T} 
                   - \frac{n \nabla^{\alpha}T}{T} + \frac{n}{c^2}DU^{\alpha}
            \right] \quad ,
\end{equation}
in which we have used the integrals $K^{\alpha_1 \dots\alpha_n}$ defined in Appendix~\ref{sec:appendixB},
together with the definition of the normalization factor $B(n,T)$ from Eq.~\ref{eq:phys-normalization}.

Using the balance equations \ref{eq:conservation-particle-number}, \ref{eq:conservation-energy} and \ref{eq:conservation-momentum} it is possible to rearrange Eq.~\ref{eq:projected-flow-tensor-tmp} as
\begin{equation}\label{eq:projected-flow-tensor}
  \Delta^{\alpha}_{\beta} N^\beta
  =
  -n c^2 \tau \frac{a_{22}G_{d}-1}{T} \left[  \nabla^{\alpha} T 
                                         -\frac{T}{P + \epsilon} \nabla^{\alpha} P
                                  \right] \quad .
\end{equation}
To conclude, we combine Eq.~\ref{eq:projected-flow-tensor} with Eq.~\ref{eq:qfromn}, getting:
\begin{equation}
  q^\alpha = -h_e \Delta^{\alpha}_{\beta} N^\beta
           = c^2 k_{\rm B} n \tau G_{d} \left(a_{22}G_{d} - 1 \right) \left[ \nabla^\alpha T -\frac{T}{P + \epsilon} \nabla^{\alpha} P \right]  \quad ,
\end{equation}
from which, by direct comparison with Eq.~\ref{eq:heat-flux}, we conclude that the thermal conductivity $\lambda$ is defined as
\begin{equation}
 \lambda = c^2 k_{\rm B} n \tau G_{d} \left(a_{22}G_{d} - 1 \right) \quad .
\end{equation}
To give a few examples:
\begin{align*}
  \lambda &= c^2 k_{\rm B} n \tau G_{d} \left( G_{d} -1 -G_{d} \frac{Ki_{1}}{K_1} \right)                                         & \quad \quad \quad d = 1 \\
  \lambda &= c^2 k_{\rm B} n \tau G_{d} \left( \frac{G_{d}}{2} -1 -G_{d} \frac{\zeta^2e^\zeta\Gamma(0,\zeta)}{2(\zeta+1)} \right) & \quad \quad \quad d = 2 \\
  \lambda &= c^2 k_{\rm B} n \tau G_{d} \left( \frac{G_{d}}{3}(5-G_{d}) -1 +G_{d} \zeta\frac{Ki_{1}}{3K_2} \right)                & \quad \quad \quad d = 3 \\
  \lambda &= c^2 k_{\rm B} n \tau G_{d} \left( \frac{G_{d}}{8}(5-G_{d}) -1 +G_{d} \frac{\zeta^4e^\zeta\Gamma(0,\zeta)}{8(\zeta^2+3\zeta+3)} \right)                & \quad \quad \quad d = 4
\end{align*}
In the ultra relativistic limit ($\zeta \rightarrow 0$), the above expression simplifies to
\begin{equation}
  \lambda_{\rm ur} = \frac{d+1}{d} c^2 k_{\rm B} n \tau \quad .
\end{equation}
%
\subsubsection{Bulk Viscosity}

In order to determine the analytic expression for the bulk viscosity we take the integral
definition of the energy-momentum tensor together with Eq.~\ref{eq:f-approx-ce}:

\begin{align}
  T^{\alpha \beta} &= c \int p^{\alpha} p^{\beta} f \frac{d^{d}p}{p_{0}} \notag \\
                   &= c \int p^{\alpha} p^{\beta} f^{\rm eq} \frac{\diff^d p}{p_0} + c \int p^{\alpha} p^{\beta} \phi f^{\rm eq} \frac{\diff^d p}{p_0} \quad .
\end{align}
next, we insert Eq.~\ref{eq:analytic-phi}, giving
\begin{equation}\label{eq:momentum-tensor-with-phi}
  T^{\alpha \beta} = c B Z^{\alpha \beta} -c^3 \tau 
                     \left[         \frac{\partial_{\nu} n}{n} K^{\nu \alpha \beta}
                             + (1-G_{d})\frac{\partial_{\nu} T}{T} K^{\nu \alpha \beta}
                                   +\frac{\partial_{\nu} T}{k_{\rm B} T^2} B Z^{\nu \alpha \beta} 
                                   -\frac{\partial_{\nu} U_{\delta}}{k_{\rm B} T} K^{\delta \nu \alpha \beta}
                     \right] \quad .
\end{equation} 

By applying the projector $\Delta_{\alpha\beta}$ and using the results of the integrals $K^{\alpha_1 \dots\alpha_n}$ given in Appendix~\ref{sec:appendixB}, we obtain:
\begin{equation}\label{eq:projected-momentum-tensor}
  \Delta_{\alpha \beta} T^{\alpha \beta} 
  =
  - d P + \tau d \left[ k_{\rm B} T D n + (1 - G_{d}) n k_{\rm B} D T + G_{d} n k_{\rm B} D T + P a_{43} (1 + \frac{2}{d}) \Delta^{\nu} U_{\nu} \right] \quad .
\end{equation}

By comparing Eq.~\ref{eq:projected-momentum-tensor} with Eq.~\ref{eq:omfromt} we directly get
\begin{align}
  \varpi = -\tau \left[ k_{\rm B} T D n +(1-G_{d}) n k_{\rm B} D T + G_{d} n k_{\rm B} D T + P a_{43} \left( 1 + \frac{2}{d} \right) \Delta^{\nu} U_{\nu} \right] \quad .
\end{align}
We then use balance equations \ref{eq:conservation-particle-number}, \ref{eq:conservation-energy} and \ref{eq:conservation-momentum} to remove the convective derivatives $D n$ and $D T$, leading to:
\begin{equation}
  \varpi = P \tau \left[ \left(1 + \frac{k_{\rm B}}{c_v} \right) +a_{43} \left( 1 + \frac{2}{d} \right) \right] \Delta^{\nu} U_{\nu} \quad ,
\end{equation}
which implies, by direct comparison with Eq.~\ref{eq:dynamic-pressure}, the following expression 
for the bulk viscosity $\mu$:
\begin{equation}
  \mu = P \tau \left[ a_{43} \left( 1 + \frac{2}{d} \right) - \frac{ \zeta^2 -G_{d}^2 +(d+2) G_{d}}{ \zeta^2 -G_{d}^2 +(d+2) G_{d}-1} \right] \quad .
\end{equation}
To give a few examples:
\begin{align*}
  \mu &= P \tau \left( \frac{\zeta^4 +G_{d}^3 -4 G_{d}^2 + 4 G_{d} -\zeta^2 G_{d}^2+2 \zeta^2 G_{d}}{
                            -\zeta^2+G_{d}^2-3 G_{d}+1}
                        + \zeta^2 \frac{Ki_{1}}{K_1}\right)  
                     & \quad d = 1 \\
  \mu &= P \tau \left( \frac{\zeta^4 +3 \zeta^2 + 2 G_{d}^3-12 G_{d}^2+18 G_{d}-\zeta^2 G_{d}^2+2 \zeta^2 G_{d}}{
                             4 \left(-\zeta^2+G_{d}^2-4 G_{d}+1\right)}
                        + \frac{e^{\zeta }\zeta^4 \Gamma (0,\zeta )}{4 (\zeta +1)} \right) 
                     & \quad d = 2 \\
  \mu &= P \tau \left( \frac{5 \zeta^4 +4 \zeta^2 +3 G_{d}^3 -18 G_{d}^2 +30 G_{d} + \zeta^2 G_{d}^3 -8 \zeta^2 G_{d}^2 -\zeta^4 G_{d} +13 \zeta^2 G_{d}}{
                             9 \left(-\zeta^2+G_{d}^2-3 G_{d}+1\right)}
                        - \zeta^3 \frac{Ki_{1}}{9K_2}\right) 
                     & \quad d = 3 \\
 \mu &= P \tau \left( \frac{\zeta ^4 G_d-\zeta ^2 G_d^3+13 \zeta ^2 G_d^2-35 \zeta ^2 G_d-8 G_d^3+80 G_d^2-200 G_d-7 \zeta ^4-25 \zeta ^2}{32 \left(-G_d^2+6 G_d+\zeta ^2-1\right)}
                        + \frac{e^{\zeta }\zeta^6 \Gamma (0,\zeta )}{32 (\zeta^2 +3 \zeta +3)} \right) 
                     & \quad d = 4
\end{align*}
The bulk viscosity of a monoatomic ultra-relativistic gas vanishes independently with
respect to the number of spatial dimensions:
\begin{equation}
  \mu_{\rm ur} = 0 \quad .
\end{equation}

\subsubsection{Shear Viscosity}

In order to determine the analytic expression for the shear viscosity $\eta$ we 
start from the definition of the pressure deviator given in Eq.~\ref{eq:pifromt}:
\begin{equation}\label{eq:pressure-dev}
  (\Delta_\alpha^{\gamma}\Delta_{\beta}^{\delta} -\frac{1}{d} \Delta^{\gamma \delta} \Delta_{\alpha \beta}) T^{\alpha \beta} = \pi^{<\alpha \beta>} \quad .
\end{equation}
We can determine the LHS of the above equation by using the same expression 
for $T^{\alpha\beta}$ previously identified in the calculation of the bulk viscosity 
(Eq.~\ref{eq:momentum-tensor-with-phi}). 
We split the task in two parts and start by considering the first term of the LHS, 
which consists in applying the projector $\Delta_\alpha^{\gamma}\Delta_{\beta}^{\delta}$ to
the energy-momentum tensor:
\begin{align}
  \Delta_{\alpha}^{\gamma} \Delta_{\beta}^{\delta} T^{\alpha \beta} 
  &= c \Delta_\alpha^\gamma\Delta_\beta^\delta Z^{\alpha\beta}-c^3\tau \left[\frac{\partial_\nu n}{n}\Delta_\alpha^\gamma\Delta_\beta^\delta K^{\nu \alpha \beta}+(1-G_{d})\frac{\partial_\nu T}{T}\Delta_\alpha^\gamma\Delta_\beta^\delta K^{\nu \alpha \beta}\right. \notag \\
  &~~~~~~~~~~~~~~~~~~~~~~~~~~~~\left.+\frac{\partial_{\nu}T}{k_{\rm B}T^2}\Delta_\alpha^\gamma\Delta_\beta^\delta Z^{\nu\alpha\beta}-\frac{\partial_{\nu}U_{\mu}}{k_{\rm B}T}\Delta_\alpha^\gamma\Delta_\beta^\delta K^{\mu   \nu \alpha \beta}\right]  \quad .
\end{align}
Using the integral definitions in Appendix~\ref{sec:appendixB} we obtain:
\begin{align}
  \Delta_{\alpha}^{\gamma} \Delta_{\beta}^{\delta} T^{\alpha \beta} = 
  -P d \Delta^{\gamma\delta} + \tau\left[k_{\rm B} T D n \Delta^{\gamma\delta} + n k_{\rm B} D T\Delta^{\gamma\delta}
  +P a_{43}(\nabla^\gamma U^\delta +\nabla^\delta U^\gamma+\Delta^{\delta\gamma}\partial_\nu U^\nu)\right] \quad .
\end{align}
By using the balance equations for energy and momentum we get
\begin{align}
 \Delta_{\alpha}^{\gamma} \Delta_{\beta}^{\delta} T^{\alpha \beta}  
 &=-P\Delta^{\gamma\delta}+P\tau\left[\left(a_{43}-1-\frac{k_{\rm B}}{c_v}\right)\Delta^{\delta\gamma}\nabla_\nu U^\nu+a_{43}(\nabla^\gamma U^\delta +\nabla^\delta U^\gamma)\right] \\ 
 &=-P\Delta^{\gamma\delta}+\mu\Delta^{\delta\gamma}\nabla_\nu U^\nu+P\tau a_{43}\left[-\frac{2}{d}\Delta^{\delta\gamma}\nabla_\nu U^\nu+(\nabla^\gamma U^\delta +\nabla^\delta U^\gamma)\right] \notag \quad .
\end{align}

We now consider the second term on the LHS of Eq.~\ref{eq:pressure-dev} and apply 
the projection $-\frac{1}{d}\Delta^{\gamma\delta}\Delta_{\alpha\beta}$ to the energy momentum tensor, giving:
\begin{equation}
  -\frac{1}{d}\Delta^{\gamma \delta} \Delta_{\alpha \beta} T^{\alpha \beta}
  =
  \Delta^{\gamma \delta} (P + \varpi) 
  =
  \Delta^{\gamma \delta} (P - \mu \nabla^\mu U_\mu) \quad .
\end{equation}

Putting all the pieces together we get
\begin{equation}\label{eq:stress-tensor-ce}
 \pi^{<\gamma\delta>} = (\Delta_\alpha^{\gamma}\Delta_\beta^{\delta}-\frac{1}{d}\Delta^{\gamma\delta}\Delta_{\alpha\beta})T^{\alpha\beta}
                      = P\tau a_{43} \left[ \nabla^\gamma U^\delta + \nabla^\delta U^\gamma 
                                            -\frac{2}{d}\Delta^{\delta\gamma}\nabla_\nu U^\nu
                                     \right] \quad.
\end{equation}

To conclude, we expand the RHS of Eq.~\ref{eq:stress-tensor}:
\begin{align}
  \pi^{<\gamma\delta>} 
  = 
  2\eta\nabla^{<\gamma}U^{\delta>} 
  = 
  \eta\left[\nabla^\gamma U^\delta +\nabla^\delta U^\gamma-\frac{2}{d}\Delta^{\delta\gamma}\nabla_\nu U^\nu\right] \quad .
\end{align}
which we compare with Eq.~\ref{eq:stress-tensor-ce} to identify the
analytical  expression for the shear viscosity $\eta$:
\begin{equation}
  \eta = P \tau a_{43} \quad .
\end{equation}
To give a few examples:
\begin{align*}
  \eta &= \frac{P \tau}{3}  \left( G_{d} -  \zeta^2 + \zeta^2 \frac{Ki_{1}}{K_1}\right)                              & \quad \quad \quad d = 1 \\
  \eta &= \frac{P \tau}{8}  \left(2G_{d} -  \zeta^2 + \frac{e^{\zeta} \zeta^4 \Gamma (0,\zeta)}{\zeta +1} \right)    & \quad \quad \quad d = 2 \\
  \eta &= \frac{P \tau}{15} \left(3G_{d} - 5\zeta^2 + G_{d} \zeta^2 - \zeta^3 \frac{Ki_{1}}{K_2}\right)              & \quad \quad \quad d = 3 \\
  \eta &= \frac{P \tau}{48} \left(8G_{d} - 7\zeta^2 + G_{d} \zeta^2 + \frac{e^{\zeta} \zeta^6 \Gamma (0,\zeta)}{\zeta^2 +3 \zeta + 3}\right)              & \quad \quad \quad d = 4
\end{align*}
In the ultra relativistic limit ($\zeta \rightarrow 0$), the above expression simplifies to
\begin{equation}
  \eta_{\rm ur} = \frac{d+1}{d+2} P \tau \quad .
\end{equation}

\subsection{Grad's Method of Moments}

The derivation of the transport coefficients following Grad's method is based on the moments of 
the particle distribution function $f$. Likewise for the Chapman Enskog procedure, the starting point
consists in defining the non-equilibrium part of $f$, which follows from an expansion in the Hilbert Space of momenta
around the equilibrium:
\begin{align}\label{eq:f_expansion}
  f = f^{\rm eq}(1+a_\alpha p^\alpha + a_{\alpha\beta} p^{\alpha}p^{\beta}+ \dots) \quad .
\end{align}

Most of the analytical derivation revolves around determining the expansion coefficients $a_{\alpha_1 \dots\alpha_N}$ 
by imposing suitable constraints on the fields of interest; in the most standard approach one
includes the fourteen fields $n, U^{\alpha}, T, q^{\alpha}, \varpi, \pi^{<\alpha \beta>}$ which
are used in the decomposition of the tensors $N^{\alpha}$ and $T^{\alpha \beta}$ 
(Eq.~\ref{eq:first-moment-neq} and Eq.~\ref{eq:second-moment-neq}).
Note that, as already discussed in the main text, it has been shown \cite{denicol-prd-2012} that this representation
is not based on irreducible tensors, and cannot guarantee that the exact form of the expansion coefficients
are obtained once the expansion is truncated. 

We follow the same procedure described in \cite{cercignani-book-2002}, where the constraints on 
the fourteen fields stem from the maximization of the entropy density $s$, and consisting of the 
following steps:

\begin{enumerate}
  \item Using Lagrange multipliers method, we find an expansion for $f$, in the form of Eq.~\ref{eq:f_expansion}, 
        that extremizes the entropy density $s$, with the constraints given by definition of 
        $N^\alpha U_\alpha$, $U_\alpha T^{\alpha\beta}$ and $U_\alpha T^{\alpha<\beta\gamma>}$. 
  \item Using Grad's ansatz for $f$ we compute the third order moment $T^{\alpha\beta\gamma}$.
  \item The above expression is then plugged into Eq.~\ref{eq:rbe-rta} to determine the non-equilibrium components
        of the energy-momentum tensor.
  \item By applying the projectors $\Delta^{\delta}_\beta U_\gamma$, $\Delta_{\beta\gamma}$, 
        and $\Delta^{(\delta}_\beta\Delta^{\epsilon )}_\gamma-\frac{1}{d}\Delta_{\beta\gamma}\Delta^{\delta\epsilon}$ 
        is it then possible to derive the constitutive equations for the heat-flux, 
        dynamic pressure and the pressure deviator. 
        The expressions for the transport coefficients are finally derived 
        by comparison with Eq.~\ref{eq:heat-flux}, Eq.~\ref{eq:stress-tensor}, and Eq.~\ref{eq:dynamic-pressure}.     
\end{enumerate}

\subsubsection{Moments expansion of the particle distribution function}

We start from the definition of the entropy per particle $s$ \cite{cercignani-book-2002}:
\begin{align}
  s[f]=-\frac{k_{\rm B}}{nc} U^\alpha \int p_\alpha f \ln{f} \frac{d^dp}{p_0} \quad .
\end{align}

Next, we define the following constraints which follow from 
the definitions of $U_\alpha N^\alpha$, $U_\alpha T^{\alpha\beta}$ and $U_\alpha T^{\alpha<\beta\gamma>}$:
\begin{align}
  g[f] 
  &= 
  U_\alpha N^\alpha - c U_\alpha \int p^\alpha f  \frac{d^dp}{p_0}                \quad , \\
  g^\beta[f] 
  &= 
  U_\alpha T^{\alpha\beta} - c U_\alpha \int p^\alpha p^\beta f  \frac{d^dp}{p_0} \quad ,  \\
  g^{<\gamma\beta>}[f] 
  &= 
  U_\alpha T^{\alpha<\beta\gamma>} - c U_\alpha \int p^\alpha p^{<\beta} p^{\gamma>} f  \frac{d^dp}{p_0} \quad .
\end{align}
We then apply Lagrange multipliers method, and look for the expression of $f$ that extremizes the functional $F[f]$, defined as
\begin{align}
  F[f] = s[f] -\lambda g[f] -\lambda_\beta g^\beta[f] -\lambda_{<\gamma\beta>} g^{<\gamma\beta>}[f] \quad ,
\end{align}
and from which follows
\begin{align}
  0 = \frac{\partial F}{\partial f} 
    = - \int p_\alpha U^\alpha \left[
                                          \frac{k_{\rm B}}{nc}\left(\ln{f}+1 \right)
                                      + c \left(\lambda + \lambda_\beta p^\beta + \lambda_{<\gamma\beta>} p^{<\gamma}p^{\beta>}\right)
                               \right] \frac{d^dp}{p_0} \quad ,
\end{align} 
\begin{align}
  f = \exp{ \left(-1-\frac{nc^2}{k_{\rm B}}(\lambda + \lambda_\beta p^\beta + \lambda_{<\gamma\beta>} p^{<\gamma}p^{\beta>})\right)} \quad .
\end{align}
Since we know that at the equilibrium the above expression needs to reduce to the Maxwell-J\"uttner distribution, 
we can split the coefficients into two parts, the equilibrium (which we label as $(\rm E)$), and the non-equilibrium part 
(which we label as $(\rm NE)$):
\begin{align}
  \lambda                 = \lambda^{(\rm E)} + \lambda^{(\rm NE)}                                  \quad ,\\
  \lambda_\beta           = \lambda^{(\rm E)}_\beta + \lambda^{(\rm NE)}_\beta                      \quad ,\\
  \lambda_{<\gamma\beta>} = \lambda^{(\rm E)}_{<\gamma\beta>} + \lambda^{(\rm NE)}_{<\gamma\beta>}  \quad ,
\end{align}
to then write:
\begin{align}
  f &= f^{\rm eq} \exp \left(-\frac{nc^2}{k_{\rm B}}(\lambda^{(\rm NE)} 
                             + \lambda_\beta^{(\rm NE)} p^\beta 
                             + \lambda_{<\gamma\beta>}^{(\rm NE)} p^{<\gamma}p^{\beta>})
                       \right) \quad .
\end{align}
Under the assumption of processes not far from equilibrium 
($\lambda^{(\rm NE)}\ll 1$, $\lambda_\beta^{(\rm NE)}\ll 1$, $\lambda_{<\gamma\beta>}^{(\rm NE)}\ll 1$) 
it is possible to linearly expand the exponential:
\begin{align}
  f \sim f^{\rm eq} \left[ 1 - \frac{nc^2}{k_{\rm B}}(\lambda^{(\rm NE)} 
                             + \lambda_\beta^{(\rm NE)} p^\beta 
                             + \lambda_{<\gamma\beta>}^{(\rm NE)} p^{<\gamma}p^{\beta>})
                    \right] \quad .
\end{align}
Applying the projectors defined in Appendix~\ref{sec:appendixA}, we can decompose the vector $\lambda_\beta^{(\rm NE)}$ 
and the Lagrange coefficients tensor, $\lambda_{\gamma\beta}^{(\rm NE)}$ in the following way:
\begin{align}
  \lambda_\beta^{(\rm NE)}           &= \tilde{\lambda} U_\beta + \tilde{\lambda}_\gamma \Delta_\beta^\gamma \quad , \\
  \lambda_{<\gamma\beta>}^{(\rm NE)} &=   \Lambda U_\gamma U_\beta 
                                    + \frac{1}{2}\Lambda_\alpha(\Delta_\gamma^\alpha U_\beta + \Delta_\beta^\alpha U_\gamma) 
                                    + \Lambda_{\alpha\delta}(\Delta_\gamma^\alpha \Delta_\beta^\delta  - \frac{1}{d}\Delta^{\alpha\delta}\Delta_{\gamma\beta})
  \quad .
\end{align}

The expression for Grad's distribution function now reads as:
\begin{align}\label{eq:grad-ansatz-v1}
  f = f^{\rm eq} & \left[ 1 - \frac{nc^2}{k_{\rm B}} \left( \lambda^{(\rm NE)} 
                          + \tilde{\lambda} U_\beta p^\beta 
                          + \tilde{\lambda}_\gamma \Delta_\beta^\gamma p^\beta + \Lambda U_\gamma U_\beta p^{\gamma}p^{\beta}  
                          \right.\right. \notag \\
                 & \left. \left.
                          + \Lambda_\alpha\Delta_\gamma^\alpha U_\beta p^{\gamma}p^{\beta} 
                          + \Lambda_{\alpha\delta}\Delta_\gamma^{<\alpha} \Delta_\beta^{\delta>}p^{\gamma}p^{\beta}\right)
                  \right] \quad ,
\end{align}
where we have used the identity 
$\lambda_{<\gamma\beta>}^{(\rm NE)} p^{<\gamma}p^{\beta>} = \lambda_{<\gamma\beta>}^{(\rm NE)} p^{\gamma}p^{\beta}$.

The next step consists in finding the unknowns $\lambda^{(\rm NE)}$, $\tilde{\lambda}$, 
$\tilde{\lambda}_\gamma$, $\Lambda$, $\Lambda_\alpha$, $\Lambda_{\alpha\delta}$.
In order to do so, we plug Eq.~\ref{eq:grad-ansatz-v1} in the integral definition of $N^\epsilon$ and $T^{\mu\nu}$ (Eq.~\ref{eq:pdf-moments-N} and Eq.~\ref{eq:pdf-moments-T}):
\begin{align}
  N^{\epsilon} = c B Z^{\epsilon} - B \frac{nc^3}{k_{\rm B}} &\left[\lambda^{(\rm NE)}Z^{\epsilon} + \tilde{\lambda} U_\beta Z^{\beta\epsilon} + \tilde{\lambda}_\gamma \Delta_\beta^\gamma Z^{\beta\epsilon} + \Lambda U_\gamma U_\beta Z^{\epsilon\gamma\beta} \right. \notag \\
  & \left. + \Lambda_\alpha\Delta_\gamma^\alpha U_\beta Z^{\epsilon\gamma\beta} + \Lambda_{\alpha\delta}\Delta_\gamma^{<\alpha} \Delta_\beta^{\delta>}Z^{\epsilon\gamma\beta}\right] \quad , \\
  T^{\mu\nu} = c B Z^{\mu\nu} - B \frac{nc^3}{k_{\rm B}} &\left[\lambda^{(\rm NE)}Z^{\mu\nu} + \tilde{\lambda} U_\beta Z^{\beta\mu\nu} + \tilde{\lambda}_\gamma \Delta_\beta^\gamma Z^{\beta\mu\nu} + \Lambda U_\gamma U_\beta Z^{\mu\nu\gamma\beta}  \right. \notag \\
  & \left. + \Lambda_\alpha\Delta_\gamma^\alpha U_\beta Z^{\mu\nu\gamma\beta} + \Lambda_{\alpha\delta}\Delta_\gamma^{<\alpha} \Delta_\beta^{\delta>} Z^{\mu\nu\gamma\beta}\right] \quad ,
\end{align}
with integrals $Z^{\alpha_1 \dots\alpha_n}$ computed in appendix~\ref{sec:appendixB}. 
Combining the above equations with Eq.~\ref{eq:landau-constraint1}, and Eq.~\ref{eq:landau-constraint2}, and together with the 
constitutive equations Eq.~\ref{eq:qfromn}, Eq.~\ref{eq:omfromt}, Eq.~\ref{eq:pifromt} it is possible to obtain
the following linear system of equations for the Lagrange coefficients:
\begin{align*}
  \begin{cases}
    0      &= \lambda ^{(\rm NE)}+\tilde{\lambda}k_{\rm B} T(G_{d}-1)  +\Lambda k_{\rm B}^2 T^2 \left(d G_{d}+\zeta^2\right)                                                                            \\
    0      &=  \Lambda_\alpha \Delta ^{\alpha\epsilon}k_{\rm B}^2 T^2\left[(d+2) G_{d}+\zeta^2\right]  + k_{\rm B} T G_{d} \tilde{\lambda}_\gamma \Delta ^{\gamma\epsilon}                                  \\
    0      &= \lambda^{(\rm NE)}(G_{d}-1) + k_{\rm B} T \tilde{\lambda } \left(d G_{d}+\zeta^2\right)+k_{\rm B}^2 \Lambda  T^2 \left[d^2 G_{d}+d \left(\zeta^2+2 G_{d}\right)+\zeta^2 (G_{d}-1)\right]   \\
    q^\mu  &= -k_{\rm B} c^2 n^2 T^2 G_{d}\left(\tilde{\lambda}_\gamma \Delta ^{\gamma\mu} + \Lambda_\alpha \Delta ^{\alpha\mu} G_{d} k_{\rm B} T    \right)                                                \\
    \varpi &= -c^2 n^2 T \left(G_{d} k_{\rm B} T \tilde{\lambda }+k_{\rm B}^2 \Lambda  T^2 \left[(d+2) G_{d}+\zeta^2\right]+\lambda^{(\rm NE)}\right)                                                    \\
    \pi^{<\epsilon\lambda>} &= -2n^2 k_{\rm B}^2 T^3  G_{d}  \Lambda^{<\epsilon\lambda>}
  \end{cases}
\end{align*}

The solution reads as:
\begin{align}
  \lambda^{(\rm NE)} &= -\frac{\zeta^2 \left(d (G_{d}+1)-(G_{d}-1)^2\right)+d G_{d} (d-2 G_{d}+2)+\zeta ^4}{\zeta^2 (d-2 G_{d})+G_{d} (-d+G_{d}-1) (-d+2 G_{d}-2)}\frac{\varpi}{c^2 n^2 T}                                              \quad , \\
  \tilde{\lambda}      &= -\frac{d \left(G_{d} (d-G_{d}+3)+\zeta^2\right)}{\zeta^2 (d-2 G_{d})+G_{d} (-d+G_{d}-1) (-d+2 G_{d}-2)}\frac{\varpi}{c^2 k_{\rm B} n^2 T^2}                                                                   \quad , \\
  \Lambda               &= \frac{\omega  \left(G_{d} (d-G_{d}+2)+\zeta^2-1\right)}{\zeta^2 (d-2 G_{d})+G_{d} (-d+G_{d}-1) (-d+2 G_{d}-2)} \frac{\varpi}{c^2 k_{\rm B}^2 n^2 T^3}                                                        \quad , \\
  \tilde{\lambda }^{\nu } \Delta _{\nu \mu } &= \frac{(d+2) G_{d}+\zeta^2}{G_{d}(-(d+2) G_{d}-\zeta^2+G_{d}^2)}  \frac{q_{\mu }}{c^2 k_{\rm B} n^2 T^2}                                                                               \quad , \\
  \Lambda ^{\nu} \Delta _{\nu \mu }          &= \frac{1}{(d+2) G_{d}+\zeta^2-G_{d}^2} \frac{q_{\mu }}{c^2 k_{\rm B}^2 n^2 T^3}                                                                               \quad , \\
  \Lambda _{<\mu \nu >} &= -\frac{\pi_{<\mu \nu >}}{2 G_{d} k_{\rm B}^2 n^2 T^3} \quad ,
\end{align}
which together with Eq.~\ref{eq:grad-ansatz-v1} completely determines Grad's ansatz for the distribution function.

\subsubsection{Third order moment}

Having determined an expression for the particle distribution function $f$, we can now calculate
its third order moment directly from the integral definition
\begin{align}
  T^{\alpha\beta\gamma} &= c \int p^\alpha p^\beta p^\gamma f \frac{d^dp}{p_0} \quad , \\
  T^{\alpha\beta\gamma} &= c B Z^{\alpha\beta\gamma} - B \frac{nc^3}{k_{\rm B}} 
                           \left[ \lambda^{(\rm NE)}Z^{\alpha\beta\gamma} 
                                  + \tilde{\lambda} U_\mu Z^{\mu\alpha\beta\gamma} 
                                  + \tilde{\lambda}_\mu \Delta_\nu^\mu Z^{\nu\alpha\beta\gamma} + 
                                  \Lambda U_\mu U_\nu Z^{\mu\nu\alpha\beta\gamma}  
                           \right. \\
  &~~~~~~~~~~~~~~~~~~~~~~~~~~~~~
                           \left.+ \Lambda_\epsilon\Delta_\nu^\epsilon U_\mu Z^{\mu\nu\alpha\beta\gamma} 
                                 + \Lambda_{\gamma\delta}\Delta_\mu^{<\gamma} \Delta_\nu^{\delta>} 
                                 Z^{\mu\nu\alpha\beta\gamma} 
                           \right]  \notag \quad , \\
  T^{\alpha\beta\gamma} &= (nC_1+\omega C_2)U^\alpha U^\beta U^\gamma+(nD_1+\omega D_2)(U^\alpha \eta^{\beta \gamma}+U^\beta \eta^{\gamma \alpha}+U^\gamma \eta^{\beta\alpha})      \label{eq:grad-third-order-moment} \\
   & +C_3\left(q^{\gamma}\eta^{\alpha\beta}+q^{\beta}\eta^{\alpha\gamma}+q^{\alpha}\eta^{\gamma\beta}\right)+D_3\left(q^{\alpha}U^\gamma U^\beta+q^{\gamma}U^\alpha U^\beta+q^{\beta}U^\gamma U^\alpha\right)                                                          \notag \\
   & +C_4\left[U^\alpha \pi^{<\beta\gamma>}+U^\beta \pi^{<\alpha\gamma>}+U^\gamma \pi^{<\alpha\beta>}\right] \quad , \notag                           
\end{align}
with coefficients
\begin{align*}
  C_1    =&      ((d+3) G_{d}+\zeta^2) \left(\frac{k_{\rm B}T}{c^2}\right)^2                    \quad , \\
  C_2    =& \frac{(d+3)G_{d} \left(d^2+(d+6) G_{d}^2-(d+2) (d+6) G_{d}+5 d+6\right)}{\zeta^2 (d-2 G_{d})+G_{d} (-d+G_{d}-1) (-d+2 G_{d}-2)} \\
         +& \frac{\zeta^2 (-3 d G_{d}+d+2 (G_{d}-5) G_{d}+2)-2 \zeta ^4}{\zeta^2 (d-2 G_{d})+G_{d} (-d+G_{d}-1) (-d+2 G_{d}-2)} 
           \left(\frac{k_{\rm B} T}{c^4}\right)                                                 \quad , \\
  C_3    =&    \frac{G_{d} }{G_{d}^2-(d+2) G_{d}-\zeta^2} \left(\frac{k_{\rm B} T}{c^2}\right)  , \quad
  C_4    =     \frac{(d+3) G_{d}+\zeta^2}{G_{d}}\left(\frac{k_{\rm B} T}{c^2}\right)            \quad , \\
  D_1    =& -\frac{c^2}{d+3}(C_1-m^2) , \quad
  D_2    = -\frac{c^2}{d+3}C_2        , \quad
  D_3    =  \frac{-G_d^2 \left(d+\zeta ^2+3\right)+(d+2) \zeta ^2 G_d+\zeta ^4}{c^2 G_d^2}C_3  \quad .
\end{align*}

\subsubsection{Grad's transport coefficients}

In order to determine the transport coefficients we now plug the third order moment, defined in the 
previous section, into the relativistic Boltzmann equation, multiply left and right by $ c p^\beta p^\gamma$ 
and integrate in momentum space, leading to:
\begin{align}
  c \int p^\beta p^\gamma p^{\alpha} \frac{\partial f}{\partial x^{\alpha}} \frac{d^dp}{p_0}
  = 
  - \frac{U_{\alpha}}{c \tau} \int p^{\alpha} p^\beta p^\gamma\left( f - f^{\rm eq} \right) \frac{d^dp}{p_0}
  \quad .
\end{align}
Next, we take the derivative out the integral
\begin{align}
  U_\alpha (T^{\alpha\beta\gamma} - T^{\alpha\beta\gamma}_E) = -c^2 \tau \partial_\alpha T^{\alpha\beta\gamma} \quad ,
\end{align}
and use the Maxwellian iteration method so that only the equilibrium part of $T^{\alpha\beta\gamma}$ is left in the derivative:
\begin{align} \label{eq:grad-third-order-eq}
  U_\alpha (T^{\alpha\beta\gamma}-T^{\alpha\beta\gamma}_E) = -c^2\tau\partial_\alpha T^{\alpha\beta\gamma}_E \quad .
\end{align}
Note that the third order moment at the equilibrium,  $T^{\alpha\beta\gamma}_E$, 
can be obtained directly from Eq.~\ref{eq:grad-third-order-moment} 
by setting to zero the non-equilibrium quantities $q^\alpha = 0$, $\varpi = 0$ and $\pi^{<\alpha\beta>} = 0$:
\begin{align*}
  T^{\alpha\beta\gamma}_E = n C_1 U^\alpha U^\beta U^\gamma 
                          + n D_1(
                                  U^\alpha \eta^{\beta  \gamma}
                                 +U^\beta  \eta^{\gamma \alpha}
                                 +U^\gamma \eta^{\beta  \alpha}
                                 ) \quad .
\end{align*}

We now multiply both sides of equation Eq.~\ref{eq:grad-third-order-eq} by respectively 
$\Delta^{\delta}_\beta U_\gamma$, $\Delta_{\beta\gamma}$, 
and $\Delta^{(\delta}_\beta\Delta^{\epsilon )}_\gamma-\frac{1}{d}\Delta_{\beta\gamma}\Delta^{\delta\epsilon}$
to obtain:
\begin{align*}
  c^2 q^\delta (C_3+c^2D_3)       &= n c^4 \tau (D_1+\zeta \partial_\zeta D_1) \frac{\nabla^\delta T}{T}-c^2\tau \left( \frac{D_1}{k_{\rm B} T} + \frac{n(c^2C_1+D_1)}{P+\epsilon}\right)\nabla^\delta P         \quad , \\
  - c^4 \frac{d}{d+3} \varpi C_2  &= -n\tau c^2 \left( 2D_1 + \frac{k_{\rm B}}{c_v}d\zeta \partial_\zeta D_1 \right) \nabla_\alpha U^\alpha                                                                 
  \quad , \\
  c^2 C_4 \pi^{<\delta\epsilon>}  &= - 2 c^2 n \tau D_1 \nabla^{<\delta}U^{\epsilon>}
  \quad .
\end{align*}
By direct comparison of the above equations with respectively Eq.~\ref{eq:heat-flux}, Eq.~\ref{eq:stress-tensor}, and Eq.~\ref{eq:dynamic-pressure},
we can define the analytic form for the relativistic transport coefficients $\lambda$, $\mu$, and $\eta$:
\begin{align}
  \lambda &= -c^2 k_{\rm B} n \tau G_{d} \frac{ \left((d+2) G_{d}+\zeta^2-G_{d}^2\right)^2}{-G_{d}^2 \left(d+\zeta^2+2\right)+(d+2) \zeta^2 G_{d}+\zeta ^4} \quad , \\
  \mu     &= P \tau \frac{\left(\zeta^2 (d-2 G_{d})+G_{d} (-d+G_{d}-1) (-d+2 G_{d}-2)\right)^2}{d \left(G_{d} (d-G_{d}+2)+\zeta^2-1\right)} \times \\ 
          &  \frac{1}{G_{d}^2 \left(d^2+8 d-2 \zeta^2+12\right)-G_{d} 
             \left(d^2+d \left(5-3 \zeta^2\right)-10 \zeta^2+6\right)+\zeta^2 \left(-d+2 \zeta^2-2\right)-(d+6) G_{d}^3} \quad , \notag \\
  \eta    &= \frac{G_{d}^2 P \tau }{(d+3) G_{d}+\zeta^2} \quad .
\end{align}

The ultra-relativistic limit ($\zeta \rightarrow 0$) of the above expressions writes as:
\begin{align}
  \lambda_{\rm ur} & = \frac{d+1}{d+2}c^2\tau n k_{\rm B} \quad , \\
  \mu_{\rm ur}     & = 0                                  \quad , \\
  \eta_{\rm ur}    & = \frac{d+1}{d+3}P\tau               \quad .
\end{align}

\section{Sod's shock tube: analytic solution in $(d+1)$ dimensions}\label{sec:appendixD}

In this appendix section, we present details on the derivation of a semi-analytical 
solution for the Sod's shock tube problem. We follow the same approach used in \cite{thompson-jofm-1986},
where the analytical solution to the special relativistic shock-tube problem was presented for the first time
(see also \cite{ambrus-prc-2018} for a comprehensive summary). We extend the original calculations to
account for the evolution of an inviscid ultra-relativistic gas in $(d+1)$-dimensions.

The initial conditions of the benchmark write as follows:
\begin{equation}
  (P, n, \beta) = 
    \begin{cases}
      (P_{\rm L}, n_{\rm L}, 0) & x < 0 \\
      (P_{\rm R}, n_{\rm R}, 0) & x > 0
    \end{cases}
  \quad ,
\end{equation}  
where we assume $n_{\rm L} > n_{\rm R}$ and $P_{\rm L} > P_{\rm R}$, with the subscript $\rm L$ and $\rm R$ 
referring respectively to the left and right sides of a membrane placed at the origin.
Once the membrane is removed the discontinuities decay, and the 
time evolution of the flow is characterized by two distinct components: 
a rarefaction wave traveling from the initial field discontinuity to the left, 
and a shock wave traveling from the initial field discontinuity to the right.
Therefore, at a generic time $t > 0$ we can identify four distinct zones,
portrayed in the example in Fig.~\ref{fig:analytic-inviscid-sod} in the main text:
\begin{itemize}
  \item The unperturbed left zone (L), where all fields keep their initial values.
  \item The rarefaction zone (*), with a rarefaction wave traveling to the left.
        We define $x_{\rm H}$ as the coordinate corresponding to the head of the
        rarefaction wave, separating zones (L) and (*), while $x_{\rm T}$, corresponding
        to the tail of the rarefaction zone separates (*) from the shock zone (C).
  \item The shock zone (C), characterized by the presence of the shock wave. 
        In this area some fields present a discontinuity, such as for example the density and temperature fields. 
        We distinguish by labeling with $I$ and $II$ the different values at the two sides of the discontinuity, 
        thus writing $n_I$ and  $n_{II}$ for the density, and $T_I$, $T_{II}$ for the temperature. 
        For fields that do not exhibit a discontinuity, such as pressure and velocity, we have
        $P_I = P_{II} = P_{\rm C}$ and $\beta_I = \beta_{II} = \beta_{\rm C}$.
        The coordinate of the discontinuity is given by $x_{\rm C}$, while $x_{\rm S}$
        corresponds to the front of the shock wave.
  \item The unperturbed left zone (R), where, again, all fields keep their initial values.
\end{itemize} 

In what follows we will determine the unknown fields by solving the relativistic Euler equations,
starting by taking into consideration the rarefaction zone.

\subsection{Rarefaction wave}

We start from the conservation equations, which we find convenient to rewrite 
here for the specific case of a ultra-relativistic gas:
\begin{align}
  Dn + n \nabla^\alpha U_\alpha                   & = 0  \quad , \\
  \nabla^\beta P -\frac{P+\epsilon}{c^2} DU^\beta & = 0  \quad , \\
  DP + P \Gamma \nabla_\alpha U^\alpha            & = 0  \quad .    
\end{align}
where
\begin{align*}
  c_v    &= k_{\rm B} d           \quad , \\
  c_P    &= k_{\rm B} (d+1)            \quad , \\
  \Gamma &= \frac{c_P}{c_v} = \left(1+\frac{k_{\rm B}}{c_v}\right) = \left(1+\frac{1}{d}\right)\quad , 
\end{align*}
and we recall the ultra-relativistic equation of state: 
\begin{align*}
  P        &= n   k_{\rm B} T \quad ,\\
  \epsilon &= d n k_{\rm B} T \quad .
\end{align*}

The rarefaction wave is self-similar with respect to the variable $w = x/t$.
We can therefore express the conservation equations in terms of $w$, getting:
\begin{align}
  \left(\beta-\frac{w}{c} \right) \partial_w n     &= -n \gamma^2 \left(1-\beta\frac{w}{c} \right) \partial_w \beta           
              \label{EQ:MAS}  \quad   , \\
  \left(\beta-\frac{w}{c} \right) \partial_w \beta &= -\frac{1}{(P+\epsilon) \gamma^2} \left(1-\beta\frac{w}{c} \right) \partial_w P        
              \label{EQ:MOM}  \quad   , \\ 
  \left(\beta-\frac{w}{c} \right) \partial_w P     &= -\gamma^2 \Gamma  \left(1-\beta\frac{w}{c} \right) \partial_w \beta            \label{EQ:ENE}   \quad  .
\end{align}
Combining the last two equations it is possible to eliminate the pressure derivative:
\begin{align}
  \left(\beta-\frac{w}{c} \right)^2 = \left(1-\beta\frac{w}{c} \right)\left(\frac{c_s}{c} \right)^2 \quad ,
\end{align}
where 
\begin{equation}
  c_s= c \sqrt{\frac{\Gamma P}{P+\epsilon}} = \frac{c}{\sqrt{d}} \quad ,
\end{equation}
is the adiabatic speed of sound for an ultra-relativistic fluid. 

It is then possible to calculate $w$:
\begin{align}
  w_{\pm} = c \left( \frac{\beta \mp 1 / \sqrt{d}}{1 \mp \beta / \sqrt{d}} \right) \quad .
\end{align}
In what follows we only take into account $w_{+}$, since $w_{-}$ represents a rarefaction wave moving to the right.
We therefore plug $w_{+}$ into equations \ref{EQ:MAS} and \ref{EQ:ENE}, giving:
\begin{align}
  \frac{1}{\sqrt{d}} \frac{\partial_w n}{n}                &= -\gamma^2\partial_w \beta \quad ,\\
  \frac{\sqrt{d}}{d+1} \frac{\partial_w P}{P} &= -\gamma^2\partial_w \beta \quad .
\end{align}

Next, we integrate both equations in $\diff w$, which allows to identify the following
Riemann invariant quantities:
\begin{align}
  n \left(\frac{1+\beta}{1-\beta}\right)^{\frac{\sqrt{d}}{2}}     = k_n = \text{cost.} \quad , \\
  P \left(\frac{1+\beta}{1-\beta}\right)^{\frac{d+1}{2\sqrt{d}}} = k_P = \text{cost.} \quad .
\end{align}

At the head of the rarefaction wave, where $\beta = \beta_L = 0$, we have 
$k_n = n_{\rm L}$ and $k_p = P_{\rm L}$. It directly follows that along the rarefaction wave 
(we label quantities in this area with a $*$) density and pressure are given by
\begin{align}
  n_{*} &= n_{\rm L}~\left(\frac{1-\beta_{*}}{1+\beta_{*}}\right)^{\frac{\sqrt{d}}{2}}     \quad , \label{EQ:NSTAR} \\
  P_{*} &= P_{\rm L}~\left(\frac{1-\beta_{*}}{1+\beta_{*}}\right)^{\frac{d+1}{2\sqrt{d}}} \quad , \label{EQ:PSTAR}
\end{align}
with 
\begin{align}
  \beta_{*} &=\frac{ \frac{w_{+}}{c} + \frac{1} {\sqrt{d}}}{1 + \frac{w_{+} }{ c \sqrt{d}} }                                                  \label{EQ:BETASTAR} \quad .
\end{align}
%
\subsection{Shock wave}

In the shock wave zone the hydrodynamic fields present a discontinuity. 
It is useful to consider a reference frame $K'$ in which the shock front is at rest.
In what follows we label with a $'$ quantities evaluated in the reference frame $K'$.
The conditions that relate the states on either side of the shock are defined
by the Rankine Hugoniot junction conditions \cite{taub-prl-1948}.
These conditions stem from imposing the continuity of the normal 
component of the energy momentum tensor $T^{\alpha\beta}$ across the interfaces.

The Rankine Hugoniot conditions applied on the $C(II)-R$ interface write al follows:
\begin{align}
  n_{II}\gamma^{'}_{\rm C} \beta^{'}_{\rm C}                                             &= n_{\rm R} \gamma^{'}_{\rm R} \beta^{'}_{\rm R}                                \quad , \\
  P_{\rm C} + (n_{II}\epsilon_{\rm C} + P_{\rm C}) \gamma_{\rm C}^{'2}\beta_{\rm C}^{'2} &= P_{\rm R} + (n_{\rm R} \epsilon_{\rm R} + P_{\rm R}) \gamma_{\rm R}^{'2}\beta_{\rm R}^{'2}  \quad , \\
  (n_{II}\epsilon_{\rm C}+P_{\rm C})\gamma_{\rm C}^{'2}\beta^{'}_{\rm C}                 &= (n_{\rm R} \epsilon_{\rm R}+P_{\rm R})\gamma_{\rm R}^{'2}\beta^{'}_{\rm R}          \quad . 
\end{align}

Simple manipulations allow to recast the above equations into
\begin{align}
  n_{II} \gamma^{'}_{\rm C} \beta^{'}_{\rm C}                  &= n_{r} \gamma^{'}_{\rm R}  \beta^{'}_{\rm R}                     \quad , \label{EQ:DENS}   \\
  \gamma_{\rm C}^{'2} P_{\rm C} \left[1+\beta_{\rm C}^{'2} d \right] &=       \gamma_{\rm R}^{'2}P_{\rm R} \left[1+\beta_{\rm R}^{'2}d\right] \quad , \label{EQ:BETA1}  \\
  P_{\rm C} \gamma_{\rm C}^{'2} \beta^{'}_{\rm C}                  &= P_{\rm R} \gamma_{\rm R}^{'2} \beta^{'}_{\rm R}                     \quad , \label{EQ:BETA2}
\end{align}
and by solving Eq.~\ref{EQ:BETA1} and Eq.~\ref{EQ:BETA2} for $\beta^{'}_{\rm C}$ and $\beta^{'}_{\rm R}$ we obtain:
\begin{align}
  \beta^{'}_{\rm C} = - \sqrt{\frac{P_{\rm C}+dP_{\rm R}}{d (P_{\rm R}+dP_{\rm C})}}  
  \quad , \quad &  
  \beta^{'}_{\rm R} = - \sqrt{\frac{P_{\rm R}+dP_{\rm C}}{d (P_{\rm C}+dP_{\rm R})}} \quad .
\end{align}

The above expressions need to be converted back to the frame $K$ in which the unperturbed fluid is at rest.
We do so by applying a Lorenz boost with velocity $\beta_s$, where $\beta_s$ represents the shock front velocity:
\begin{gather*}
\begin{bmatrix} 
  (U^0_{\rm C})^{'} \\
  (U^z_{\rm C})^{'} 
\end{bmatrix}
=
\begin{bmatrix}
   \gamma_s           &   -\gamma_s \beta_s \\
  -\gamma_s \beta_s   &    \gamma_s
\end{bmatrix}
\begin{bmatrix} 
  U^0_{\rm C} \\
  U^z_{\rm C} 
\end{bmatrix}
\quad , \quad 
\begin{bmatrix} 
  (U^0_{\rm R})^{'} \\
  (U^z_{\rm R})^{'} 
\end{bmatrix}
=
\begin{bmatrix}
   \gamma_s           &   -\gamma_s \beta_s \\
  -\gamma_s \beta_s   &    \gamma_s
\end{bmatrix}
\begin{bmatrix} 
  U^0_{\rm R} \\
  U^z_{\rm R} 
\end{bmatrix}
\end{gather*}
from which it follows
\begin{align}
  \beta^{'}_{\rm C} = \frac{\beta_{\rm C}-\beta_s}{1-\beta_s\beta_{\rm C}} 
  \quad , \quad &    
  \beta^{'}_{\rm R} = \frac{\beta_{\rm R}-\beta_s}{1-\beta_s\beta_{\rm R}} \quad .
\end{align}

In the frame $K$ the velocity at the right of the shock front is zero ($\beta_{\rm R} = 0$), therefore
we have
\begin{align*}
  \beta_s &= -\beta^{'}_{\rm R}                       \\
  \beta_{\rm C} &= \frac{\beta^{'}_{\rm C}-\beta^{'}_{\rm R}}{1-\beta^{'}_{\rm R}\beta^{'}_{\rm C}}      \label{EQ:BETAC}
\end{align*}
and finally
\begin{align}
  \beta_s = \sqrt{\frac{P_{\rm R}+dP_{\rm C}}{d[P_{\rm C}+dP_{\rm R}]}}  \quad , \quad &  \beta_{\rm C} = \sqrt{\frac{d(P_{\rm C}-P_{\rm R})^2}{[P_{\rm R}+dP_{\rm C}][P_{\rm C}+dP_{\rm R}]}} \quad .
\end{align}

From Eq.~\ref{EQ:DENS} we can calculate $n_{II}$:
\begin{align}
  n_{II} = n_{r} \frac{\gamma^{'}_{\rm R} \beta^{'}_{\rm R}}{\gamma^{'}_{\rm C} \beta^{'}_{\rm C}} = n_{\rm R} \sqrt{\frac{P_{\rm C}[P_{\rm R}+dP_{\rm C}]}{P_{\rm R}[P_{\rm C}+dP_{\rm R}]}} \quad .
\end{align}

Next, we consider the contact point between zones $*$ and $C(I)$, having spatial coordinate $x_{\rm T}$. 
From Eq.~\ref{EQ:PSTAR} determine the self-similarity variable $w_{\rm T} = x_{\rm T} / t$:
\begin{align}
  w_{\rm T} 
  = 
  c \frac{1- 1 / \sqrt{d}-(1+ 1 / \sqrt{d})\left(\frac{P_{\rm C}}{P_{\rm L}}\right)^\frac{2\sqrt{d}}{d+1}}{1- 1 / \sqrt{d}+(1+ 1 / \sqrt{d})\left(\frac{P_{\rm C}}{P_{\rm L}}\right)^\frac{2\sqrt{d}}{d+1}} \quad .
\end{align}

The pressure on the central plateau $P_{\rm C}$ can be found by numerically solving the equation $\beta_*(w_{\rm T}) = \beta_{\rm C} $, thus:
%
%
\begin{align}
   \frac{\left(\frac{P_{\rm L}}{P_{\rm C}}\right)^\frac{2\sqrt{d}}{d+1}-1}{\left(\frac{P_{\rm L}}{P_{\rm C}}\right)^\frac{2\sqrt{d}}{d+1}+1}
  -\sqrt{\frac{d(P_{\rm C}-P_{\rm R})^2}{[P_{\rm R}+dP_{\rm C}][P_{\rm C}+dP_{\rm R}]}} 
  = 0 \quad .
\end{align}

Finally, the density field $n_{I}$ can be obtained from Eq.~\ref{EQ:NSTAR}:
\begin{align}
  n_{I} = n_*(w_{\rm T}) = n_{\rm L} \left(\frac{P_{\rm C}}{P_{\rm L}}\right)^\frac{d}{d+1} \quad .
\end{align}

\subsection{Full Solution}

To conclude this section we summarize the full solution of the relativistic
Sod's shock tube for a inviscid ultra-relativistic gas in $(d+1)$-dimensions.

\begin{align*}
    x_{\rm H} &= -\frac{c~t}{\sqrt{d}}                                          \\
    x_{\rm T} &= \frac{\beta_{\rm C}- 1 / \sqrt{d}}{1 - \beta_{\rm C} / \sqrt{d}} c~t \\
    x_{\rm C} &= \beta_{\rm C}~c~t                           \\
    x_{\rm S} &= \beta_s~c~t
\end{align*}
\begin{align*}
  \beta(x,t) = 
  \begin{cases}
    \beta_L                                                           & x < x_{\rm H}                   \\
    \beta_* =  \frac{w / c + 1 / \sqrt{d}}{1 + \frac{w}{c\sqrt{d}}}                              & x_{\rm H} < x < x_{\rm T} \\
    \beta_{\rm C} =  \sqrt{\frac{d(P_{\rm C}-P_{\rm R})^2}{[P_{\rm R}+dP_{\rm C}][P_{\rm C}+dP_{\rm R}]}} & x_{\rm T} < x < x_{\rm S} \\
    \beta_{\rm R}                                                     & x > x_{\rm S}
  \end{cases}
\end{align*}
\begin{align*}
  n(x,t) = 
  \begin{cases}
    n_L                                                                                 & x < x_{\rm H}       \\
    n_*    = n_L \left( \frac{(1- 1 / \sqrt{d})(1-w/c)}{(1+ 1 / \sqrt{d})(1+w/c)} \right)^{\frac{\sqrt{d}}{2}} & x_{\rm H} < x < x_{\rm T} \\
    n_I    = n_L \left( \frac{P_{\rm C}}{P_{\rm L}} \right)^\frac{d}{d+1}                   & x_{\rm T} < x < x_{\rm C} \\
    n_{II} = n_{\rm R} \sqrt{ \frac{P_{\rm C}[P_{\rm R}+dP_{\rm C}]}{P_{\rm R}[P_{\rm C}+dP_{\rm R}]}}                            & x_{\rm C} < x < x_{\rm S} \\
    n_{\rm R}                                                                                 & x > x_{\rm S}         
  \end{cases}
\end{align*}
\begin{align*}
  P(x,t) = 
  \begin{cases}
    P_{\rm L}                                                                                                                & x < x_{\rm H}       \\
    P_* =  P_{\rm L} \left( \frac{(1- 1 / \sqrt{d})(1-w/c)}{(1+ 1 / \sqrt{d})(1+w/c)} \right)^{\frac{d+1}{2\sqrt{d}}}                               & x_{\rm H} < x < x_{\rm T} \\
    P_{\rm C} 
                                                                                                                       & x_{\rm T} < x < x_{\rm S} \\
    P_{\rm R}                                                                                                                & x > x_{\rm S}
  \end{cases}
\end{align*}
where $w = \frac{x}{t}$, $\beta_s$ is
\begin{align*}
\beta_s &=   \sqrt{\frac{P_{\rm R}+dP_{\rm C}}{d[P_{\rm C}+dP_{\rm R}]}}  
\end{align*}
and the pressure value $P_{\rm C}$ can be found by numerically solving the equation: 
\begin{equation}
  \frac{\left(\frac{P_{\rm L}}{P_{\rm C}}\right)^\frac{2 \sqrt{d} }{d+1}-1}{\left(\frac{P_{\rm L}}{P_{\rm C}}\right)^\frac{2 \sqrt{d} }{d+1}+1}
  = 
  \sqrt{\frac{d(P_{\rm C}-P_{\rm R})^2}{[P_{\rm R}+dP_{\rm C}][P_{\rm C}+dP_{\rm R}]}}
\end{equation}

\section{Numerical Evaluation of Bessel Functions}\label{sec:appendixE}

The implementation of a RLBM simulation in mildly relativistic regimes requires
the evaluation of several coefficients containing Bessel functions. 
An example is the quantity:
\begin{equation}
  G_{d} =  \zeta \frac{ K_{\frac{d+3}{2}}(\zeta) }{ K_{\frac{d+1}{2}}(\zeta) } \quad .
\end{equation}

For even dimensions, the above expression simplifies to ratios of well behaved polynomials,
since Bessel functions of fractional order reduce to reverse Bessel polynomials \cite{grosswald-book-1979}.
For odd dimensions, we employ the approximations provided by Abramowitz and Stegun \cite{abramowitz-book-1965},
which we report here for completeness:

For $0 < x \le 2$, with $t=x/2$
\begin{align}
  K_0 (x)   =& -\log(t)I_0(x) - 0.57721566 + 0.42278420 t^2 + 0.23069756 t^4 + 0.03488590 t^6             \notag \\
             &+ 0.00262698 t^8 + 0.00010750 t^{10} + 0.00000740 t^{12}  + \epsilon                               \\
  x K_1 (x) =& x \log(t) I_1(x) + 1 + 0.15443 t^2 -0.67278579 t^4 -0.18156897 t^6                         \notag \\
             &-0.01919402 t^8 -0.00110404 t^{10} - 0.00004686 t^{12}  + \epsilon   
\end{align}
For $ x \ge 2 $, with $t=2/x$
\begin{align}
  x^{\frac{1}{2}} e^x K_0 (x) =& 1.25331414 - 0.07832358 t + 0.02189568 t^2 
                                - 0.01062446 t^3 + 0.00587872 t^4 \notag \\
                               &- 0.00251540 t^5 + 0.00053208 t^6 
                                + \epsilon   \\
  x^{\frac{1}{2}} e^x K_1 (x) =& 1.25331414 +0.23498619 t -  0.03655620 t^2 
                                + 0.01504268 t^3 - 0.00780353 t^4 \notag \\
                               &+ 0.00325614 t^5 - 0.00068245 t^6 + \epsilon
\end{align}

In the above $I_0$ and $I_1$ are modified Bessel function of the first kind, which can be 
approximated for $0 \le x \le 3.75$, with $t=x/3.75$
\begin{align}
    I_0 (x)            =& 1 + 3.5156229 t^2 + 3.0899424 t^4 + 1.20674 t^6 + 0.2659732 t^8 + 0.03607 t^{10}  
                        \notag\\
                        & + 0.00458 t^{12} + \epsilon     \\
    \frac{I_1 (x)}{x}  =& 0.5 + 0.87890594 t^2 + 0.51498869 t^4 + 0.15084934 t^6 + 0.2658733 t^8 + 0.00301532 t^{10}
                        \notag\\
                        & + 0.00032411 t^{12} + \epsilon 
\end{align}
For $x \ge 3.75$, with $t=3.75/x$
\begin{align}
  x^{\frac{1}{2}} e^x I_0 (x) =& 0.39894228 + 0.01328592 t + 0.00225319 t^2 -0.00157565 t^3 + 0.00916281 t^4 \notag  \\
                               &- 0.02057706 t^5 + 0.02635537 t^6 - 0.01647633 t^7 + 0.00392377t^8  + \epsilon       \\
  x^{\frac{1}{2}} e^x I_1 (x) =& 0.39894228 - 0.03988 t -  0.00362018 t^2 + 0.00163801 t^3 - 0.01031555 t^4 \notag   \\
                               &+ 0.02282967 t^5 - 0.02895312 t^6 + 0.01787654 t^7 - 0.00420059 t^8 + \epsilon
\end{align}

Bessel functions of higher indexes can be calculated starting from $K_0$ and $K_1$ using the recurrence relation
\begin{align}
  K_{\nu+1} (x) = K_{\nu-1} (x) - \frac{2\nu}{x} K_{\nu} (x) \quad .
\end{align}

We remark that since in general we are mostly interested in approximated ratios of Bessel functions,
the above expressions can be further optimized, allowing in particular to avoid the calculation of
exponential functions.

The calculation of transport coefficient requires the evaluation of Bickley-Naylor functions,
in odd dimensions, and of the incomplete Euler Gamma function, in even dimensions.
We report a few useful expressions for their numerical approximation:

For $x > 4$ 
\begin{align}
  Ki_1(x) = e^{-x} (1.253314 x^{-3.5} (-2.592773 + x(1.007812 + x(-0.625 + x))))
\end{align}
And for $x < 4$

\begin{align}
Ki_1(x) &= 1.570796 - 1.115931x - 0.120772x^3 - 0.005674x^5 - 0.000129x^7                             \\
      &- 1.740915\cdot10^{-6} x^9 - 1.535233\cdot10^{-8}x^{11} - 9.574254\cdot10^{-11}x^{13} \notag \\ 
      &+ x\log(x)(1 + 0.083333x^2 + 0.003125x^4 + 0.000062x^6 + 7.535204\cdot10^{-7}x^8      \notag \\
      &+ 6.165167\cdot10^{-9}x^{10} + 3.622694\cdot10^{-11}x^{12})                           \notag
\end{align}

For $x < 2.27238$:
\begin{align}
\Gamma(0,x) &= -8.193389712664089\cdot10^{-13} x^{14} +1.2353110643708935\cdot10^{-11} x^{13}  \notag\\
            &  -1.7397297489890083\cdot10^{-10} x^{12}+2.27746439867652\cdot10^{-9} x^{11}    \notag\\ 
            &  -2.755731922398589\cdot10^{-8} x^{10}+3.0619243582206544\cdot10^{-7} x^9       \notag\\
            &  -3.1001984126984127\cdot10^{-6} x^8+0.0000283447 x^7-0.000231481 x^6+0.00166667\notag\\ 
            &  x^5-0.0104167 x^4+0.0555556 x^3-0.25 x^2+x-1. \log (x)-0.577216
\end{align}
For $2.27238 < x < 7.43645$
\begin{align}
\Gamma(0,x) &= 4.178709619181551\cdot10^{-11} x^{14}-3.0381052760184365\cdot10^{-9} x^{13}          \notag\\
              &+1.0237477782471583\cdot10^{-7} x^{12}-2.121002724219454\cdot10^{-6} x^{11}           \notag\\
              &+0.0000302179 x^{10}-0.000313651 x^9+0.00245114 x^8-0.0146941 x^7                     \notag\\
              &+0.0681914 x^6-0.245306 x^5+0.67964 x^4-1.42611 x^3+2.18696 x^2-2.25643 x+1.22105           
\end{align}
For $x > 7.43645$
\begin{align}
\Gamma(0,x) &= 0.1175 e^{-x} \log \left(\frac{1}{x}+1\right)+0.44125 e^{-x} \log \left(\frac{2}{x}+1\right)
\end{align}

\newpage
\section{Relativistic Orthogonal Polynomials}\label{sec:appendixF}  
In this appendix we list the relativistic orthogonal polynomials, 
up to the second order in $(3+1)$, $(2+1)$ and  $(1+1)$ dimensions. Third order polynomials are available 
as supplemental material \cite{SOM}.
All polynomials have been derived following a Gram-Schmidt procedure starting from the set
$ \mathcal{V} = \{ 1, p^{\alpha}, p^{\alpha} p^{\beta} \dots \}$ ($\alpha, \beta \in \{ 0, x, y, z, w\}$),
and using a weighting function $\omega$, which is specified for each case.

We label polynomials of order $n$ using the notation $J^{(n)}_{m_1 \dots m_n}$, $m_i \in {0,x,y,z}$,  
with the subscript $m$ referring to the corresponding element of the generating basis $ \mathcal{V}$ .
Throughout in this appendix, $\tilde{m} = m/T_0$,
$\tilde{p}^{\alpha} = p^{\alpha}/T_0$ and $\zeta = m/T$.
\subsection{(1+1) dimensions}

We use the weighting function
\begin{equation}
  \omega(p^0,T_0) = \frac{1}{2 K_0(\tilde{m})} \exp{\left( -\tilde{p}^0\right)} \quad ,
\end{equation}
and the shorthands 
\begin{equation*}
  G = \zeta \frac{K_{1} \left( \zeta \right) }{K_{0} \left( \zeta \right)}, \quad \quad  \tilde{G} = \tilde{m} \frac{K_{1} \left( \tilde{m} \right) }{K_{0} \left( \tilde{m} \right)} \quad .
\end{equation*}

\begin{align*}
  J^{(0)}      &= 1 \\
  J^{(1)}_{0}  &= \frac{\tilde{p}^0}{\sqrt{-\tilde{G}^2+\tilde{G}+\tilde{m}^2}}-\frac{\tilde{G}}{\sqrt{-\tilde{G}^2+\tilde{G}+\tilde{m}^2}} \\
  J^{(1)}_{x}  &= \frac{\tilde{p}^x}{\sqrt{\tilde{G}}} \\
  J^{(2)}_{00} &= \frac{\left(\tilde{p}^0\right)^2}{\sqrt{\tilde{G} \left(3-\frac{\tilde{G}}{-\tilde{G}^2+\tilde{G}+\tilde{m}^2}\right)+2 \tilde{m}^2}}
    + \frac{\tilde{p}^0 \left(\tilde{m}^2-\left(\tilde{G}-2\right) \tilde{G}\right)}{\left(\left(\tilde{G}-1\right) \tilde{G}-\tilde{m}^2\right) \sqrt{\tilde{G} \left(3-\frac{\tilde{G}}{-\tilde{G}^2+\tilde{G}+\tilde{m}^2}\right)+2 \tilde{m}^2}} \\
&
    + \frac{\tilde{G} \tilde{m}^2-\tilde{G}^2 \left(\tilde{m}^2+1\right)+\tilde{m}^4}{\left(\left(\tilde{G}-1\right) \tilde{G}-\tilde{m}^2\right) \sqrt{\tilde{G} \left(3-\frac{\tilde{G}}{-\tilde{G}^2+\tilde{G}+\tilde{m}^2}\right)+2 \tilde{m}^2}} \\
  J^{(2)}_{0x} &= \frac{\tilde{p}^0 \tilde{p}^x}{\sqrt{-\frac{\tilde{m}^4}{\tilde{G}}+\left(\tilde{G}-1\right) \tilde{m}^2+2 \tilde{G}}}+\frac{\tilde{p}^x \left(-2 \tilde{G}-\tilde{m}^2\right)}{\tilde{G} \sqrt{-\frac{\tilde{m}^4}{\tilde{G}}+\left(\tilde{G}-1\right) \tilde{m}^2+2 \tilde{G}}}
\end{align*}

As discussed in the main-text, it can be useful to define the polynomials in the 1D case also
in terms of the weighting function
\begin{equation}
  \omega(p^0, T_0) = \frac{1}{2 \tilde{m} T_0 K_1(\tilde{m})} \exp{\left( -\tilde{p}^0\right)} \quad ;
\end{equation}
in this case, we define:

\begin{equation*}
  G = \zeta \frac{K_{2} \left( \zeta \right) }{K_{1} \left( \zeta \right)}, \quad \quad  
  \tilde{G} = \tilde{m} \frac{K_{2} \left( \tilde{m} \right) }{K_{1} \left( \tilde{m} \right)} \quad .
\end{equation*}

\begin{align*}
  \mathcal{J}^{(0)}      &= \frac{1}{\sqrt{\frac{\tilde{G}-2}{\tilde{m}^2}}} \\
  \mathcal{J}^{(1)}_{0}  &= \frac{\tilde{p}^0}{\sqrt{-\frac{\tilde{m}^2}{\tilde{G}-2}+\tilde{G}-1}}-\frac{\tilde{m}^2}{\left(\tilde{G}-2\right) \sqrt{-\frac{\tilde{m}^2}{\tilde{G}-2}+\tilde{G}-1}} \\
  \mathcal{J}^{(1)}_{x}  &= \tilde{p}^x \\
  \mathcal{J}^{(2)}_{00} &= \frac{\left(\tilde{p}^0\right)^2}{\sqrt{\frac{\tilde{G}-2}{-\left(\tilde{G}-3\right) \tilde{G}+\tilde{m}^2-2}+2 \tilde{G}-1}}
                + \frac{\tilde{p}^0 \left(\tilde{m}^2-\left(\tilde{G}-2\right) \tilde{G}\right)}{\left(\left(\tilde{G}-3\right) \tilde{G}-\tilde{m}^2+2\right) \sqrt{\frac{\tilde{G}-2}{-\left(\tilde{G}-3\right) \tilde{G}+\tilde{m}^2-2}+2 \tilde{G}-1}} \\
&
                + \frac{\tilde{m}^2 \left(\left(\tilde{G}-3\right) \tilde{G}-\tilde{m}^2+1\right)}{\left(-\left(\tilde{G}-3\right) \tilde{G}+\tilde{m}^2-2\right) \sqrt{\frac{\tilde{G}-2}{-\left(\tilde{G}-3\right) \tilde{G}+\tilde{m}^2-2}+2 \tilde{G}-1}} \\
  \mathcal{J}^{(2)}_{0x} &= \frac{\tilde{p}^0 \tilde{p}^x}{\sqrt{\tilde{m}^2-\left(\tilde{G}-3\right) \tilde{G}}}-\frac{\tilde{p}^x \tilde{G}}{\sqrt{\tilde{m}^2-\left(\tilde{G}-3\right) \tilde{G}}} 
\end{align*}

\subsection{(2+1) dimensions}

For the polynomials in 2 space dimensions,  we use the weighting function
\begin{equation}
  \omega(p^0,T_0) = \frac{e^{\tilde{m}}}{2 \pi T_0} \exp{\left( -\tilde{p}^0\right)} \quad ,
\end{equation}
and the shorthand 

\begin{equation*}
  G = \zeta + 1, \quad \quad  \tilde{G} = \tilde{m} + 1 \quad .
\end{equation*}

\begin{align*}
  J^{(0)}      &= 1 \\
  J^{(1)}_{0}  &= \tilde{p}^0-\tilde{G} \\
  J^{(1)}_{x}  &= \frac{\tilde{p}^x}{\sqrt{\tilde{G}}} \\
  J^{(1)}_{y}  &= \frac{\tilde{p}^y}{\sqrt{\tilde{G}}} \\
  J^{(2)}_{00} &= \tilde{p}^0 \left(-\tilde{m}-2\right)+\frac{1}{2} \tilde{m} \left(\tilde{m}+4\right)+\frac{\left(\tilde{p}^0\right)^2}{2}+1 \\
  J^{(2)}_{0x} &= \frac{\tilde{p}^0 \tilde{p}^x \left(\tilde{m}+1\right)}{\tilde{G} \sqrt{\frac{2 \tilde{m} \left(\tilde{m}+3\right)+3}{\tilde{G}}}}+\frac{\tilde{p}^x \left(-\tilde{m} \left(\tilde{m}+3\right)-3\right)}{\tilde{G} \sqrt{\frac{2 \tilde{m} \left(\tilde{m}+3\right)+3}{\tilde{G}}}} \\
  J^{(2)}_{xx} &= -\frac{\left(\tilde{p}^0\right)^2}{2 \sqrt{3 \tilde{G}+\tilde{m}^2}}+\frac{\left(\tilde{p}^x\right)^2}{\sqrt{3 \tilde{G}+\tilde{m}^2}}+\frac{\tilde{m}^2}{2 \sqrt{3 \tilde{G}+\tilde{m}^2}} \\
  J^{(2)}_{0y} &= \frac{\tilde{p}^0 \tilde{p}^y \left(\tilde{m}+1\right)}{\tilde{G} \sqrt{\frac{2 \tilde{m} \left(\tilde{m}+3\right)+3}{\tilde{G}}}}+\frac{\tilde{p}^y \left(-\tilde{m} \left(\tilde{m}+3\right)-3\right)}{\tilde{G} \sqrt{\frac{2 \tilde{m} \left(\tilde{m}+3\right)+3}{\tilde{G}}}} \\
  J^{(2)}_{xy} &= \frac{\tilde{p}^x \tilde{p}^y}{\sqrt{3 \tilde{G}+\tilde{m}^2}}
\end{align*}

\subsection{(3+1) dimensions}

In 3D we use the weighting function
\begin{equation}
  \omega(p^0,T_0) = \frac{1}{4 \pi \tilde{m} T_0^2 K_1(\tilde{m})} \exp{\left( -\tilde{p}^0\right)} \quad ,
\end{equation}
and the shorthands 

\begin{equation*}
  G = \zeta \frac{K_{2} \left( \zeta \right) }{K_{1} \left( \zeta \right)}, \quad \quad  \tilde{G} = \tilde{m} \frac{K_{2} \left( \tilde{m} \right) }{K_{1} \left( \tilde{m} \right)} \quad .
\end{equation*}

\begin{align*}
  J^{(0)}      &= 1 \\
  J^{(1)}_{0}  &= \frac{\tilde{p}^0}{\sqrt{\tilde{m}^2-\left(\tilde{G}-3\right) \tilde{G}}}-\frac{\tilde{G}}{\sqrt{\tilde{m}^2-\left(\tilde{G}-3\right) \tilde{G}}} \\
  J^{(1)}_{x}  &= \frac{\tilde{p}^x}{\sqrt{\tilde{G}}} \\
  J^{(1)}_{y}  &= \frac{\tilde{p}^y}{\sqrt{\tilde{G}}} \\
  J^{(1)}_{z}  &= \frac{\tilde{p}^z}{\sqrt{\tilde{G}}} \\
  J^{(2)}_{00} &= \frac{\left(\tilde{p}^0\right)^2}{\sqrt{3 \tilde{G} \left(\frac{3 \tilde{G}}{\left(\tilde{G}-3\right) \tilde{G}-\tilde{m}^2}+5\right)+6 \tilde{m}^2}}
                 +\frac{\sqrt{3} \tilde{p}^0 \left(\tilde{m}^2-\left(\tilde{G}-4\right) \tilde{G}\right)}{\left(\left(\tilde{G}-3\right) \tilde{G}-\tilde{m}^2\right) \sqrt{\tilde{G} \left(\frac{3 \tilde{G}}{\left(\tilde{G}-3\right) \tilde{G}-\tilde{m}^2}+5\right)+2 \tilde{m}^2}} \\
               &
                 +\frac{3 \tilde{G} \tilde{m}^2-\tilde{G}^2 \left(\tilde{m}^2+3\right)+\tilde{m}^4}{\left(\left(\tilde{G}-3\right) \tilde{G}-\tilde{m}^2\right) 
                  \sqrt{3 \tilde{G} \left(\frac{3 \tilde{G}}{\left(\tilde{G}-3\right) \tilde{G}-\tilde{m}^2}+5\right)+6 \tilde{m}^2}} \\
  J^{(2)}_{0x} &= \frac{\tilde{p}^0 \tilde{p}^x}{\sqrt{-\frac{\tilde{m}^4}{\tilde{G}}+\left(\tilde{G}-3\right) \tilde{m}^2+4 \tilde{G}}}+\frac{\tilde{p}^x \left(-4 \tilde{G}-\tilde{m}^2\right)}{\tilde{G} \sqrt{-\frac{\tilde{m}^4}{\tilde{G}}+\left(\tilde{G}-3\right) \tilde{m}^2+4 \tilde{G}}} \\
  J^{(2)}_{xx} &= -\frac{\left(\tilde{p}^0\right)^2}{2 \sqrt{3} \sqrt{4 \tilde{G}+\tilde{m}^2}}+\frac{\sqrt{3} \left(\tilde{p}^x\right)^2}{2 \sqrt{4 \tilde{G}+\tilde{m}^2}}+\frac{\tilde{m}^2}{2 \sqrt{3} \sqrt{4 \tilde{G}+\tilde{m}^2}} \\
  J^{(2)}_{0y} &= \frac{\tilde{p}^0 \tilde{p}^y}{\sqrt{-\frac{\tilde{m}^4}{\tilde{G}}+\left(\tilde{G}-3\right) \tilde{m}^2+4 \tilde{G}}}+\frac{\tilde{p}^y \left(-4 \tilde{G}-\tilde{m}^2\right)}{\tilde{G} \sqrt{-\frac{\tilde{m}^4}{\tilde{G}}+\left(\tilde{G}-3\right) \tilde{m}^2+4 \tilde{G}}} \\
  J^{(2)}_{xy} &= \frac{\tilde{p}^x \tilde{p}^y}{\sqrt{4 \tilde{G}+\tilde{m}^2}} \\
  J^{(2)}_{yy} &= -\frac{\left(\tilde{p}^0\right)^2}{2 \sqrt{4 \tilde{G}+\tilde{m}^2}}+\frac{\left(\tilde{p}^x\right)^2}{2 \sqrt{4 \tilde{G}+\tilde{m}^2}}+\frac{\left(\tilde{p}^y\right)^2}{\sqrt{4 \tilde{G}+\tilde{m}^2}}+\frac{\tilde{m}^2}{2 \sqrt{4 \tilde{G}+\tilde{m}^2}} \\
  J^{(2)}_{0z} &= \frac{\tilde{p}^0 \tilde{p}^z}{\sqrt{-\frac{\tilde{m}^4}{\tilde{G}}+\left(\tilde{G}-3\right) \tilde{m}^2+4 \tilde{G}}}+\frac{\tilde{p}^z \left(-4 \tilde{G}-\tilde{m}^2\right)}{\tilde{G} \sqrt{-\frac{\tilde{m}^4}{\tilde{G}}+\left(\tilde{G}-3\right) \tilde{m}^2+4 \tilde{G}}} \\
  J^{(2)}_{xz} &= \frac{\tilde{p}^x \tilde{p}^z}{\sqrt{4 \tilde{G}+\tilde{m}^2}} \\
  J^{(2)}_{yz} &= \frac{\tilde{p}^y \tilde{p}^z}{\sqrt{4 \tilde{G}+\tilde{m}^2}}
\end{align*}

\section{Relativistic Orthogonal Projections}\label{sec:appendixG}

In this appendix we list the  expressions of the orthogonal projections, 
up to the second order in $(3+1)$, $(2+1)$ and $(1+1)$ dimensions. 
Third order coefficients  are available as supplemental material \cite{SOM}.

The notation used for the projection coefficients is the same previously introduced for the orthogonal polynomials; we write
\begin{equation*}
  a^{k} = b^{k} \frac{n}{T_0}
\end{equation*}

\subsection{(1+1) dimensions}
For the polynomials derived using the weighting function
\begin{equation}
  \omega(p^0,T_0) = \frac{1}{2 K_0(\tilde{m})} \exp{\left( -\tilde{p}^0\right)} \quad ,
\end{equation}
and with the shorthands 
\begin{equation*}
  G = \zeta \frac{K_{1} \left( \zeta \right) }{K_{0} \left( \zeta \right)}, \quad \quad  \tilde{G} = \tilde{m} \frac{K_{1} \left( \tilde{m} \right) }{K_{0} \left( \tilde{m} \right)} \quad .
\end{equation*}
we have:

\begin{align*}
  b^{(0)}      &= \frac{1}{G \tilde{T}} \\
  b^{(1)}_{0}  &= \frac{U^0}{\sqrt{-\tilde{G}^2+\tilde{G}+\tilde{m}^2}}-\frac{\tilde{G}}{G \tilde{T} \sqrt{-\tilde{G}^2+\tilde{G}+\tilde{m}^2}} \\
  b^{(1)}_{x}  &= \frac{U^x}{\sqrt{\tilde{G}}} \\
  b^{(2)}_{00} &= \frac{\left(U^0\right)^2 \left(2 G \tilde{T}^2+\tilde{m}^2\right)}{G \tilde{T} \sqrt{\tilde{G} \left(3-\frac{\tilde{G}}{-\tilde{G}^2+\tilde{G}+\tilde{m}^2}\right)+2 \tilde{m}^2}}
                 +\frac{U^0 \left(\tilde{m}^2-\left(\tilde{G}-2\right) \tilde{G}\right)}{\left(\left(\tilde{G}-1\right) \tilde{G}-\tilde{m}^2\right) \sqrt{\tilde{G} \left(3-\frac{\tilde{G}}{-\tilde{G}^2+\tilde{G}+\tilde{m}^2}\right)+2 \tilde{m}^2}} \\
&                 
                 + \frac{G \tilde{m}^2 \tilde{T}^2+\tilde{G} \left(G \tilde{T}^2+\tilde{m}^2\right)-\tilde{G}^2 \left(G \tilde{T}^2+\tilde{m}^2+1\right)+\tilde{m}^4}{G \tilde{T} \left(\left(\tilde{G}-1\right) \tilde{G}-\tilde{m}^2\right) \sqrt{\tilde{G} \left(3-\frac{\tilde{G}}{-\tilde{G}^2+\tilde{G}+\tilde{m}^2}\right)+2 \tilde{m}^2}} \\
  b^{(2)}_{0x} &= \frac{U^0 U^x \tilde{T} \left(\frac{\tilde{m}^2}{\tilde{T}^2}+2 G\right)}{G \sqrt{-\frac{\tilde{m}^4}{\tilde{G}}+\left(\tilde{G}-1\right) \tilde{m}^2+2 \tilde{G}}}-\frac{U^x \left(2 \tilde{G}+\tilde{m}^2\right)}{\tilde{G} \sqrt{-\frac{\tilde{m}^4}{\tilde{G}}+\left(\tilde{G}-1\right) \tilde{m}^2+2 \tilde{G}}}
\end{align*}

For the polynomials $\mathcal{J}^{(n)}_{\alpha_1 \cdots \alpha_n}$,  defined using the weighting function
\begin{equation}
  \omega(p^0,T_0) = \frac{1}{2 \tilde{m} T_0 K_1(\tilde{m})} \exp{\left( -\tilde{p}^0\right)} \quad ,
\end{equation}
we define the projections
\begin{equation*}
  \mathcal{A}^{k} = \mathcal{B}^{k} \frac{n}{T_0} \quad ,
\end{equation*}
and using the shorthands
\begin{equation*}
  G = \zeta \frac{K_{2} \left( \zeta \right) }{K_{1} \left( \zeta \right)}, \quad \quad  
  \tilde{G} = \tilde{m} \frac{K_{2} \left( \tilde{m} \right) }{K_{1} \left( \tilde{m} \right)} \quad .
\end{equation*}
we have:

\begin{align*}
  \mathcal{B}^{(0)}      &= \frac{(G-2) \tilde{T}}{\tilde{m}^2 \sqrt{\frac{\tilde{G}-2}{\tilde{m}^2}}} \\
  \mathcal{B}^{(1)}_{0}  &= \frac{U^0}{\sqrt{-\frac{\tilde{m}^2}{\tilde{G}-2}+\tilde{G}-1}}-\frac{(G-2) \tilde{T}}{\left(\tilde{G}-2\right) \sqrt{-\frac{\tilde{m}^2}{\tilde{G}-2}+\tilde{G}-1}} \\
  \mathcal{B}^{(1)}_{x}  &= U^x \\
  \mathcal{B}^{(2)}_{00} &= \frac{G \left(U^0\right)^2 \tilde{T}}{\sqrt{\frac{\tilde{G}-2}{-\left(\tilde{G}-3\right) \tilde{G}+\tilde{m}^2-2}+2 \tilde{G}-1}}
                + \frac{U^0 \left(\tilde{m}^2-\left(\tilde{G}-2\right) \tilde{G}\right)}{\left(\left(\tilde{G}-3\right) \tilde{G}-\tilde{m}^2+2\right) \sqrt{\frac{\tilde{G}-2}{-\left(\tilde{G}-3\right) \tilde{G}+\tilde{m}^2-2}+2 \tilde{G}-1}} \\
&
                - \frac{\tilde{T} \left(-G \tilde{m}^2+(G-1) \left(\tilde{G}-3\right) \tilde{G}+\tilde{m}^2+G\right)}{\left(\left(\tilde{G}-3\right) \tilde{G}-\tilde{m}^2+2\right) \sqrt{\frac{\tilde{G}-2}{-\left(\tilde{G}-3\right) \tilde{G}+\tilde{m}^2-2}+2 \tilde{G}-1}} \\
  \mathcal{B}^{(2)}_{0x} &= \frac{G U^0 U^x \tilde{T}}{\sqrt{\tilde{m}^2-\left(\tilde{G}-3\right) \tilde{G}}}-\frac{U^x \tilde{G}}{\sqrt{\tilde{m}^2-\left(\tilde{G}-3\right) \tilde{G}}} 
\end{align*}

\subsection{(2+1) dimensions}

We use the shorthands
\begin{equation*}
  G = \zeta + 1, \quad \quad  \tilde{G} = \tilde{m} + 1 \quad .
\end{equation*}

\begin{align*}
  b^{(0)}      &= \frac{1}{\tilde{m}+\tilde{T}} \\
  b^{(1)}_{0}  &= U^0-\frac{\tilde{G}}{\tilde{m}+\tilde{T}} \\
  b^{(1)}_{x}  &= \frac{U^x}{\sqrt{\tilde{G}}} \\
  b^{(1)}_{y}  &= \frac{U^y}{\sqrt{\tilde{G}}} \\
  b^{(2)}_{00} &= \frac{1}{2} \left(U^0\right)^2 \tilde{T} \left(\frac{\tilde{m}^2}{G \tilde{T}^2}+3\right)+\frac{1}{2} \left(\frac{\tilde{m} \left(\tilde{m}+4\right)+2}{\tilde{m}+\tilde{T}}-\tilde{T}\right)+U^0 \left(-\tilde{m}-2\right) \\
  b^{(2)}_{0x} &= \frac{U^0 U^x \left(\tilde{m}+1\right) \tilde{T} \left(\frac{\tilde{m}^2}{\tilde{T}^2}+3 G\right)}{G \tilde{G} \sqrt{\frac{2 \tilde{m} \left(\tilde{m}+3\right)+3}{\tilde{G}}}}-\frac{U^x \left(\tilde{m} \left(\tilde{m}+3\right)+3\right)}{\tilde{G} \sqrt{\frac{2 \tilde{m} \left(\tilde{m}+3\right)+3}{\tilde{G}}}} \\
  b^{(2)}_{xx} &= -\frac{\left(U^0\right)^2 \left(3 G \tilde{T}^2+\tilde{m}^2\right)}{2 G \tilde{T} \sqrt{3 \tilde{G}+\tilde{m}^2}}+\frac{\left(U^x\right)^2 \tilde{T} \left(\frac{\tilde{m}^2}{\tilde{T}^2}+3 G\right)}{G \sqrt{3 \tilde{G}+\tilde{m}^2}}+\frac{3 G \tilde{T}^2+\tilde{m}^2}{2 G \tilde{T} \sqrt{3 \tilde{G}+\tilde{m}^2}} \\
  b^{(2)}_{0y} &= \frac{U^0 U^y \left(\tilde{m}+1\right) \tilde{T} \left(\frac{\tilde{m}^2}{\tilde{T}^2}+3 G\right)}{G \tilde{G} \sqrt{\frac{2 \tilde{m} \left(\tilde{m}+3\right)+3}{\tilde{G}}}}-\frac{U^y \left(\tilde{m} \left(\tilde{m}+3\right)+3\right)}{\tilde{G} \sqrt{\frac{2 \tilde{m} \left(\tilde{m}+3\right)+3}{\tilde{G}}}} \\
  b^{(2)}_{xy} &= \frac{U^x U^y \tilde{T} \left(\frac{\tilde{m}^2}{\tilde{T}^2}+3 G\right)}{G \sqrt{3 \tilde{G}+\tilde{m}^2}}
\end{align*}

\subsection{(3+1) dimensions}

We use the shorthands
\begin{equation*}
  G = \zeta \frac{K_{2} \left( \zeta \right) }{K_{1} \left( \zeta \right)}, \quad \quad  \tilde{G} = \tilde{m} \frac{K_{2} \left( \tilde{m} \right) }{K_{1} \left( \tilde{m} \right)} \quad .
\end{equation*}

\begin{align*}
  b^{(0)}      &= \frac{1}{G \tilde{T}} \\
  b^{(1)}_{0}  &= \frac{U^0}{\sqrt{\tilde{m}^2-\left(\tilde{G}-3\right) \tilde{G}}}-\frac{\tilde{G}}{G \sqrt{\tilde{m}^2-\left(\tilde{G}-3\right) \tilde{G}} \tilde{T}} \\
  b^{(1)}_{x}  &= \frac{U^x}{\sqrt{\tilde{G}}} \\
  b^{(1)}_{y}  &= \frac{U^y}{\sqrt{\tilde{G}}} \\
  b^{(1)}_{z}  &= \frac{U^z}{\sqrt{\tilde{G}}} \\
  b^{(2)}_{00} &= \frac{\left(\tilde{m}^2+4 G \tilde{T}^2\right) \left(U^0\right)^2}{G \sqrt{6 \tilde{m}^2+3 \tilde{G} \left(\frac{3 \tilde{G}}{\left(\tilde{G}-3\right) \tilde{G}-\tilde{m}^2}+5\right)} \tilde{T}}
                 +\frac{3 \left(\tilde{m}^2-\left(\tilde{G}-4\right) \tilde{G}\right) U^0}{\left(\left(\tilde{G}-3\right) \tilde{G}-\tilde{m}^2\right) \sqrt{6 \tilde{m}^2+3 \tilde{G} \left(\frac{3 \tilde{G}}{\left(\tilde{G}-3\right) \tilde{G}-\tilde{m}^2}+5\right)}} \\
&                 
                 +\frac{\tilde{m}^4+G \tilde{T}^2 \tilde{m}^2+3 \tilde{G} \left(\tilde{m}^2+G \tilde{T}^2\right)-\tilde{G}^2 \left(\tilde{m}^2+G \tilde{T}^2+3\right)}{G \left(\left(\tilde{G}-3\right) \tilde{G}-\tilde{m}^2\right) \sqrt{6 \tilde{m}^2+3 \tilde{G} \left(\frac{3 \tilde{G}}{\left(\tilde{G}-3\right) \tilde{G}-\tilde{m}^2}+5\right)} \tilde{T}} \\
  b^{(2)}_{0x} &= \frac{U^0 U^x \left(\frac{\tilde{m}^2}{\tilde{T}^2}+4 G\right) \tilde{T}}{G \sqrt{-\frac{\tilde{m}^4}{\tilde{G}}+\left(\tilde{G}-3\right) \tilde{m}^2+4 \tilde{G}}}-\frac{U^x \left(\tilde{m}^2+4 \tilde{G}\right)}{\tilde{G} \sqrt{-\frac{\tilde{m}^4}{\tilde{G}}+\left(\tilde{G}-3\right) \tilde{m}^2+4 \tilde{G}}} \\
  b^{(2)}_{xx} &= -\frac{\left(\tilde{m}^2+4 G \tilde{T}^2\right) \left(U^0\right)^2}{2 \sqrt{3} G \sqrt{\tilde{m}^2+4 \tilde{G}} \tilde{T}}+\frac{\tilde{m}^2+4 G \tilde{T}^2}{2 \sqrt{3} G \sqrt{\tilde{m}^2+4 \tilde{G}} \tilde{T}}+\frac{\left(U^x\right)^2 \left(\tilde{m}^2+4 G \tilde{T}^2\right) \sqrt{3}}{2 G \sqrt{\tilde{m}^2+4 \tilde{G}} \tilde{T}} \\
  b^{(2)}_{0y} &= \frac{U^0 U^y \left(\frac{\tilde{m}^2}{\tilde{T}^2}+4 G\right) \tilde{T}}{G \sqrt{-\frac{\tilde{m}^4}{\tilde{G}}+\left(\tilde{G}-3\right) \tilde{m}^2+4 \tilde{G}}}-\frac{U^y \left(\tilde{m}^2+4 \tilde{G}\right)}{\tilde{G} \sqrt{-\frac{\tilde{m}^4}{\tilde{G}}+\left(\tilde{G}-3\right) \tilde{m}^2+4 \tilde{G}}} \\
  b^{(2)}_{xy} &= \frac{U^x U^y \left(\frac{\tilde{m}^2}{\tilde{T}^2}+4 G\right) \tilde{T}}{G \sqrt{\tilde{m}^2+4 \tilde{G}}} \\
  b^{(2)}_{yy} &= -\frac{\left(\tilde{m}^2+4 G \tilde{T}^2\right) \left(U^0\right)^2}{2 G \sqrt{\tilde{m}^2+4 \tilde{G}} \tilde{T}}+\frac{\left(U^y\right)^2 \left(\tilde{m}^2+4 G \tilde{T}^2\right)}{G \sqrt{\tilde{m}^2+4 \tilde{G}} \tilde{T}}+\frac{\left(U^x\right)^2 \left(\tilde{m}^2+4 G \tilde{T}^2\right)}{2 G \sqrt{\tilde{m}^2+4 \tilde{G}} \tilde{T}}+\frac{\tilde{m}^2+4 G \tilde{T}^2}{2 G \sqrt{\tilde{m}^2+4 \tilde{G}} \tilde{T}} \\
  b^{(2)}_{0z} &= \frac{U^0 U^z \left(\frac{\tilde{m}^2}{\tilde{T}^2}+4 G\right) \tilde{T}}{G \sqrt{-\frac{\tilde{m}^4}{\tilde{G}}+\left(\tilde{G}-3\right) \tilde{m}^2+4 \tilde{G}}}-\frac{U^z \left(\tilde{m}^2+4 \tilde{G}\right)}{\tilde{G} \sqrt{-\frac{\tilde{m}^4}{\tilde{G}}+\left(\tilde{G}-3\right) \tilde{m}^2+4 \tilde{G}}} \\
  b^{(2)}_{xz} &= \frac{U^x U^z \left(\frac{\tilde{m}^2}{\tilde{T}^2}+4 G\right) \tilde{T}}{G \sqrt{\tilde{m}^2+4 \tilde{G}}} \\
  b^{(2)}_{yz} &= \frac{U^y U^z \left(\frac{\tilde{m}^2}{\tilde{T}^2}+4 G\right) \tilde{T}}{G \sqrt{\tilde{m}^2+4 \tilde{G}}}
\end{align*}

\section{Quadratures}\label{sec:appendixH}

In this appendix we present a collection of Gauss-type quadratures that can be used
to implement an RLBM on a Cartesian grid, in the massive 
and massless cases in $(3+1)$,  $(2+1)$ and  $(1+1)$ dimensions. 
Examples are given at several order $N$; the order of a quadrature  coincides with the maximum 
order of the polynomials for which the orthonormality conditions are satisfied:
\begin{equation}
  \int \omega(p^0,T_0) J_l( (\tilde{p}^{\mu}) ) J_k( (\tilde{p}^{\mu}) ) \frac{\diff \bm{\tilde{p}}}{\tilde{p}^0} 
  = 
  \sum_i w_i J_l( (\tilde{p}^{\mu}_{i}) )J_k( (\tilde{p}^{\mu}_{i}) ) = \delta_{lk} \quad ,
\end{equation}
where $J^{(k)}$ are the orthogonal polynomials introduced in the previous appendix, 
$(\tilde{p}^\mu_{i})$ the discrete $(D+1)$ momentum vectors, and $w_i$ the quadrature weights.
\subsection{Mildly relativistic regime}
We use the following parametrization of the momentum vectors:
\begin{equation}
  (\tilde{p}^\mu_{i}) = \tilde{m} \gamma_i (1, v_0 n_{i}) \quad ,
\end{equation}
where $n_{i} \in \mathbb{Z}^D$ are the vectors forming the stencil $G = \{ n_i ~|~ i = 1,2,\dots,i_{max} \}$ 
defined by the (on-lattice) quadrature points, $v_0$ is a free parameter that can be freely chosen such 
that $v_i = v_0 || n_i || < 1, \forall i$, $\tilde{m}$ is the non-dimensional rest mass in terms of a reference 
temperature $T_0$, and $\gamma_i$ is the Lorentz factor associated to $v_i$.

In the following we present a few selected stencils, alongside a graphical view of their correspondent
working range in terms of the parameter $\bar{m}$, that can be used to
build a numerically stable quadrature at both 2nd and 3rd order. 
The list of weights for each specific stencil is available as supplemental material \cite{SOM}.

\subsubsection{$(3 + 1)$ dimensions}
\begin{table}[H]
  \centering
  \includegraphics[width=.75\columnwidth]{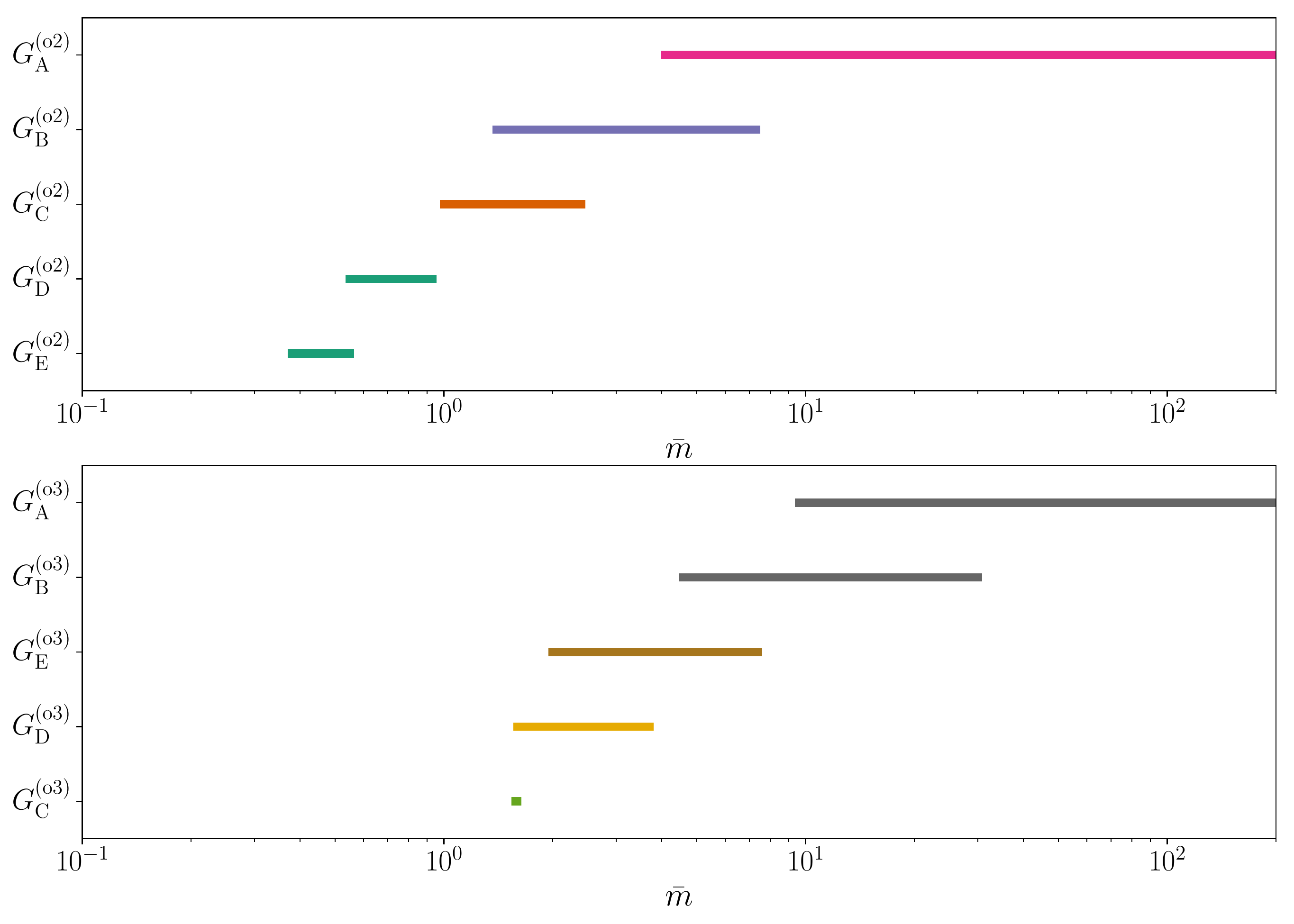}
\centering
\resizebox{.7\columnwidth}{!}{
\begin{tabular}{|c|c|c|c|c|}
\toprule
\multicolumn{5}{|c|}{Order 2} \\
\midrule 
$G_{\rm A}^{(\rm o2)} = \{\bigcup n_i \}$ &
$G_{\rm B}^{(\rm o2)} = \{\bigcup n_i \}$ &
$G_{\rm C}^{(\rm o2)} = \{\bigcup n_i \}$ &
$G_{\rm D}^{(\rm o2)} = \{\bigcup n_i \}$ &
$G_{\rm E}^{(\rm o2)} = \{\bigcup n_i \}$ \\
\midrule 
$(\phantom{\pm} 0, \phantom{\pm} 0, \phantom{\pm} 0)_{\phantom{\texttt{FS}}}$ & $(\phantom{\pm} 0, \phantom{\pm} 0, \phantom{\pm} 0)_{\phantom{\texttt{FS}}}$ & $(\phantom{\pm} 0, \phantom{\pm} 0, \phantom{\pm} 0)_{\phantom{\texttt{FS}}}$ & $(\phantom{\pm} 0, \phantom{\pm} 0, \phantom{\pm} 0)_{\phantom{\texttt{FS}}}$ & $(\phantom{\pm} 0, \phantom{\pm} 0, \phantom{\pm} 0)_{\phantom{\texttt{FS}}}$ \\
$(         \pm  1, \phantom{\pm} 0, \phantom{\pm} 0)_{\texttt{FS          }}$ & $(         \pm  2,          \pm  1,          \pm  1)_{\texttt{FS          }}$ & $(         \pm  1, \phantom{\pm} 1, \phantom{\pm} 0)_{\texttt{FS          }}$ & $(         \pm  1, \phantom{\pm} 1, \phantom{\pm} 0)_{\texttt{FS          }}$ & $(         \pm  2,          \pm  2, \phantom{\pm} 0)_{\texttt{FS          }}$ \\
$(         \pm  1, \phantom{\pm} 1, \phantom{\pm} 0)_{\texttt{FS          }}$ & $(         \pm  2,          \pm  2,          \pm  1)_{\texttt{FS          }}$ & $(         \pm  2,          \pm  2,          \pm  2)_{\texttt{FS          }}$ & $(         \pm  4,          \pm  4,          \pm  2)_{\texttt{FS          }}$ & $(         \pm  4,          \pm  4,          \pm  2)_{\texttt{FS          }}$ \\
$(         \pm  1, \phantom{\pm} 1, \phantom{\pm} 1)_{\texttt{FS          }}$ & $(         \pm  3,          \pm  1, \phantom{\pm} 0)_{\texttt{FS          }}$ & $(         \pm  3,          \pm  2,          \pm  1)_{\texttt{FS          }}$ & $(         \pm  5,          \pm  4, \phantom{\pm} 0)_{\texttt{FS          }}$ & $(         \pm  5,          \pm  4, \phantom{\pm} 0)_{\texttt{FS          }}$ \\
$(         \pm  2, \phantom{\pm} 0, \phantom{\pm} 0)_{\texttt{FS          }}$ & $(         \pm  3,          \pm  2, \phantom{\pm} 0)_{\texttt{FS          }}$ & $(         \pm  3,          \pm  3,          \pm  1)_{\texttt{FS          }}$ & $(         \pm  6,          \pm  2, \phantom{\pm} 0)_{\texttt{FS          }}$ & $(         \pm  6,          \pm  2, \phantom{\pm} 0)_{\texttt{FS          }}$ \\
$(         \pm  2,          \pm  1, \phantom{\pm} 0)_{\texttt{FS          }}$ & $(         \pm  3,          \pm  1,          \pm  1)_{\texttt{FS          }}$ & $(         \pm  4, \phantom{\pm} 0, \phantom{\pm} 0)_{\texttt{FS          }}$ & $(         \pm  4,          \pm  4,          \pm  3)_{\texttt{FS          }}$ & $(         \pm  6,          \pm  2,          \pm  1)_{\texttt{FS          }}$ \\

\midrule 
\multicolumn{5}{|c|}{Order 3} \\
\midrule 
$G_{\rm A}^{(\rm o3)} = \{\bigcup n_i \}$ &
$G_{\rm B}^{(\rm o3)} = \{\bigcup n_i \}$ &
$G_{\rm C}^{(\rm o3)} = \{\bigcup n_i \}$ &
$G_{\rm D}^{(\rm o3)} = \{\bigcup n_i \}$ &
$G_{\rm E}^{(\rm o3)} = \{\bigcup n_i \}$ \\
\midrule 
$(\phantom{\pm} 0, \phantom{\pm} 0, \phantom{\pm} 0)_{\phantom{\texttt{FS}}}$ & $(\phantom{\pm} 0, \phantom{\pm} 0, \phantom{\pm} 0)_{\phantom{\texttt{FS}}}$ & $(\phantom{\pm} 0, \phantom{\pm} 0, \phantom{\pm} 0)_{\phantom{\texttt{FS}}}$ & $(\phantom{\pm} 0, \phantom{\pm} 0, \phantom{\pm} 0)_{\phantom{\texttt{FS}}}$ & $(\phantom{\pm} 0, \phantom{\pm} 0, \phantom{\pm} 0)_{\phantom{\texttt{FS}}}$ \\
$(         \pm  1, \phantom{\pm} 0, \phantom{\pm} 0)_{\texttt{FS          }}$ & $(         \pm  1, \phantom{\pm} 0, \phantom{\pm} 0)_{\texttt{FS          }}$ & $(         \pm  1, \phantom{\pm} 0, \phantom{\pm} 0)_{\texttt{FS          }}$ & $(         \pm  2,          \pm  1,          \pm  1)_{\texttt{FS          }}$ & $(         \pm  2,          \pm  1,          \pm  1)_{\texttt{FS          }}$ \\
$(         \pm  1, \phantom{\pm} 1, \phantom{\pm} 0)_{\texttt{FS          }}$ & $(         \pm  1, \phantom{\pm} 1, \phantom{\pm} 1)_{\texttt{FS          }}$ & $(         \pm  4, \phantom{\pm} 0, \phantom{\pm} 0)_{\texttt{FS          }}$ & $(         \pm  3,          \pm  3,          \pm  1)_{\texttt{FS          }}$ & $(         \pm  3,          \pm  3,          \pm  1)_{\texttt{FS          }}$ \\
$(         \pm  1, \phantom{\pm} 1, \phantom{\pm} 1)_{\texttt{FS          }}$ & $(         \pm  2, \phantom{\pm} 0, \phantom{\pm} 0)_{\texttt{FS          }}$ & $(         \pm  4,          \pm  1, \phantom{\pm} 0)_{\texttt{FS          }}$ & $(         \pm  4,          \pm  4, \phantom{\pm} 0)_{\texttt{FS          }}$ & $(         \pm  4,          \pm  1,          \pm  1)_{\texttt{FS          }}$ \\
$(         \pm  2, \phantom{\pm} 0, \phantom{\pm} 0)_{\texttt{FS          }}$ & $(         \pm  2,          \pm  2, \phantom{\pm} 0)_{\texttt{FS          }}$ & $(         \pm  4,          \pm  4, \phantom{\pm} 0)_{\texttt{FS          }}$ & $(         \pm  4,          \pm  1,          \pm  1)_{\texttt{FS          }}$ & $(         \pm  4,          \pm  3,          \pm  1)_{\texttt{FS          }}$ \\
$(         \pm  2,          \pm  1, \phantom{\pm} 0)_{\texttt{FS          }}$ & $(         \pm  2,          \pm  1,          \pm  1)_{\texttt{FS          }}$ & $(         \pm  4,          \pm  3,          \pm  2)_{\texttt{FS          }}$ & $(         \pm  4,          \pm  3,          \pm  1)_{\texttt{FS          }}$ & $(         \pm  4,          \pm  3,          \pm  2)_{\texttt{FS          }}$ \\
$(         \pm  2,          \pm  2, \phantom{\pm} 0)_{\texttt{FS          }}$ & $(         \pm  2,          \pm  2,          \pm  1)_{\texttt{FS          }}$ & $(         \pm  4,          \pm  3,          \pm  3)_{\texttt{FS          }}$ & $(         \pm  4,          \pm  3,          \pm  2)_{\texttt{FS          }}$ & $(         \pm  4,          \pm  4,          \pm  2)_{\texttt{FS          }}$ \\
$(         \pm  2,          \pm  1,          \pm  1)_{\texttt{FS          }}$ & $(         \pm  2,          \pm  2,          \pm  2)_{\texttt{FS          }}$ & $(         \pm  5,          \pm  1, \phantom{\pm} 0)_{\texttt{FS          }}$ & $(         \pm  4,          \pm  3,          \pm  3)_{\texttt{FS          }}$ & $(         \pm  5,          \pm  2,          \pm  1)_{\texttt{FS          }}$ \\
$(         \pm  2,          \pm  2,          \pm  1)_{\texttt{FS          }}$ & $(         \pm  3, \phantom{\pm} 0, \phantom{\pm} 0)_{\texttt{FS          }}$ & $(         \pm  5,          \pm  3, \phantom{\pm} 0)_{\texttt{FS          }}$ & $(         \pm  5,          \pm  3, \phantom{\pm} 0)_{\texttt{FS          }}$ & $(         \pm  5,          \pm  3,          \pm  1)_{\texttt{FS          }}$ \\
$(         \pm  2,          \pm  2,          \pm  2)_{\texttt{FS          }}$ & $(         \pm  3,          \pm  2, \phantom{\pm} 0)_{\texttt{FS          }}$ & $(         \pm  5,          \pm  2,          \pm  1)_{\texttt{FS          }}$ & $(         \pm  5,          \pm  2,          \pm  1)_{\texttt{FS          }}$ & $(         \pm  5,          \pm  2,          \pm  2)_{\texttt{FS          }}$ \\
$(         \pm  3, \phantom{\pm} 0, \phantom{\pm} 0)_{\texttt{FS          }}$ & $(         \pm  3,          \pm  1,          \pm  1)_{\texttt{FS          }}$ & $(         \pm  5,          \pm  2,          \pm  2)_{\texttt{FS          }}$ & $(         \pm  5,          \pm  2,          \pm  2)_{\texttt{FS          }}$ & $(         \pm  6, \phantom{\pm} 0, \phantom{\pm} 0)_{\texttt{FS          }}$ \\ 
\bottomrule
\end{tabular}
}
  \caption{ Example of stencils that can be used to construct a numerically stable quadrature,
            both at the second and third order, for a RLBM in $(3+1)$ dimensions. 
            In the figure horizontal bars represent the working range of values $\bar{m}$
            of each quadrature.
           }\label{fig:quadrature-3D}
\end{table}

\newpage
\subsubsection{$(2 + 1)$ dimensions}

\begin{table}[H]
  \centering
  \includegraphics[width=.75\textwidth]{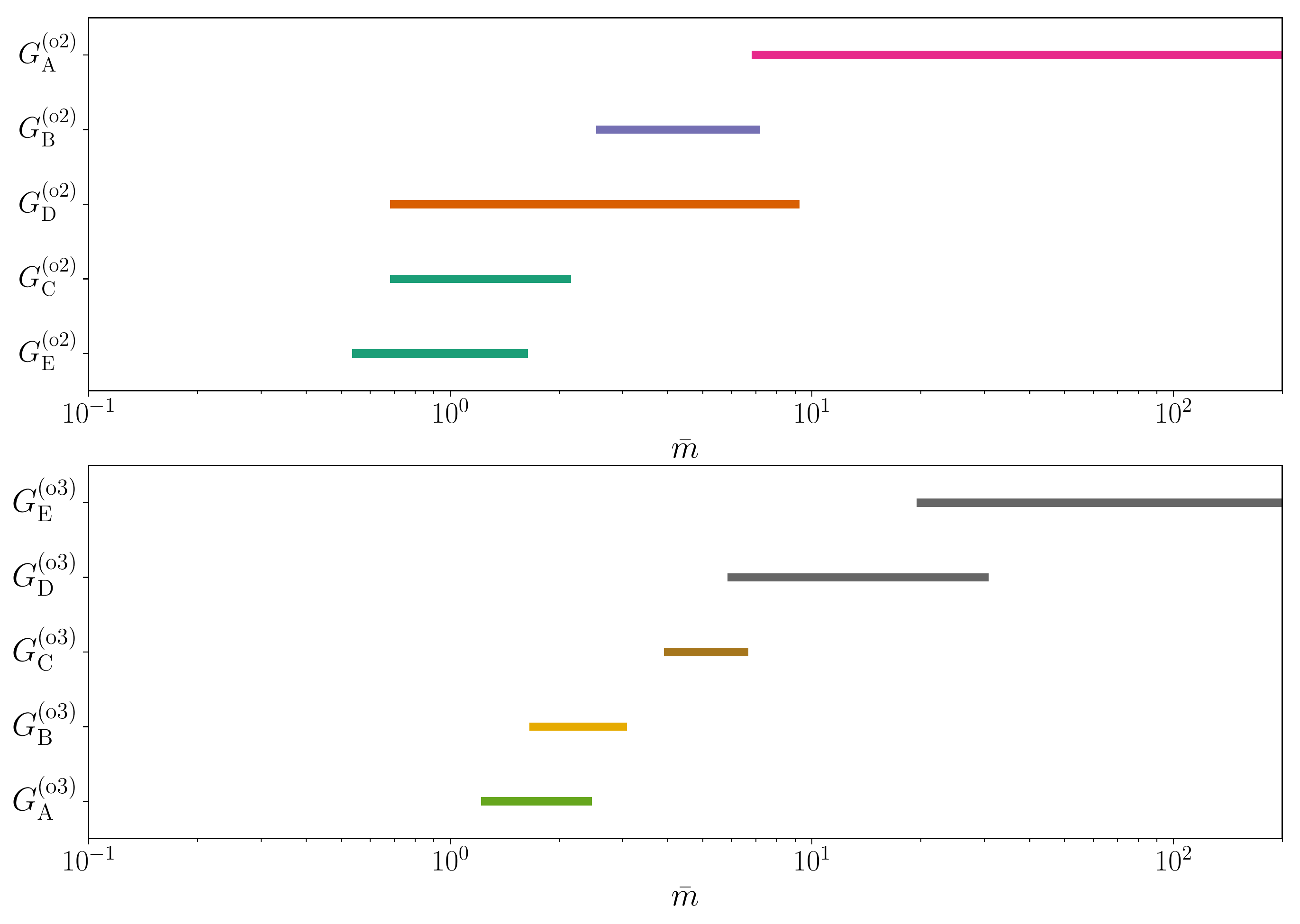}
\centering
\resizebox{.7\columnwidth}{!}{
\begin{tabular}{|c|c|c|c|c|}
\toprule
\multicolumn{5}{|c|}{Order 2} \\
\midrule 
$G_{\rm A}^{(\rm o2)} = \{\bigcup n_i \}$ &
$G_{\rm B}^{(\rm o2)} = \{\bigcup n_i \}$ &
$G_{\rm C}^{(\rm o2)} = \{\bigcup n_i \}$ &
$G_{\rm D}^{(\rm o2)} = \{\bigcup n_i \}$ &
$G_{\rm E}^{(\rm o2)} = \{\bigcup n_i \}$ \\
\midrule 
$(\phantom{\pm} 0, \phantom{\pm} 0)_{\phantom{\texttt{FS}}}$ & $(\phantom{\pm} 0, \phantom{\pm} 0)_{\phantom{\texttt{FS}}}$ & $(\phantom{\pm} 0, \phantom{\pm} 0)_{\phantom{\texttt{FS}}}$ & $(\phantom{\pm} 0, \phantom{\pm} 0)_{\phantom{\texttt{FS}}}$ & $(\phantom{\pm} 0, \phantom{\pm} 0)_{\phantom{\texttt{FS}}}$ \\
$(         \pm  1, \phantom{\pm} 0)_{\texttt{FS          }}$ & $(         \pm  1, \phantom{\pm} 0)_{\texttt{FS          }}$ & $(         \pm  3,          \pm  3)_{\texttt{FS          }}$ & $(         \pm  3,          \pm  1)_{\texttt{FS          }}$ & $(         \pm  5,          \pm  2)_{\texttt{FS          }}$ \\
$(         \pm  1,          \pm  1)_{\texttt{FS          }}$ & $(         \pm  1,          \pm  1)_{\texttt{FS          }}$ & $(         \pm  4,          \pm  2)_{\texttt{FS          }}$ & $(         \pm  4,          \pm  2)_{\texttt{FS          }}$ & $(         \pm  5,          \pm  3)_{\texttt{FS          }}$ \\
$(         \pm  2, \phantom{\pm} 0)_{\texttt{FS          }}$ & $(         \pm  2,          \pm  2)_{\texttt{FS          }}$ & $(         \pm  4,          \pm  3)_{\texttt{FS          }}$ & $(         \pm  4,          \pm  3)_{\texttt{FS          }}$ & $(         \pm  5,          \pm  4)_{\texttt{FS          }}$ \\
$(         \pm  2,          \pm  1)_{\texttt{FS          }}$ & $(         \pm  3,          \pm  2)_{\texttt{FS          }}$ & $(         \pm  5, \phantom{\pm} 0)_{\texttt{FS          }}$ & $(         \pm  5, \phantom{\pm} 0)_{\texttt{FS          }}$ & $(         \pm  6, \phantom{\pm} 0)_{\texttt{FS          }}$ \\
$(         \pm  2,          \pm  2)_{\texttt{FS          }}$ & $(         \pm  4, \phantom{\pm} 0)_{\texttt{FS          }}$ & $(         \pm  5,          \pm  1)_{\texttt{FS          }}$ & $(         \pm  5,          \pm  1)_{\texttt{FS          }}$ & $(         \pm  6,          \pm  2)_{\texttt{FS          }}$ \\
\midrule 
\multicolumn{5}{|c|}{Order 3} \\
\midrule 
$G_{\rm A}^{(\rm o3)} = \{\bigcup n_i \}$ &
$G_{\rm B}^{(\rm o3)} = \{\bigcup n_i \}$ &
$G_{\rm C}^{(\rm o3)} = \{\bigcup n_i \}$ &
$G_{\rm D}^{(\rm o3)} = \{\bigcup n_i \}$ &
$G_{\rm E}^{(\rm o3)} = \{\bigcup n_i \}$ \\
\midrule 
$(\phantom{\pm} 0, \phantom{\pm} 0)_{\phantom{\texttt{FS}}}$ & $(\phantom{\pm} 0, \phantom{\pm} 0)_{\phantom{\texttt{FS}}}$ & $(\phantom{\pm} 0, \phantom{\pm} 0)_{\phantom{\texttt{FS}}}$ & $(\phantom{\pm} 0, \phantom{\pm} 0)_{\phantom{\texttt{FS}}}$ & $(\phantom{\pm} 0, \phantom{\pm} 0)_{\phantom{\texttt{FS}}}$ \\
$(         \pm  1,          \pm  1)_{\texttt{FS          }}$ & $(         \pm  1,          \pm  1)_{\texttt{FS          }}$ & $(         \pm  1, \phantom{\pm} 0)_{\texttt{FS          }}$ & $(         \pm  1, \phantom{\pm} 0)_{\texttt{FS          }}$ & $(         \pm  1, \phantom{\pm} 0)_{\texttt{FS          }}$ \\
$(         \pm  4,          \pm  2)_{\texttt{FS          }}$ & $(         \pm  2,          \pm  2)_{\texttt{FS          }}$ & $(         \pm  2, \phantom{\pm} 0)_{\texttt{FS          }}$ & $(         \pm  1,          \pm  1)_{\texttt{FS          }}$ & $(         \pm  1,          \pm  1)_{\texttt{FS          }}$ \\
$(         \pm  5,          \pm  4)_{\texttt{FS          }}$ & $(         \pm  4,          \pm  3)_{\texttt{FS          }}$ & $(         \pm  2,          \pm  2)_{\texttt{FS          }}$ & $(         \pm  2, \phantom{\pm} 0)_{\texttt{FS          }}$ & $(         \pm  2, \phantom{\pm} 0)_{\texttt{FS          }}$ \\
$(         \pm  5,          \pm  5)_{\texttt{FS          }}$ & $(         \pm  5, \phantom{\pm} 0)_{\texttt{FS          }}$ & $(         \pm  3, \phantom{\pm} 0)_{\texttt{FS          }}$ & $(         \pm  2,          \pm  1)_{\texttt{FS          }}$ & $(         \pm  2,          \pm  1)_{\texttt{FS          }}$ \\
$(         \pm  6,          \pm  2)_{\texttt{FS          }}$ & $(         \pm  5,          \pm  5)_{\texttt{FS          }}$ & $(         \pm  3,          \pm  2)_{\texttt{FS          }}$ & $(         \pm  3,          \pm  1)_{\texttt{FS          }}$ & $(         \pm  2,          \pm  2)_{\texttt{FS          }}$ \\
$(         \pm  6,          \pm  3)_{\texttt{FS          }}$ & $(         \pm  6,          \pm  2)_{\texttt{FS          }}$ & $(         \pm  3,          \pm  3)_{\texttt{FS          }}$ & $(         \pm  3,          \pm  2)_{\texttt{FS          }}$ & $(         \pm  3, \phantom{\pm} 0)_{\texttt{FS          }}$ \\
$(         \pm  6,          \pm  4)_{\texttt{FS          }}$ & $(         \pm  6,          \pm  3)_{\texttt{FS          }}$ & $(         \pm  4, \phantom{\pm} 0)_{\texttt{FS          }}$ & $(         \pm  3,          \pm  3)_{\texttt{FS          }}$ & $(         \pm  3,          \pm  1)_{\texttt{FS          }}$ \\
$(         \pm  7,          \pm  1)_{\texttt{FS          }}$ & $(         \pm  6,          \pm  4)_{\texttt{FS          }}$ & $(         \pm  4,          \pm  1)_{\texttt{FS          }}$ & $(         \pm  4, \phantom{\pm} 0)_{\texttt{FS          }}$ & $(         \pm  3,          \pm  2)_{\texttt{FS          }}$ \\
$(         \pm  7,          \pm  2)_{\texttt{FS          }}$ & $(         \pm  7,          \pm  1)_{\texttt{FS          }}$ & $(         \pm  4,          \pm  2)_{\texttt{FS          }}$ & $(         \pm  4,          \pm  2)_{\texttt{FS          }}$ & $(         \pm  3,          \pm  3)_{\texttt{FS          }}$ \\
\bottomrule
\end{tabular}
}
  \caption{ Example of stencils that can be used to construct a numerically stable quadrature,
            both at the second and third order, for a RLBM in $(2+1)$ dimensions.  
            In the figure horizontal bars represent the working range of values $\bar{m}$
            of each quadrature.
           }\label{fig:quadrature-2D}
\end{table}

\newpage
\subsubsection{$(1 + 1)$ dimensions}

\begin{table}[H]
  \centering
  \includegraphics[width=.75\columnwidth]{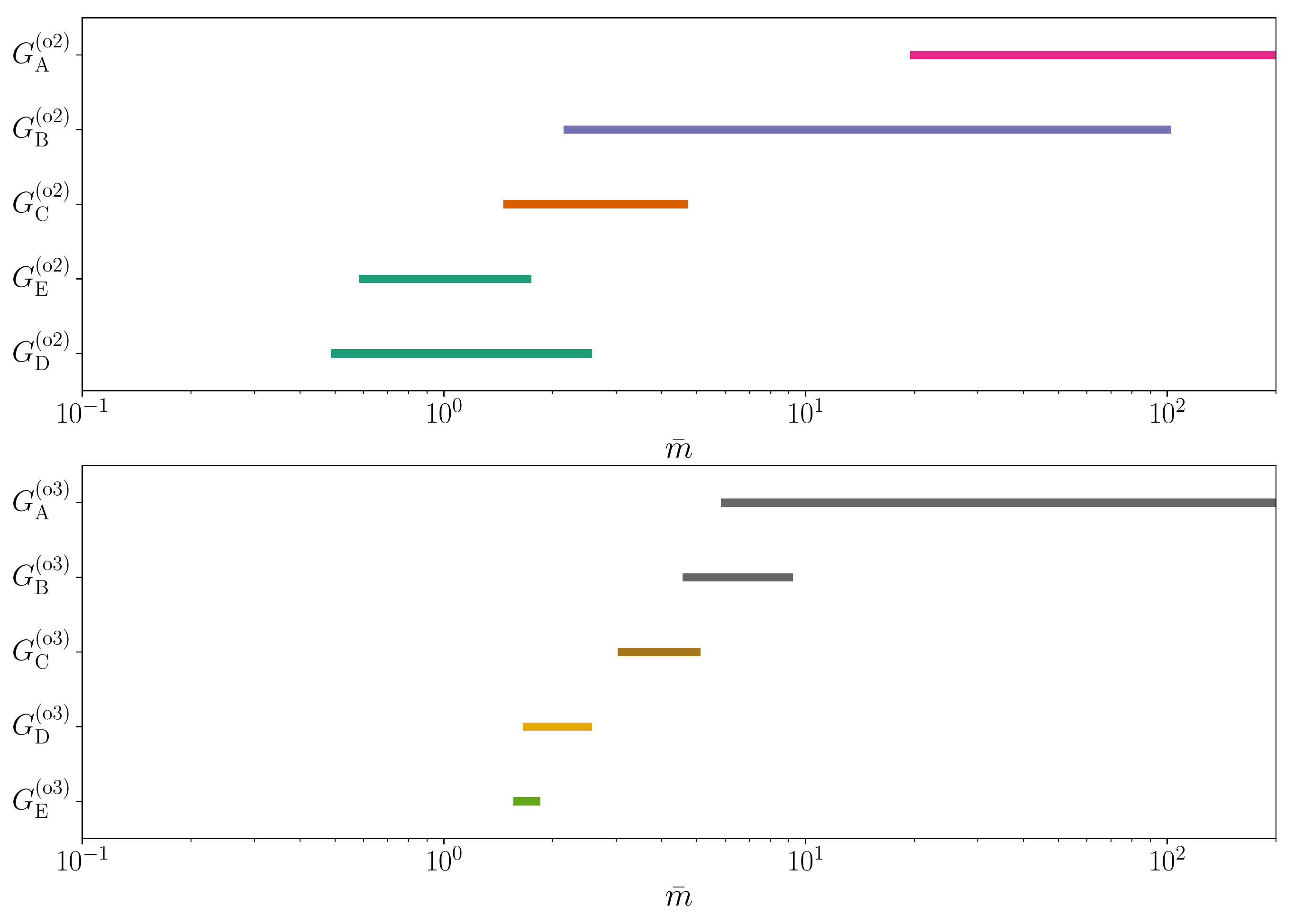}
\centering
\resizebox{.7\columnwidth}{!}{
\begin{tabular}{|c|c|c|c|c|}
\toprule
\multicolumn{5}{|c|}{Order 2} \\
\midrule 
$G_{\rm A}^{(\rm o2)} = \{\bigcup n_i \}$ &
$G_{\rm B}^{(\rm o2)} = \{\bigcup n_i \}$ &
$G_{\rm C}^{(\rm o2)} = \{\bigcup n_i \}$ &
$G_{\rm D}^{(\rm o2)} = \{\bigcup n_i \}$ &
$G_{\rm E}^{(\rm o2)} = \{\bigcup n_i \}$ \\
\midrule 
$ (\phantom{\pm} 0) $ & $ (\phantom{\pm} 0) $ & $ (\phantom{\pm} 0) $ & $ (\phantom{\pm} 0) $ & $ (\phantom{\pm} 0) $ \\
$ (         \pm  1) $ & $ (         \pm  1) $ & $ (         \pm  1) $ & $ (         \pm 13) $ & $ (         \pm 14) $ \\
$ (         \pm  2) $ & $ (         \pm  2) $ & $ (         \pm  2) $ & $ (         \pm 16) $ & $ (         \pm 15) $ \\
$ (         \pm  3) $ & $ (         \pm  3) $ & $ (         \pm  3) $ & $ (         \pm 17) $ & $ (         \pm 17) $ \\
$ (         \pm  7) $ & $ (         \pm  4) $ & $ (         \pm  5) $ & $ (         \pm 18) $ & $ (         \pm 18) $ \\
\midrule 
\multicolumn{5}{|c|}{Order 3} \\
\midrule 
$G_{\rm A}^{(\rm o3)} = \{\bigcup n_i \}$  &
$G_{\rm B}^{(\rm o3)} = \{\bigcup n_i \}$  &
$G_{\rm C}^{(\rm o3)} = \{\bigcup n_i \}$  &
$G_{\rm D}^{(\rm o3)} = \{\bigcup n_i \}$  &
$G_{\rm E}^{(\rm o3)} = \{\bigcup n_i \}$  \\
\midrule 
$ (\phantom{\pm} 0) $ & $ (\phantom{\pm} 0) $ & $ (\phantom{\pm} 0) $ & $ (\phantom{\pm} 0) $ & $ (\phantom{\pm} 0) $ \\
$ (         \pm  1) $ & $ (         \pm  1) $ & $ (         \pm  1) $ & $ (         \pm 11) $ & $ (         \pm 12) $ \\
$ (         \pm  2) $ & $ (         \pm  2) $ & $ (         \pm  5) $ & $ (         \pm 13) $ & $ (         \pm 13) $ \\
$ (         \pm  3) $ & $ (         \pm  4) $ & $ (         \pm  6) $ & $ (         \pm 15) $ & $ (         \pm 15) $ \\
$ (         \pm  4) $ & $ (         \pm  5) $ & $ (         \pm  8) $ & $ (         \pm 16) $ & $ (         \pm 16) $ \\
$ (         \pm  5) $ & $ (         \pm  6) $ & $ (         \pm  9) $ & $ (         \pm 17) $ & $ (         \pm 17) $ \\
$ (         \pm  6) $ & $ (         \pm  7) $ & $ (         \pm 10) $ & $ (         \pm 18) $ & $ (         \pm 18) $ \\
\bottomrule
\end{tabular}
}
  \caption{ Example of stencils that can be used to construct a numerically stable quadrature,
            both at the second and third order, for a RLBM in $(1+1)$ dimensions. 
            In the figure horizontal bars represent the working range of values $\bar{m}$
            of each quadrature.
           }\label{fig:quadrature-1D}
\end{table}

\newpage
\subsection{Ultra relativistic regime}\label{sec:ur-quadratures}

For the special case of massless particles we have an extra degree of freedom given by the fact 
that in this case velocity does not depend on energy. We then associate several energy shells
to each vector, thus adding a second index in the definition of the discrete momentum vectors: 
\begin{equation}
  (\tilde{p}^\mu_{i,j}) = \tilde{p}^0_j (1, \frac{n_i}{||n_i||}) \quad ,
\end{equation}
where the index $j$ labels different energy shells, and it is clear that $||n_i||$
has to be the same for all the stencil vectors since all the particles travel
at the same speed $v_i = c = 1, \forall i$. 

In the following tables we list the stencil vectors, the energy shells and the quadrature weights 
defining Gauss-type quadratures up to order $5$, in both $(3 + 1)$,  $(2 + 1)$ and  $(1 + 1)$
dimensions. In the third column of the tables, weights are listed for all energy shells of the first stencil, followed by all shells of the second stencil and so on.

\subsubsection{$(3 + 1)$ dimensions}
~~~~
\begin{longtable}[!h]{|c|c|c|}
\toprule
\multicolumn{3}{|c|}{Order 2} \\
\midrule
$G = \{\bigcup n_i \}$  & $\bar{p}_j^0$ & $w_{i j}$ \\
\midrule
  $(\pm  2,          \pm  1,          \pm  1)_{\texttt{FS}} $ & 3.3054072893322786 & 0.0245283950433191\\
  $(\pm  3, \phantom{\pm} 0, \phantom{\pm} 0)_{\texttt{FS}} $ & 0.9358222275240878 & 0\\
                                                              & 7.7587704831436335 & 0.0163006691342629\\
                                                              &                    & 0\\
                                                              &                    & 0.0003891858228425\\
                                                              &                    & 0.0017936666649682\\
\midrule
\multicolumn{3}{|c|}{Order 3} \\
\midrule
$G = \{\bigcup n_i \}$  & $\bar{p}_j^0$ & $w_{i j}$ \\
\midrule
  $(\pm  4,          \pm  4,          \pm  3)_{\texttt{FS}} $ & 0.7432919279814314 & 0\\
  $(\pm  5,          \pm  4, \phantom{\pm} 0)_{\texttt{FS}} $ & 2.5716350076462784 & 0\\
  $(\pm  6,          \pm  2,          \pm  1)_{\texttt{FS}} $ & 5.7311787516890996 & 0.0093098040253911\\
                                                              & 10.953894312683190 & 0.0085195569675087\\
                                                              &                    & 0\\
                                                              &                    & 0.0056909667738262\\
                                                              &                    & 0.0013041770173120\\
                                                              &                    & 0\\
                                                              &                    & 0.0008932820065742\\
                                                              &                    & 0.0000029126213348\\
                                                              &                    & 0.0000338363537565\\
                                                              &                    & 0.0000090390475856\\

\midrule
\multicolumn{3}{|c|}{Order 4} \\
\midrule
$G = \{\bigcup n_i \}$  & $\bar{p}_j^0$ & $w_{i j}$ \\
\midrule
  $(\pm  6,          \pm  6,          \pm  3)_{\texttt{FS}} $ & 0.6170308532782703 & 0.0035940787317887 \\
  $(\pm  7,          \pm  4,          \pm  4)_{\texttt{FS}} $ & 2.1129659585785241 & 0 \\
  $(\pm  8,          \pm  4,          \pm  1)_{\texttt{FS}} $ & 4.6108331510175324 & 0.0054532635512587 \\
  $(\pm  9,          \pm  0,          \pm  0)_{\texttt{FS}} $ & 8.3990669712048421 & 0 \\
                                                              & 14.260103065920830 & 0.0051872438667849 \\
                                                              &                    & 0 \\
                                                              &                    & 0.0078705587777011 \\
                                                              &                    & 0 \\
                                                              &                    & 0.0023465096932558 \\
                                                              &                    & 0 \\
                                                              &                    & 0.0014234434124415 \\
                                                              &                    & 0.0027124005098564 \\
                                                              &                    & 0.0001406124343838 \\
                                                              &                    & 0 \\
                                                              &                    & 0.0000921239034238 \\
                                                              &                    & 0.0001538745394240 \\
                                                              &                    & 0.0000004165349753 \\
                                                              &                    & 0.0000006474891627 \\
                                                              &                    & 0.0000008063189502 \\
                                                              &                    & 0.0000007889057767 \\

\midrule
\multicolumn{3}{|c|}{Order 5} \\
\midrule
$G = \{\bigcup n_i \}$  & $\bar{p}_j^0$ & $w_{i j}$ \\
\midrule
  $(\pm  9,          \pm  7,          \pm  4)_{\texttt{FS}} $ & 0.5276681217111288 & 0.0021976619314893 \\
  $(\pm  9,          \pm  8,          \pm  1)_{\texttt{FS}} $ & 1.7962998096434089 & 0 \\
  $(\pm 11,          \pm  4,          \pm  3)_{\texttt{FS}} $ & 3.8766415204769122 & 0.0035867160274663 \\
  $(\pm 11,          \pm  5, \phantom{\pm} 0)_{\texttt{FS}} $ & 6.9188165667047218 & 0 \\
  $(\pm 12,          \pm  1,          \pm  1)_{\texttt{FS}} $ & 11.234610429083115 & 0 \\
                                                              & 17.645963552380712 & 0.0030360121467997 \\
                                                              &                    & 0.0011645316435194 \\
                                                              &                    & 0.0060892600232298 \\
                                                              &                    & 0 \\
                                                              &                    & 0 \\
                                                              &                    & 0.0018738906904627 \\
                                                              &                    & 0.0011933134012061 \\
                                                              &                    & 0 \\
                                                              &                    & 0 \\
                                                              &                    & 0.0023241041809842 \\
                                                              &                    & 0.0001734349866413 \\
                                                              &                    & 0.0000829805751350 \\
                                                              &                    & 0.0002171069160008 \\
                                                              &                    & 0 \\
                                                              &                    & 0.0000808225598283 \\
                                                              &                    & 0.0000073565421488 \\
                                                              &                    & 0.0000007284387015 \\
                                                              &                    & 0 \\
                                                              &                    & 0.0000125232075787 \\
                                                              &                    & 0.0000031002793646 \\
                                                              &                    & 0.0000000245062369 \\
                                                              &                    & 0.0000000077527228 \\
                                                              &                    & 0.0000000187275854 \\
                                                              &                    & 0.0000000173590551 \\
                                                              &                    & 0.0000000104611619 \\
\bottomrule
\caption{Definition of quadratures up to the fifth order for a ultra-relativistic
          RLBM in (3+1) dimensions, following the parametrization for 
          the discrete momentum vectors introduced in Eq.~\ref{eq:ur-discrete-momentum-vectors}.
        }
\end{longtable}

\subsubsection{$(2 + 1)$ dimensions}
~~~~
\begin{longtable}[!h]{|c|c|c|}
\toprule
\multicolumn{3}{|c|}{Order 2} \\
\midrule
$G = \{\bigcup n_i \}$  & $\bar{p}_j^0$ & $w_{i j}$ \\
\midrule
  $(\pm  3,          \pm  4)_{\texttt{FS}} $ & 0.4157745567834790 & 0\\
  $(\pm  5, \phantom{\pm} 0)_{\texttt{FS}} $ & 2.2942803602790417 & 0.0888866262411466\\
                                             & 6.2899450829374791 & 0\\
                                             &                    & 0.0348147166961551\\
                                             &                    & 0.0017535654166088\\
                                             &                    & 0.0004218743543938\\
\midrule
\multicolumn{3}{|c|}{Order 3} \\
\midrule
$G = \{\bigcup n_i \}$  & $\bar{p}_j^0$ & $w_{i j}$ \\
\midrule
  $(\pm  3,          \pm  4)_{\texttt{FS}} $ & 0.3225476896193923 & 0\\
  $(\pm  5, \phantom{\pm} 0)_{\texttt{FS}} $ & 1.7457611011583465 & 0.0753942630427042\\
                                             & 4.5366202969211279 & 0.0410206173754781\\
                                             & 9.3950709123011331 & 0.0241670278669858\\
                                             &                    & 0.0044457884155769\\
                                             &                    & 0.0026380943565871\\
                                             &                    & 0.0000616926157132\\
                                             &                    & 0.0000365655303385\\
\midrule
\multicolumn{3}{|c|}{Order 4} \\
\midrule
$G = \{\bigcup n_i \}$  & $\bar{p}_j^0$ & $w_{i j}$ \\
\midrule
  $(\pm 15,          \pm 10)_{\texttt{FS}} $ & 0.2635603197181409 & 0.0378774109856788\\
  $(\pm 17,          \pm  6)_{\texttt{FS}} $ & 1.4134030591065167 & 0\\
  $(\pm 18,          \pm  1)_{\texttt{FS}} $ & 3.5964257710407220 & 0.0273420403371722\\
                                             & 7.0858100058588375 & 0.0289416469003179\\
                                             & 12.640800844275782 & 0\\
                                             &                    & 0.0208917044850790\\
                                             &                    & 0.0055131239981112\\
                                             &                    & 0\\
                                             &                    & 0.0039796822121021\\
                                             &                    & 0.0002621995147262\\
                                             &                    & 0\\
                                             &                    & 0.0001892703202640\\
                                             &                    & 0.0000007577431574\\
                                             &                    & 0.0000020654153959\\
                                             &                    & 0.0000000980879948\\
\midrule
\multicolumn{3}{|c|}{Order 5} \\
\midrule
$G = \{\bigcup n_i \}$  & $\bar{p}_j^0$ & $w_{i j}$ \\
\midrule
  $(\pm 15,          \pm 10)_{\texttt{FS}} $ & 0.2228466041792606 & 0.0333190352542491 \\
  $(\pm 17,          \pm  6)_{\texttt{FS}} $ & 1.1889321016726230 & 0 \\
  $(\pm 18,          \pm  1)_{\texttt{FS}} $ & 2.9927363260593140 & 0.0240515489894963 \\
                                             & 5.7751435691045105 & 0.0302726248231082 \\
                                             & 9.8374674183825899 & 0 \\
                                             & 15.982873980601701 & 0.0218524790234068 \\
                                             &                    & 0.0032794546554440 \\
                                             &                    & 0.0108922181038115 \\
                                             &                    & 0 \\
                                             &                    & 0.0006330401002278 \\
                                             &                    & 0.0002681818675293 \\
                                             &                    & 0.0003986777138864 \\
                                             &                    & 0.0000146443834593 \\
                                             &                    & 0.0000094697718432 \\
                                             &                    & 0.0000085129950491 \\
                                             &                    & 0.0000000530695690 \\
                                             &                    & 0.0000000267552665 \\
                                             &                    & 0.0000000324936526 \\                                    
\bottomrule
\caption{Definition of quadratures up to the fifth order for a ultra-relativistic
          RLBM in (2+1) dimensions, following the parametrization for 
          the discrete momentum vectors introduced in Eq.~\ref{eq:ur-discrete-momentum-vectors}.
        }
\end{longtable}

\subsubsection{$(1 + 1)$ dimensions}
~~~~
\begin{longtable}[!h]{|c|c|c|}
\toprule
\multicolumn{3}{|c|}{Order 2} \\
\midrule
$G = \{\bigcup n_i \}$  & $\bar{p}_j^0$ & $w_{i j}$ \\
\midrule
$  (\pm  1)          $  & 3.414213562373095048 & 0.042893218813452475 \\
                        & 0.585786437626904951 & 1.457106781186547524 \\
\midrule
\multicolumn{3}{|c|}{Order 3} \\
\midrule
$G = \{\bigcup n_i \}$  & $\bar{p}_j^0$ & $w_{i j}$ \\
\midrule
$  (\pm  1)          $  & 6.289945082937479196 & 0.001651724516604877 \\
                        & 0.415774556783479083 & 1.710285053107483686 \\
                        & 2.294280360279041719 & 0.121396555709244769 \\
\multicolumn{3}{|c|}{Order 4} \\
\midrule
$G = \{\bigcup n_i \}$  & $\bar{p}_j^0$ & $w_{i j}$ \\
\midrule
$  (\pm  1)          $  & 1.869968763544262523 & 0.322547689619392311 \\
                        & 0.008571999852268322 & 4.536620296921127983 \\
                        & 0.000057401877068880 & 9.395070912301133129 \\
                        & 0.204735168059733606 & 1.745761101158346575 \\
\midrule
\multicolumn{3}{|c|}{Order 5} \\
\midrule
$G = \{\bigcup n_i \}$  & $\bar{p}_j^0$ & $w_{i j}$ \\
\midrule
$  (\pm  1)          $  & 0.26356031971814091020 & 1.97964401902679930651 \\
                        & 3.59642577104072208122 & 0.02111608983931054815 \\
                        & 1.41340305910651679221 & 0.28206165857260368669 \\
                        & 7.08581000585883755692 & 0.00050971712153383997 \\
                        & 12.6408008442757826594 & 0.00000184877308595198 \\
\bottomrule
\caption{ Definition of quadratures up to the fifth order for a ultra-relativistic
          RLBM in (1+1) dimensions, following the parametrization for 
          the discrete momentum vectors introduced in Eq.~\ref{eq:ur-discrete-momentum-vectors}.
        }
\end{longtable}

\newpage
\section*{References}

\bibliography{biblio}


\end{document}